\documentclass[final,5p,times,twocolumn]{elsarticle}
\pdfoutput=1

\usepackage{graphicx}
\usepackage{subfig}

\usepackage{amsmath}
\usepackage{amssymb}

\journal{Nuclear Instruments and Methods in Physics Research A}

\begin{document}


\begin{frontmatter}



\title{Precision analysis of the photomultiplier response to ultra low signals}


\author{Pavel~Degtiarenko \corref{cor1}}
\address{Jefferson Lab, Newport News, Virginia, USA}
\ead{pavel@jlab.org}
\cortext[cor1]{ Corresponding author Tel: +1 757 269 6274, Fax: +1 757 269 6050}

\begin{abstract}

A new computational model for the description of the photon detector
response functions measured in conditions of low light is presented,
together with examples of the observed photomultiplier signal
amplitude distributions, successfully described using the
parameterized model equation. In extension to the previously known
approximations, the new model describes the underlying discrete
statistical behavior of the photoelectron cascade multiplication
processes in photon detectors with complex non-uniform gain structure
of the first dynode. Important features of the model include the
ability to represent the true single-photoelectron spectra from
different photomultipliers with a variety of parameterized shapes,
reflecting the variability in the design and in the individual
parameters of the detectors. The new software tool is available for
evaluation of the detectors' performance, response, and efficiency
parameters that may be used in various applications including the
ultra low background experiments such as the searches for Dark Matter
and rare decays, underground neutrino studies, optimizing operations
of the Cherenkov light detectors, help in the detector selection
procedures, and in the experiment simulations.

\end{abstract}

\begin{keyword}
Photon detector \sep Photomultiplier  \sep
Photoelectron  \sep  Signal amplitude spectra \sep 
Photon detection efficiency



\end{keyword}

\end{frontmatter}


%
%
%
%
\section{Introduction}

This work has been initiated by the new large-scale RICH
detector~\cite{RICH} development undertaken as a part of the CLAS12
detector upgrade~\cite{CLAS12} at Jefferson Lab, during which a large
number (more than 27 thousand) of the ultra low light detector
channels needed to be studied, selected, and characterized. Solving
this problem helped us to realize the importance of the new approach
to a wider set of applications involving the multitude of the ultra low
light detection systems.

The study revisits the problem of description and parameterization of
the photomultiplier tube (PMT) response functions measured in the
conditions of low light when only a few photoelectrons contribute to
each measured signal. Correct evaluation of the single photoelectron
(SPE) response is of significant interest for the photon detector
science and metrology. It is also critical for many applications in
the particle detector field where characterization of the detector
response and efficiency is required for data analysis, and in
astrophysics where precise photon flux measurements are vital, see,
for example, Refs.~\cite{Vicic, Blaksley}.

Several approaches to this problem have been developed and utilized,
see
Refs.~\cite{Lombard_Martin,Gale_Gibson,Rademacker,Chirikov-Zorin,Bellamy,Dossi},
and references therein. The common feature of the previous work in
this field is the use of a rather rigid functional form for the
description of the SPE spectra, such as the Poisson distribution form
in \cite{Lombard_Martin,Gale_Gibson,Rademacker,Chirikov-Zorin}, the
Gaussian form in \cite{Bellamy}, or a more complicated form of a
weighted sum of Gaussian and exponential distribution in \cite{Dossi}.
Certain types of photon detectors exhibit, however, more complicated
behavior of the spectra, see
Refs.~\cite{Wright,Bogdan,Pope,McCann,Photonis,Hamamatsu}.
Qualitatively it may be understood, for example, if the properties of
the first amplification cascade of the device (the first dynode of a
PMT) are non-uniform. This effect may be expected more visible in the
multianode photomultiplier tubes (MAPMTs) in which the area along the
edges of the first dynodes, possibly exhibiting different gain
compared to the central parts, may be relatively large. Other physics
effects and PMT design and construction features may contribute to the gain
non-uniformity.  SPE spectra in such cases can be expected to require
a larger number of parameters for their description compared to the
standard approach.

It is possible in principle to measure the SPE spectra experimentally
at very low light conditions, and then use the data to predict the
amplitude spectrum at any light \cite{Wright}.  The method is,
however, resource consuming as the measurements at a really low light
are difficult.  Attempts to extract such detailed SPE spectra
information from measurements in realistic conditions require
complicated deconvolution algorithms \cite{Pope}.

This study presents a new method of describing the SPE spectra of
virtually any reasonable complexity, therefore providing the tools for
the understanding and characterization of the photon detector response
in general. Finding a suitable structure of the SPE spectra and the
set of parameters describing experimental signal amplitude
distributions measured by the PMT photon detectors is the challenge
that this work addresses. A systematic approach and successful
solution to this problem opens better opportunities to characterize
and calibrate such photon detectors, make an educated selection of
sample devices that would work best for a particular purpose, create
new software tools simulating behavior of the photon detectors in real
installations.

%
%
%
%

\section{General definitions}

An amplitude response function of a photon detector in general, PMT in
particular, may be defined in terms of probability distributions as
described, for example, in Ref.~\cite{PDG_Prob}.  Following the
notation and terminology of \cite{PDG_Prob}, the function
$f_{PMT}(s;parameters)$ represents the parameterized probability
density function (p.d.f.) of signal amplitude (or charge) $s$.

The parameterized p.d.f. describes and may be used to approximate the
probability distribution of the observed value of $s$ in experiments
in which multiple repeatable measurements are performed in stable
conditions of constant low light delivered to the photon detector.  A
typical generalized setup for such measurements assumes that large and
stable pulses of light are generated, short enough to be measured
within the timing gates of the signal measurement system (and the
gates in turn are selected as short as reasonably possible to minimize
the noise contributions). The light pulse is then subjected to a heavy
and stable filtering such that only a few photons per pulse reached
the detector. Photons reaching the photon detector have a probability
of knocking out the photoelectrons at the detector's first stage
(photocathode in the case of a PMT), in accordance with its
photoelectron emission efficiency. The number of the photoelectrons
produced in one event is the integer random variable $m \geq 0$. The
average number of photoelectrons in one event $\langle m \rangle
\equiv \mu$ may be also defined as the ratio of the total number of
photoelectrons generated to the number of triggers.

Every measurement in such setup is triggered externally, the
resulting signal amplitude or charge is recorded, and thus the
experimentally measured distribution is accumulated as a normalized
function of $s$: $W(s) = (1/N_{\mathrm{tot}})
\mathrm{d}N/\mathrm{d}s$, where $N_{\mathrm{tot}}$ is the total
number of triggers in the run, and $\mathrm{d}N/\mathrm{d}s$ is the
accumulated experimental histogram with bin width
$\mathrm{d}s$. Function $W(s)$ ($ \equiv \mathrm{d}N/\mathrm{d}s
\ \mathrm{p.d.f.}$) constitutes, therefore, the probability density
function of observed value of $s$ during the measurements.

Such normalized experimental distributions can be compared with
$f_{PMT}(s;parameters)$, also normalized to unit area by
definition. Then the set of parameters may
be found, corresponding to the best description of the data by the
parameterized function, using, for example, the method of maximum
likelihood as described in \cite{PDG_Stat}.

The signal values $s_{\mathrm{meas}}$ are generally measured by a
signal measurement system such as the Analog (or Charge) to Digital
Converter (ADC, or QDC) devices, in units of their output
(channels). The average pedestal value of the measured signal $\langle
s_{\mathrm{ped}} \rangle$ is obtained from the events with zero number
of photoelectrons observed: $\langle s_{\mathrm{ped}} \rangle \equiv
{\langle s_{\mathrm{meas}} \rangle}_{m=0}$.  In a typical setup as
described above, a noticeable portion of the events may produce no
photoelectrons, satisfying the condition $m=0$. The resulting
measured random variable distribution on $s_{\mathrm{meas}}$ will
exhibit corresponding peak at $s_{\mathrm{meas}} = \langle
s_{\mathrm{ped}} \rangle$. The spread of the pedestal peak
corresponds to the experimental resolution of the signal measurement
system, and ideally is described by a Gaussian with the standard
deviation $\sigma$ (in channels). The pedestal spread may be
also measured in separate runs with the light source turned off, or
the light completely filtered out.

The true signal value is defined here as
\begin{equation}
\label{true_s}
  s = s_{\mathrm{meas}} - \langle s_{\mathrm{ped}} \rangle ,
\end{equation}
such that ${\langle s \rangle}_{m=0} = 0$ for events with $m = 0$. If
$m > 0$, the average signal amplitude $\langle s \rangle$ is expected
to be above zero.  By definition, at $m=1$ when only one photoelectron
is produced, the $s$ random variable will be distributed according to
the SPE spectrum $p_1(s)$ p.d.f. Average $s$ over the $p_1(s)$
p.d.f. spectrum defines the $scale$ parameter, corresponding to the
average signal value of the SPE signals:
\begin{equation}
\label{scale}
  scale = {\langle s \rangle}_{m=1} .
\end{equation}
In linear systems the parameter $scale$ is directly proportional to the 
value of the photon detector $gain$, that is, the ratio of the
measured output current to the measured current from the photocathode.

Another convenient variable for use in the further discussion is the
value of the normalized signal amplitude $a = s/scale$, such that
${\langle a \rangle}_{m=1} = 1$. The probability distribution of the $a$
random variable, $f(a;parameters)$ p.d.f., can be linked to the 
$f_{PMT}(s;parameters)$ p.d.f. through the relation 
\begin{equation}
\label{f_a_pdf}
  f(a) = scale \cdot f_{PMT}(a \cdot scale;parameters) , 
\end{equation}
to satisfy the normalization requirement
\begin{equation}
\label{Normal}
  \int\limits_{-\infty}^{\infty} f(a)\mathrm{d}a = 1.
\end{equation}
The dependence on the vector of parameters is omitted for brevity in
the $f(a)$ definition of Eq.~(\ref{f_a_pdf}), but assumed implicitly.

The probability distributions of the $a$ random variable in the events
with fixed number of photoelectrons $m \geq 0$ are defined as $p_m(a)$
p.d.f., with $p_0(a)$ characterizing the pedestal measurement, and
$p_1(a)$ being the SPE spectrum,
characteristic for the setup comprised of the photon detector and the 
signal measurement system.

The functions $p_m(a)$ are the result of the convolutions of the
intrinsic photodetector response probability distribution functions
$\rho_m(a)$ and the normalized signal measurement
system resolution function $R(a)$ such that
\begin{equation}
\label{convpm}
  p_m(a)= \int\limits_{-\infty}^{\infty} \mathrm{d}x\ R(x)\ \rho_m(a-x) \equiv \rho_m * R ,
\end{equation}
with $\rho_0(a) = \delta(a)$, and, correspondingly, $p_0(a)=R(a)$.
%
%
%
%

\section{Photomultiplier response model}
\label{Section2}

In the typical experimental setups as explained above, 
the random variable $m$ is distributed according to the Binomial p.d.f. 
\cite{PDG_Prob}. The two model assumptions of 

\begin{enumerate}[(a)]
\item stable and extremely small probability for an initial photon from the light
source to pass the heavy filtering and knock out a
photoelectron during one event, and

\item the absence of inter-dependency between
the photoelectrons
\end{enumerate}

-- guarantee that the probabilities of observing $m$
photoelectrons in one event will be distributed according to the
Poisson distribution (see Refs.~\cite{Wright, PDG_Prob}):
\begin{equation}
 P(m;\mu) = \frac{\mu^{m} e^{-\mu}}{m!}.
\end{equation}

The conditions (a) and (b) above, along with the model assumptions of

\begin{enumerate}[(a)]
\setcounter{enumi}{2}
\item negligible noise contribution,

\item linearity of the signal measurement system, and 

\item non-biased signal measurement system resolution function, 
corresponding to the condition ${\langle R(a) \rangle} = 0$
\end{enumerate}

-- allow us to unambiguously establish the relation between the value
of the $scale$ parameter, the average signal amplitude measured $
\langle s \rangle $, and the average number of photoelectrons per one
event $\mu$:
\begin{equation}
\label{scalemu}
 scale = \langle s \rangle/\mu ,
\end{equation}
which follows from the property of the Poisson distribution to have
its mean value equal to $\mu$ and the assumptions of independence,
negligible noise, linearity, and non-biased measurement. 
Correspondingly, $\langle a \rangle = \mu$.

The five model assumptions (a) through (e), which seem to be realistic
in many practical cases, are used in further discussions and model
descriptions.

From these conditions it also follows that the intrinsic p.d.f. $\rho_2(a)$,
being the distribution of the sum of two random variables each
corresponding to the intrinsic SPE spectrum $\rho_1(a)$ p.d.f., can be
explicitly expressed as their convolution
\begin{equation}
 \rho_2(a) = \rho_1 * \rho_1 \equiv {\rho_1}^{*2}(a) ,
\end{equation}
and, generally, for $m \geq 1$ photoelectrons the explicit expression
\begin{equation}
 \rho_m(a) = {\rho_1}^{*m}(a)
\end{equation}
is the convolution of $m$ intrinsic SPE functions.

Thus, the p.d.f. for the intrinsic amplitude distribution from the photodetector
becomes 
\begin{equation}
\label{phigeneral}
 \phi(a) = \sum\limits_{m=0}^{\infty} P(m;\mu) \rho_m(a),
\end{equation}
and the expression for the model p.d.f. $f(a)$ becomes
\begin{equation}
\label{fgeneral}
 f(a) =  \phi * R
 = e^{-\mu} R(a) + \sum\limits_{m=1}^{\infty} P(m;\mu)\ {\rho_1}^{*m} * R . 
\end{equation}

The function $f(a)$ satisfies the normalization requirement
(\ref{Normal}) following the normalization of the Poisson p.d.f. and
the normalization properties of the convolution algebra.  A general
textbooks on Probability and Mathematical Statistics such as
Refs.~\cite{SumOfPoissons} and \cite{MathStat} may be consulted for
the definitions and for the discussion of the convolution properties.

To find an appropriate functional form for the possible
parameterization of the intrinsic function $\rho_1(a)$, we consider the
process of electron multiplication at the second stage of the
photodetector (at the first dynode of a PMT).  Every photoelectron
hitting the first dynode has a probability of knocking one or more
second-stage electrons, which in turn will be amplified at the
following dynodes. The average number of the second-stage electrons
per one photoelectron, $\nu$ $( \equiv \nu_{\mathrm{average}})$, can
be considered a characteristic parameter of the detector. In every
event, the number of the second-stage electrons $n$ is a random
variable which characterizes the eventually measured signal $s$. Thus,
we may characterize the intrinsic SPE spectrum function $\rho_1(a)$ internally in
the model by the discrete intrinsic probability distribution of the integer
random variable $n$ with its p.d.f. being the function of $n$:
$q_1(n)$. Similarly, the discrete intrinsic functions $q_m(n)$ may be
introduced, corresponding to the continuous intrinsic signal distributions
$\rho_m(a)$.

We may also build the discrete intrinsic second stage amplitude distribution
$\phi_2(n)$ similar to Eq.~(\ref{phigeneral}):
\begin{equation}
\label{fgen_n}
 \phi_2(n) = \sum\limits_{m=0}^{\infty} P(m;\mu) q_m(n) = 
 e^{-\mu} q_0(n) + 
\sum\limits_{m=1}^{\infty} P(m;\mu) {q_1}^{*m}(n), 
\end{equation}
where $q_0(n) = 0$ for all $n$, except $n=0$, where $q_0(0) = 1$.  The
rules and properties of the convolutions of the discrete functions are
similar to the convolutions of the continuous functions, with
integration being replaced by summation.

The connection of the discrete intrinsic $\phi_2(n)$ p.d.f. to the continuous
function $f(a)$ may be derived as follows. If we assume that the
signal measurement system measures the number of second-stage
electrons $n$ directly, then the measured discrete signal distribution
can be represented as a function of the normalized amplitude $a$ in
the form of the infinite sum of correspondingly weighted
delta-functions, one per each value of $n \geq 0$:
\begin{equation}
\label{deltaf}
  D(a)= \sum\limits_{n=0}^{\infty}
  \delta \left (a - \frac{n}{\nu} \right )
   \sum\limits_{m=0}^{\infty} P(m;\mu) q_m(n) ,
\end{equation}
where $n$ in the argument of the delta-function is normalized to the
average multiplicity $\nu$ of electrons produced by a
single photoelectron at the first dynode, to provide the proper scale
for the $a$ variable, that is, to make average $a$ to be equal to one
in events with only one photoelectron.

The output of the signal measurement system, corresponding to the
resulting model function $f(a)$, constitutes the convolution of the
discrete intrinsic input spectrum of Eq.~(\ref{deltaf}) with a realistic signal
measurement resolution function $R(a)$ (often it is a Gaussian with the
standard deviation parameter $\sigma_a$). The convolution may be
performed by integrating the equation
\begin{multline}
\label{convolution}
  f(a) = \int\limits_{-\infty}^{\infty} \mathrm{d}x\ R(x)\ D(a-x) = \\
  = \int\limits_{-\infty}^{\infty} \mathrm{d}x \frac{1}{\sqrt{2
      \pi}\ \sigma_a} \exp{\left (- \frac{x^2}{2 \sigma_a^2} \right )}
  \ D(a-x).
\end{multline}

The result of the integration may be presented in the form
\begin{equation}
\label{PMTmodel}
  f(a)= \sum\limits_{n=0}^{\infty} G(a,n;\sigma_a)
   \sum\limits_{m=0}^{\infty} P(m;\mu) q_m(n) ,
\end{equation}
in which 
the probability of observing $n$ electrons exiting the first dynode 
(the inner sum over $m$ as defined in Eq.~(\ref{fgen_n})) is multiplied by the function 
\begin{equation}
 \label{Gfunction}
 G(a,n;\sigma_a) = \frac{1}{\sqrt{2 \pi} \ \sigma_a} 
  \exp{\left [- \frac{(a - n/ \nu)^{2}}{2 
\ \sigma_a^{2}} \right ]}.
\end{equation}

The connection between the continuous and discrete intrinsic signal
distributions for events at fixed $m$ may be written correspondingly:
\begin{equation}
\label{pmqm}
  p_m(a)= \sum\limits_{n=0}^{\infty} G(a,n;\sigma_a) q_m(n) .
\end{equation}

Eq. (\ref{PMTmodel}) with $G(a,n;\sigma_a)$ in the form of
Eq. (\ref{Gfunction}) corresponds to the model of a hypothetical
photon detector consisting of only two stages of multiplication.  For
a PMT, it would be the photocathode and the first dynode. Such a
device would be connected to a signal measurement system with a linear
response and the Gaussian measurement function, measuring
signals from any number of secondary electrons with the same
resolution (standard deviation parameter of the Gaussian)
$\sigma$ in channels ADC. The standard deviation $\sigma$ can
be determined from the experimental data by fitting the pedestal
amplitude distribution with a parameterized Gaussian, and the $\sigma_a$
parameter in Eq. (\ref{Gfunction}) would then be determined as 
\begin{equation}
\label{sigmaa}
\sigma_a = \sigma/scale .
\end{equation}

Realistic PMTs generally have more stages.  The third one, and the
stages that follow, may introduce extra statistical spread in the
charge collected by the ADC at any given n. This can be modeled in the
way similar to Eq.~(\ref{fgen_n}) by choosing the number of the
third-stage electrons $k$ as the integer random variable
characterizing the signal $s$. The new discrete intrinsic third stage amplitude
distribution $\phi_3(k)$ will look as follows:
\begin{equation}
\label{fgen_k}
 \phi_3(k) = \sum\limits_{m=0}^{\infty} P(m;\mu) 
 \sum\limits_{n=0}^{\infty} q_m(n) r_n(k) , 
\end{equation}
where $r_n(k)$ is the p.d.f. for the probability of observing $k$
electrons at the third stage of the PMT if the number of electrons at
the second stage is $n$. Further stages can in principle be considered
by building corresponding functions $\phi_4(l)$ ($l$ being the number of
electrons at the fourth stage), etc. Practically, however, they would
be of interest only if the signal measurement system was capable of
resolving extremely small signals corresponding to a single electron
from the corresponding stage.  In this work we limit the model at the
second stage, corresponding to the $n$ random variable. Further stages
help to define and model the additional statistical spread in
conversion of the integer scale $n$ into continuous scale $s$ of the
measured signal amplitude.

We approximate the extra statistical spread in the measured value of
$n$, assuming there is another variable in the model, parameter $\xi$,
corresponding roughly to the average number of the electrons knocked
off at the third stage (from the second dynode of the PMT) by the
electrons coming from the second stage (first dynode).  The spread in
the number of these third stage electrons is assumed to be purely
statistical when the number of third-stage electrons is reasonably
high ($ n \xi > 10 $), and is simulated using Eq.~(\ref{fgen_k})
otherwise, assuming that the gain at the fourth stage is equal to
$\xi$ also, and the statistical spread there is purely
statistical. Such approximation cannot be used at a very small $\xi <
1$. In practice the PMT cascade multiplication factors at the second 
and third dynodes are generally well above 1.

The statistical spread is implemented in the model by substituting the
$\sigma_a$ parameter in Eq.~(\ref{Gfunction}) with the new parameter
$\sigma_{\mathrm{eff}}$ which may depend on $n$,
\begin{equation}
\label{sigma_eff}
\sigma_{\mathrm{eff}}(n) = \sqrt{\sigma_a^2 + \sigma_{\xi}^2} = \sqrt{\left
  ({\frac{\sigma}{scale}}\right )^{2} +
  \frac{n}{{\nu}^2 \xi}},
\end{equation}
which is the result of adding in quadrature the normalized sigma 
as defined in Eq.~(\ref{sigmaa}), and the
parameterized spread of the measured amplitude by the third and
further amplification stages of the detector. The relative statistical
error for the value of the scaling term $n/\nu$ in
Eq.~(\ref{Gfunction}) is assumed to be
\begin{equation}
\frac{\sigma_{\xi}(n/ \nu)}{n/ \nu} =
\frac{1}{\sqrt{n \xi}}.
\end{equation}
Correspondingly, the quadrature contribution of this uncertainty to
the overall standard deviation becomes
\begin{equation}
\sigma_{\xi}^2(n/ \nu) = \frac{n}{{\nu}^2 \xi}.
\end{equation}

We note here that this approach will result in the pedestal peak in
the spectrum (at $n = 0$) being described by the Gaussian with
$\sigma_{\mathrm{eff}}(0) = \sigma_a$.  In experiments at low
light where a significant portion of events results in no
photoelectrons (corresponding to the values of $m = 0$ and $n = 0$),
say, at $\mu$ less than 2-3, the pedestal peak can be used in the
independent fit procedure to determine the value of $\sigma$.

So far we have introduced five parameters in the attempt to link the
measured experimental signal amplitude distribution
$\mathrm{d}N/\mathrm{d}s \ \mathrm{p.d.f.}$ and the parameterized
function $f(a)$, namely, $scale$, $\sigma$, $\mu$, $\nu$, and
$\xi$. The problem will fully be solved when we find appropriate
parameterized form for the function $q_1(n)$ for use in
Eq.~(\ref{fgen_n}) such that the resulting parameterized function
$f(a)$ of Eq.~(\ref{PMTmodel}) could approximate experimental data
successfully.

The simplest practical model for the production of the second-stage
electrons is the model of independent Poissonian production with
average $\nu$, assuming that every photoelectron produces the
secondary electrons independently and uniformly, as it was suggested
in the earlier studies, see
Refs.~\cite{Lombard_Martin,Gale_Gibson,Rademacker,Chirikov-Zorin}. The
explicit form for the function $q_1(n)$ in such case is the Poisson
distribution
\begin{equation}
 q_1(n) = P(n;\nu) \equiv \frac{\nu^{n} e^{-\nu}}{n!}.
\end{equation}
Rules for adding random Poissonian variables and convolution algebra
(see, for example, Refs. \cite{SumOfPoissons,MathStat}) result also in
the explicit form for the functions $q_m(n)$ at any $m \geq 1$:
\begin{equation}
\label{qmnnu}
 q_m(n;\nu) = \frac{(m\nu)^{n} e^{-m\nu}}{n!} \equiv P(n;m\nu),
\end{equation}
and the expression for the function $f(a)$ of Eq.~(\ref{PMTmodel})
becomes the double sum on $n$ and $m$:
\begin{equation}
\label{PMT_5par}
  f(a) = \sum\limits_{n=0}^{\infty} \left \{ G(a,n;\sigma_{\mathrm{eff}})
  \left [ e^{-\mu} q_0(n) + \sum\limits_{m=1}^{\infty} P(m;\mu) P(n;m\nu)
   \right ] \right \}.
\end{equation}

For a given set of parameters and at any $a$ the sum (\ref{PMT_5par})
may be evaluated numerically. Resulting function $f(a;parameters)$
could be a reasonable approximation for the
$\mathrm{d}N/\mathrm{d}a(a)$ p.d.f. for some
photodetectors. In the case of PMTs we have found that we need more
flexibility and more than one parameter to describe the second-stage
production function $q_1(n)$ satisfactorily.

Building on the above approach, we may increase the complexity and
variability of the model approximation for the function $q_1(n)$ by
assuming that several Poisson distributions with different averages
can contribute to it. Qualitatively, such pattern could be observed,
for example, in the case of a photomultiplier having a non-uniform
first dynode with distinct areas of different first dynode
gain. Generally, more parameters allow to investigate more complicated
shapes of the function $q_1(n)$. Arguably, given enough gain
components and corresponding extra free parameters, we could claim
ultimately good description of any measured spectrum by decomposing it
into a series of constituent Poisson distributions.

Assume that the discrete intrinsic SPE distribution function $q_1$ is composed
of $L \geq 1$ elementary Poissonian components such that it can be
presented in the form
\begin{equation}
\label{SPE_L}
q_1(n;{\bf v_L}) =  \sum\limits_{u=1}^L \alpha_u P(n;\nu_u) ,
\end{equation}
wherein the corresponding partial gains, or average multiplicities of
the Poissonian components are $\nu_1, ..., \nu_L $, their relative
contributions to the SPE function are $\alpha_1, ..., \alpha_L $,
satisfying the equation
\begin{equation}
\sum\limits_{u=1}^{L} \alpha_u = 1 ,
\end{equation}
and 
${\bf v_L} = (\nu_1, \alpha_2, \nu_2, ... , \alpha_L, \nu_L)$ is the
vector of parameters, with ${\bf v_1} \equiv (\nu_1)$, ${\bf v_2}
\equiv (\nu_1, \alpha_2, \nu_2)$, ${\bf v_3} \equiv (\nu_1, \alpha_2,
\nu_2, \alpha_3, \nu_3)$, etc.

In general, at any $m \geq 1$, $q_m(n;{\bf v_L})$ may be written as
\begin{equation}
\label{mPE_PDF_L}
q_m(n;{\bf v_L})= {q_1}^{*m}(n;{\bf v_L}) \equiv M_L(n,m;{\bf v_L}).
\end{equation}
The equation for the multinomial $M_L(n,m;{\bf v_L})$
function then follows from the properties of convolution powers 
(see Ref.~\cite{MathStat}) applied to $q_1(n;{\bf v_L})$:
\begin{multline}
  \label{M_L_funct}
  M_L(n,m;{\bf v_L}) 
  = \left [ \sum\limits_{u=1}^L \alpha_u P(n;\nu_u) \right ] ^{*m} = \\
  =  \sum\limits_{\substack{ i_1+...+i_L=m \\ i_1, ..., i_L\geq  0 }}^{ } 
  m! \prod\limits_{u=1}^{L} \left ( \frac{1}{i_u!} \alpha_{u}^{\,i_u}
  \right ) P(n;\nu_{cL}) ,
\end{multline}
wherein 
\begin{equation}
  \nu_{cL} = \sum\limits_{u=1}^{L} \nu_u i_u
\end{equation}
is the average multiplicity of the secondary electrons in each of the
$(i_1, ..., i_L)$ combinatorial elements contributing to the sum in
Eq. (\ref{M_L_funct}). The combinatorial sum is performed for all
$L$-dimensional combinatorial elements $(i_1, ..., i_L)$ satisfying
the conditions $i_u \geq 0$ for each $u$, and $\sum\limits_{u=1}^L i_u
= m$. See Ref.~\cite{NIST} for the definitions and for the discussion
of the multinomial coefficients in the sum.

The number $L$ of elementary Poisson distributions in the
decomposition of Eq. (\ref{SPE_L}) can be chosen to accommodate
expected or observed complexity in the SPE spectra. Larger $L$ values
would provide for more complicated spectral shapes, but require more
computing resources, as well as increase the number of variable
parameters, making the approximation process more difficult.

The explicit form for the function $M_1(n,m;{\bf v_1})$ at $L = 1$ 
is similar to that of Eq. (\ref{qmnnu}):
\begin{equation}
  M_1(n,m;{\bf v_1}) =  P(n;\nu_1 m),
\end{equation}
at $L = 2$ it can be represented as the binomial sum:
\begin{multline}
  \label{Bfunction}
  M_2(n,m;{\bf v_2}) \equiv B(n,m;{\bf b}) = \\
  = \sum\limits_{i=0}^{m} 
  \frac{m!}{i!(m-i)!} (1-\alpha_2)^i (\alpha_2)^{m-i} 
  P(n;\nu_1 i + \nu_2 m - \nu_2 i) ,
\end{multline}
and at $L = 3$ it corresponds to the trinomial sum:
\begin{multline}
  \label{Tfunction}
  M_3(n,m;{\bf v_3}) \equiv \\ \equiv T(n,m;{\bf t})
  = \sum\limits_{\substack{i_1+i_2+i_3=m \\ i_1,i_2,i_3\geq  0}}^{ } 
  \frac{m! }{i_1!\  i_2!\  i_3!} \alpha_{1}^{\,i_1} \alpha_{2}^{i_2}
  \alpha_{3}^{i_3} P(n;\nu_{c}) ,
\end{multline}
wherein 
\begin{equation}
\nu_{c} = \nu_1 i_1 + \nu_2 i_2 + \nu_3 i_3 
\end{equation}
is the average multiplicity of the secondary electrons in each of the
$(i_1,i_2,i_3)$ combinatorial elements,
and
\begin{equation}
 P(n;\nu_{c}) = \frac{(\nu_{c})^{n} \ \exp(-\nu_{c})}{n!} .
\end{equation}

The trinomial sum of Eq.~(\ref{Tfunction}) proved to be sufficient in
characterizing the measured SPE amplitude distributions in this study.

Thus, for the purpose of the approximation of the amplitude
distributions experimentally measured in PMT photon detectors we use
the following equation:
\begin{equation}
\label{PMT_9par}
  f(a;{\bf d}) = \sum\limits_{n=0}^{\infty} \left \{ G(a,n;\sigma_{\mathrm{eff}})
  \left [ e^{-\mu} q_0(n) + \sum\limits_{m=1}^{\infty} P(m;\mu) T(n,m;{\bf t})
   \right ] \right \}.
\end{equation}

The set of parameters {\bf d} includes $scale$, $\sigma$, $\mu$,
$\xi$, and the vector ${\bf t} = (\nu_1, \alpha_2, \nu_2, \alpha_3,
\nu_3)$.  The average multiplicity of the secondary electrons produced
by one photoelectron (average second stage gain) in this case will be
\begin{equation}
\label{nu_average}
 \nu = \nu_1 ( 1 - \alpha_2 - \alpha_3) + \nu_2  \alpha_2 + \nu_3  \alpha_3.
\end{equation}

The list of parameters taking full advantage of the PMT spectra
approximation by Eq.~(\ref{PMT_9par}) is given in
Table~\ref{tablepar}.  Parameter forms $\nu_2 / \nu_1$, $\alpha_3 / (1
- \alpha_2)$, and $\nu_3 / \nu_1$ are used in the computations to
simplify the fit procedure as the limits on these parameter forms can
be set universally. The original equation's symmetry between
parameters $\nu_1, \nu_2$, and $\nu_3$, and between $\alpha_1, \alpha_2$,
and $\alpha_3$ is broken in the fitting procedure to make it more
stable. The model parameters may be extracted from their table forms
unambiguously.

\begin{table}
\centering
\makeatletter \c@table=0 \makeatother
\caption{List of PMT model fit parameters}
\label{tablepar}
\begin{tabular}{llcr}
        &           &    &           \\
Name    & Limits    &    & Brief Description \\
        &           &    &            \\
$scale$ & $> 0$     & -- & average amplitude of SPE \\
        &           &    & signals (channels ADC) \\
$\sigma$& $> 0$     & -- & standard deviation of the \\
        &           &    & pedestal fit (channels ADC) \\
$\mu$   & $> 0$     & -- & average multiplicity  \\
        &           &    & of photoelectrons \\
$\nu_1$ & $> 0$     & -- & average multiplicity of the \\
        &           &    & first gain component in (\ref{SPE_L})  \\
$\alpha_2$&$[0,1]$ & -- & portion of second gain \\
        &           &    & component in (\ref{SPE_L}) \\
$\nu_2/\nu_1$ & $[0,1]$ & -- & relative multiplicity of the \\
        &           &    & second gain component in (\ref{SPE_L})\\
$\alpha_3/(1-\alpha_2)\ \ $ & $[0,1]$ & -- & relative portion of third gain \\
        &           &    & component in (\ref{SPE_L}) \\
$\nu_3/\nu_1$ & $[0,1]$ & -- & relative multiplicity of the \\
        &           &    & third  gain component in (\ref{SPE_L}) \\
$\xi$   & $ > 1$    & -- & average multiplicity at the \\
        &           &    & second dynode  \\

\end{tabular}
\end{table}

%
%
%
%

\begin{figure*}[h!] 
 \centering 
  \subfloat[H8500 MAPMT, anode \#39, test setup at low light
    conditions corresponding to $\mu~=~0.306$]{%
    \includegraphics[clip=true,trim=0 10 0
      45,width=.48\textwidth,keepaspectratio]
                    {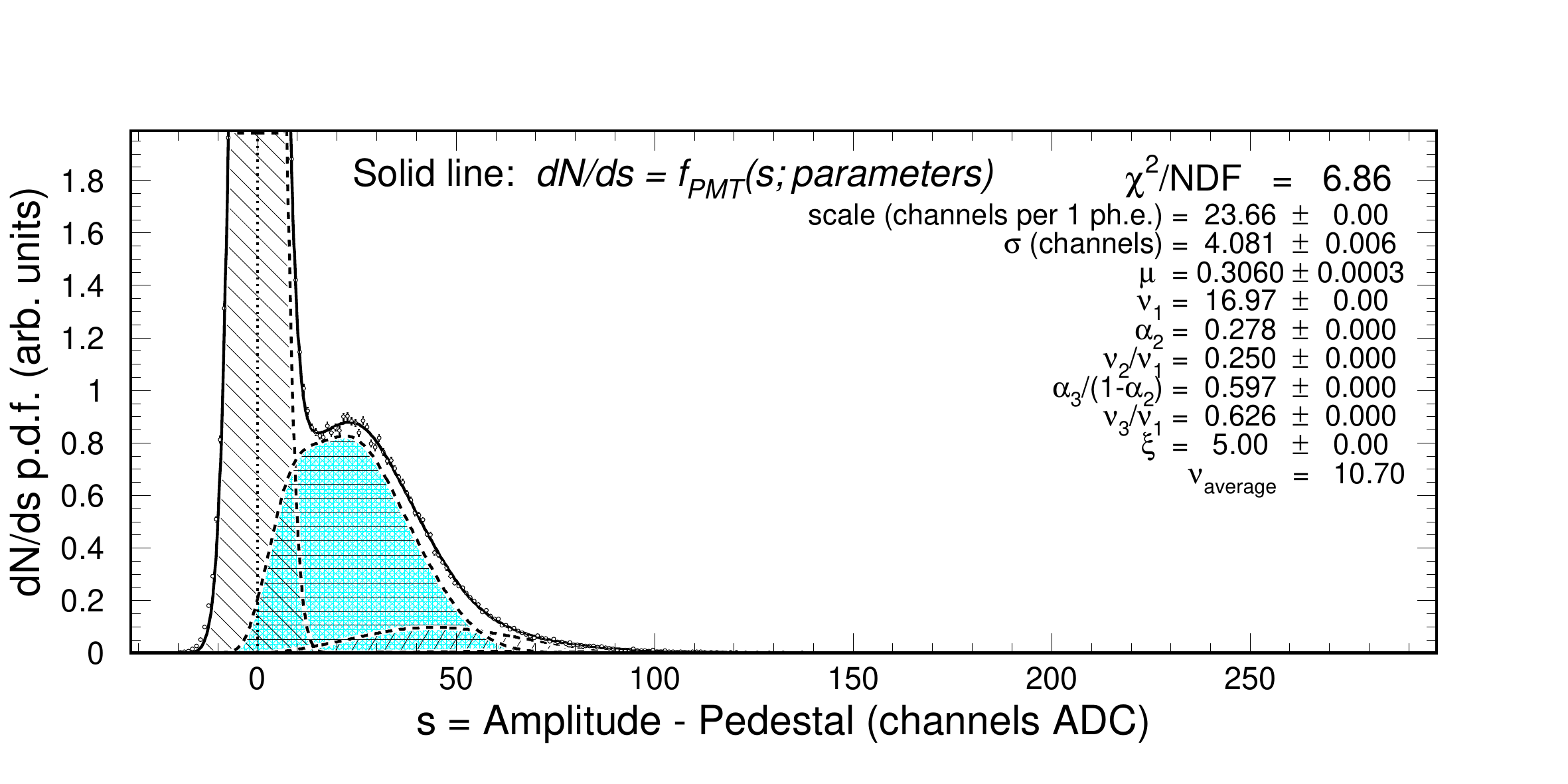}}\hfill 
  \subfloat[H8500 MAPMT, anode \#39, test setup at lower-medium light
    conditions corresponding to $\mu~=~0.869$]{%
    \includegraphics[clip=true,trim=0 10 0
      45,width=.48\textwidth,keepaspectratio]
                    {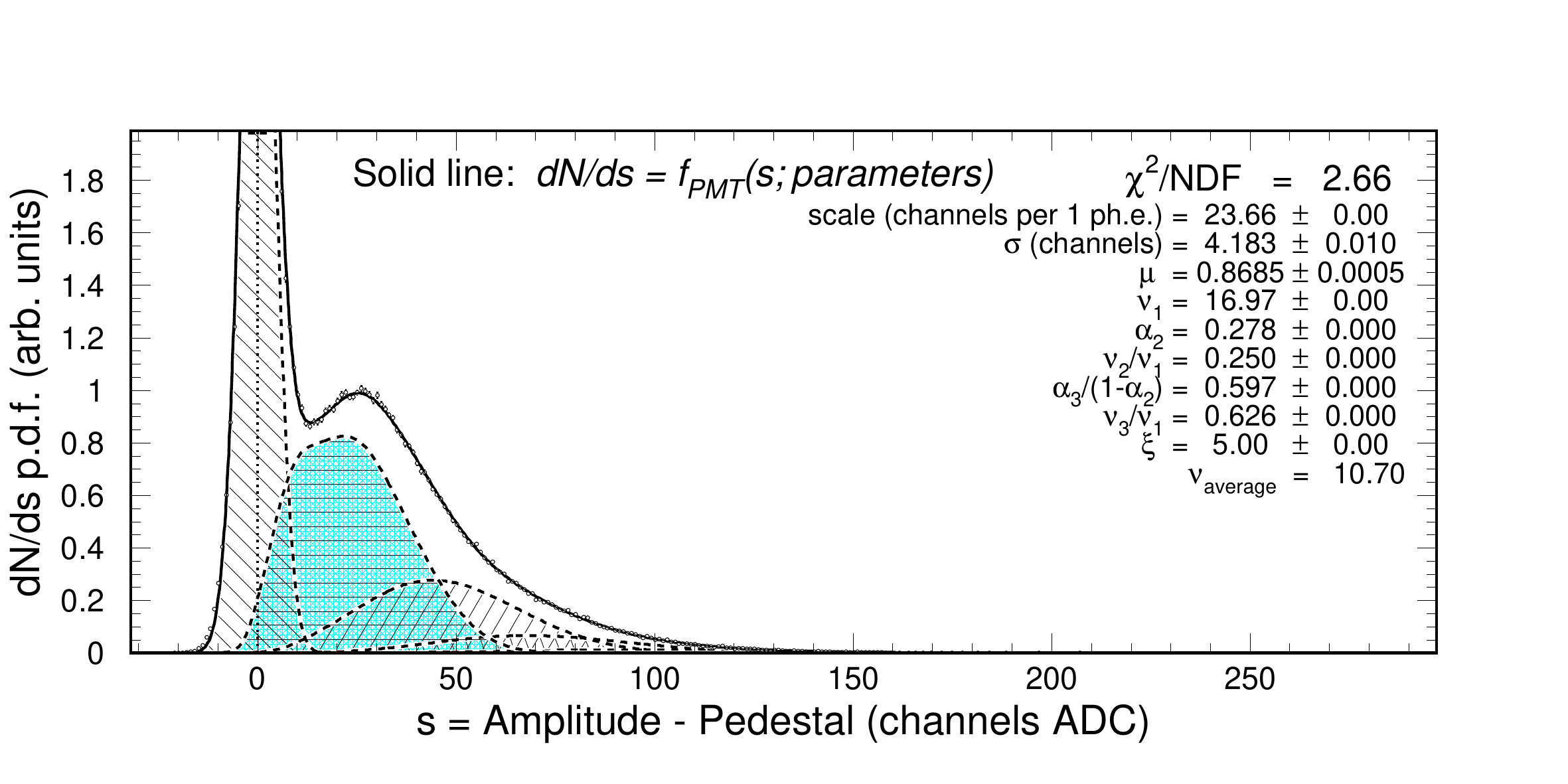}}\\
  \subfloat[H8500 MAPMT, anode \#39, test setup at upper-medium light
    conditions corresponding to $\mu~=~1.653$]{%
    \includegraphics[clip=true,trim=0 10 0
      45,width=.48\textwidth,keepaspectratio]
                    {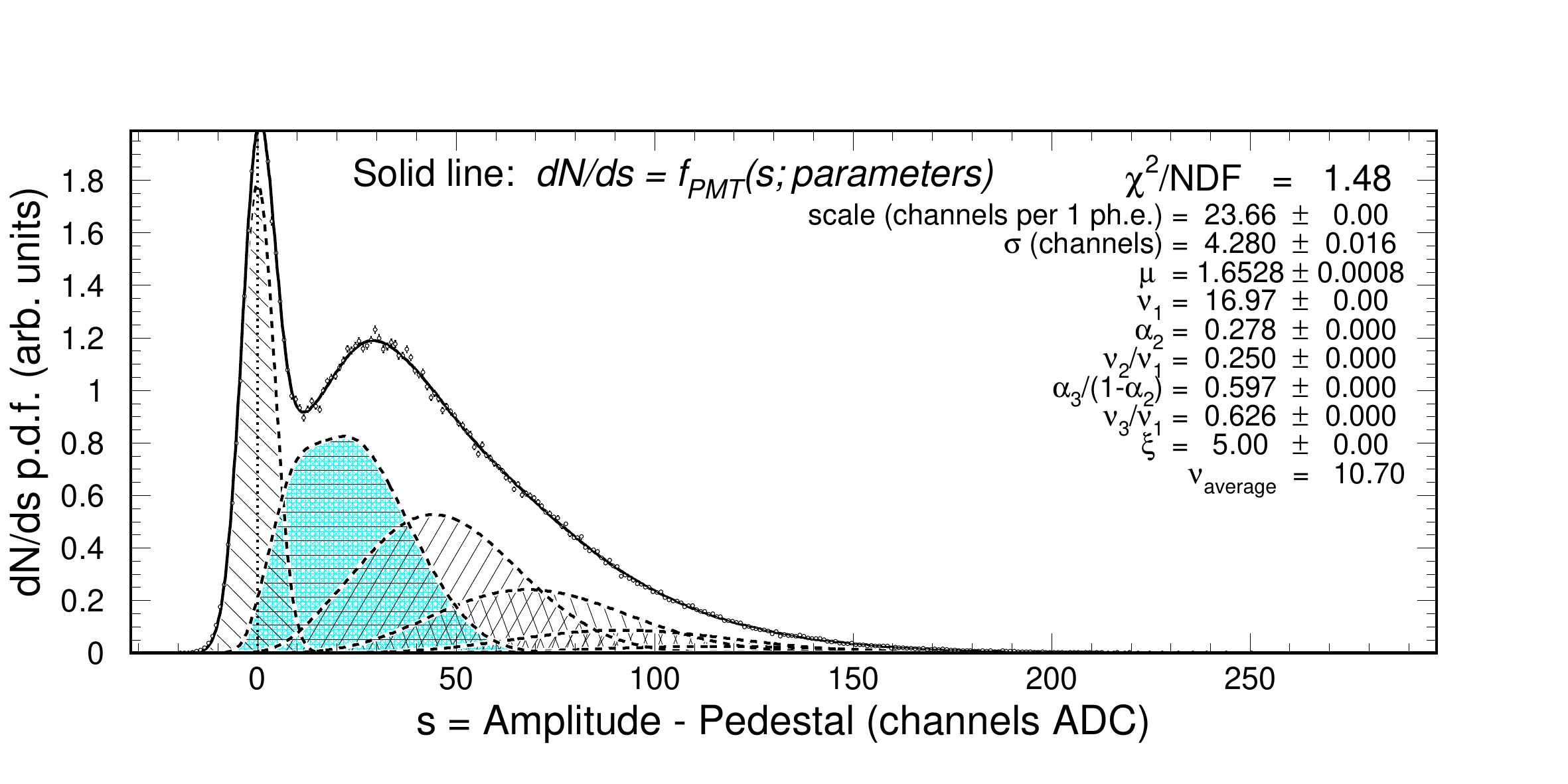}}\hfill
  \subfloat[H8500 MAPMT, anode \#39, test setup at higher light
    conditions corresponding to $\mu~=~2.734$]{%
    \includegraphics[clip=true,trim=0 10 0
      45,width=.48\textwidth,keepaspectratio]
                    {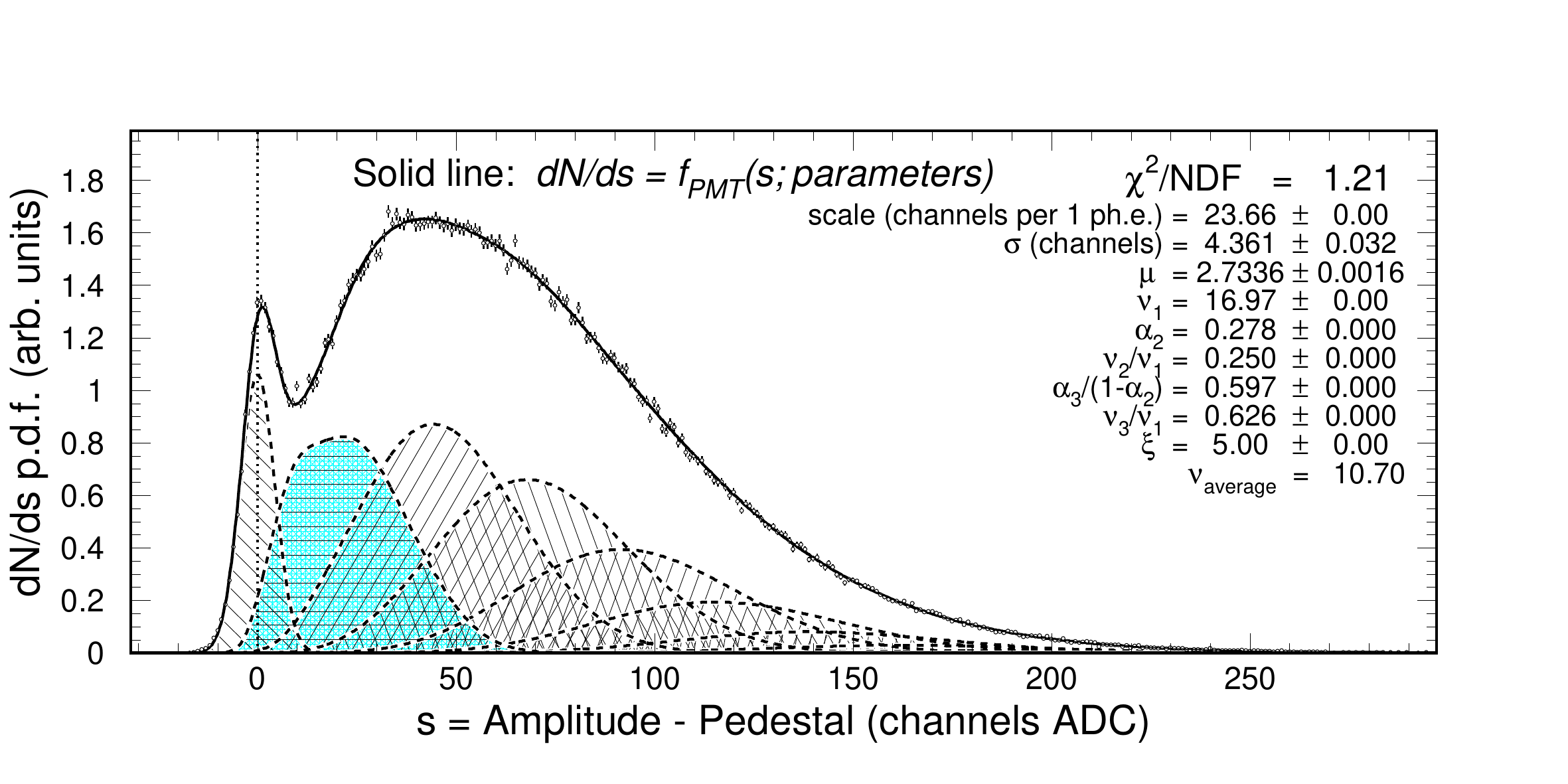}}
  \caption{A set of amplitude distributions measured with a Hamamatsu
    H8500 photomultiplier, anode \#39, at ten light conditions, four
    of which are shown.  The experimental data \cite{Malace} are shown
    as open circles with error bars, the fit function $f_{PMT}(s;
    parameters)$ is shown as a solid line, and the contributions to
    $f_{PMT}$ coming from events with zero, one, and more
    photoelectrons are shown as areas under dashed lines with
    different types of hatch.  The area corresponding to the SPE
    contribution uses horizontal lines as hatch type, and is
    highlighted. Only two parameters, $\sigma$ and $\mu$, are left variable
    in all final fits.}
\label{fig:sim_pixel_39_lin}
\end{figure*}

\section{Implementation of the model}

The fitting procedure was written in KUIP \cite{KUIP} macro language
and in FORTRAN within the framework of the Physics Analysis
Workstation (PAW) \cite{PAW} package from CERN, with the use of the
multiparametric functional minimization routine MINUIT \cite{MINUIT}.
The software development tools chosen are a bit outdated. However, the
choice of KUIP as the high-level programming language, operating
effectively with the data analysis objects, both interactively and in
the batch mode, helped significantly in the relatively quick
development of the fit algorithm and procedure. The FORTRAN code for
the fitting function and the KUIP macro language routines with the
implementation examples are available upon request.  Currently,
plans to export the code into the Root \cite{Root} environment are
under consideration.

Numerical evaluation of the function $f(a;{\bf d})$ in
Eq.~(\ref{PMT_9par}) is performed by setting finite limits of
summation over $n$ and $m$. The upper limit on $m$ in this study, at
relatively low average photoelectron multiplicities $\mu \lessapprox
3$, is set at 16. The contribution of higher values of $m$ to the sum
is negligible at such conditions. The limits of summation over $n$ are
selected dependent on the value of $a$ such that $| a - n/\nu | < 8 
\sigma_{\mathrm{eff}} $, neglecting the value of the
Gaussian $G(a,n;\sigma_{\mathrm{eff}})$ of Eq.~(\ref{Gfunction})
beyond 8 standard deviations. If the lower limit obtained from the
above condition is below zero, it is set at $n=0$. The values of the
model limiting parameters can be adjusted if needed for different
conditions, for example, higher values of $\mu$ may require using
higher upper limit on $m$.

No formal proof of the convergence of the summation method has been
developed; however, an indirect verification is done every time by
checking that the calculated function is normalized to unit area in
accordance with Eq.~(\ref{Normal}), with accuracy much better than
1\%.

As an independent verification of the implementation of the method, we
observe that the mean $a$ value for the $p_m(a)$ p.d.f. calculated
using Eq.~(\ref{pmqm}) is $\langle a \rangle = m$ as expected for all
$m \geq 0$.

In the fitting procedure, a raw measured amplitude distribution
$\mathrm{d}N/\mathrm{d}s$ is normalized to have the integral (the sum
of all channels in the histogram) to be equal to one, representing the
measured probability distribution $\mathrm{d}N/\mathrm{d}s
\ \mathrm{p.d.f.}$, to be approximated by the model probability
density function $f_{PMT}(s;{\bf d})$. The first guess of the values
of $ \langle s_{\mathrm{ped}} \rangle $ and $\sigma$ is made based on
the Gaussian fit of the left side and top of the first peak in the
distribution, representing events with $n = 0$. The average amplitude
$ \langle s \rangle $ is then calculated together with the initial
estimate of $\mu$ parameter to obtain the initial value of the $scale$
parameter, which allows us to present the probability distribution as
a function of normalized signal amplitude $a$. After that, the data
set is used in the process of finding the best set of parameters
describing it in the form of Eq.~(\ref{PMT_9par}), using MINUIT.

The stability of the multiparametric fitting procedure strongly
depends on the right choice of the parameters' initial values.  In the
following examples, different strategies were implemented to achieve
such stability, generally by splitting the process into several
stages, starting with the separate fit of the pedestal Gaussian to
determine the pedestal position and standard deviation, then setting
the initial value of $\mu$ by evaluating the portion of events in the
pedestal region and using the assumption that it is equal to
$\exp{(-\mu)}$, and then gradually increasing the number of variable
parameters in the consequent fits.

In the examples that included measurements of amplitude distributions
from the same photodetector in identical conditions, only varying the
amount of light delivered to the detector per one measurement, the
procedure included the next layer of a ``global fit''. After the best
set of parameters describing each individual measurement was found,
the parameters related to the SPE amplitude distribution were averaged
across the set and fixed at those values. The remaining ``external''
parameters (such as $scale$, $\mu$, and $\sigma$) were set free for
the subsequent fit. The cycle of fitting procedures starting with
releasing all the parameters and making a new fit, averaging the SPE
parameters and fixing them at the new values, and then re-fitting only
``external'' parameters was performed several times until final
convergence was reached.  The quality of the resulting approximation
is an indication that the parameters of the SPE distribution were
found correctly and may serve as values characterizing the
device. These data sets illustrate the ``predictive power'' of the
model, that is, its ability to predict how the amplitude distributions
would look in different experiments with the same PMT (at different
light conditions, and with different resolution of the signal
measurement, for example).

%
%
%
%

\begin{figure*}[!h] 
\centering 
  \subfloat[H8500 MAPMT, anode \#45, test setup at low light
    conditions corresponding to $\mu~=~0.256$]{%
    \includegraphics[clip=true,trim=0 10 0
      45,width=.48\textwidth,keepaspectratio]
                    {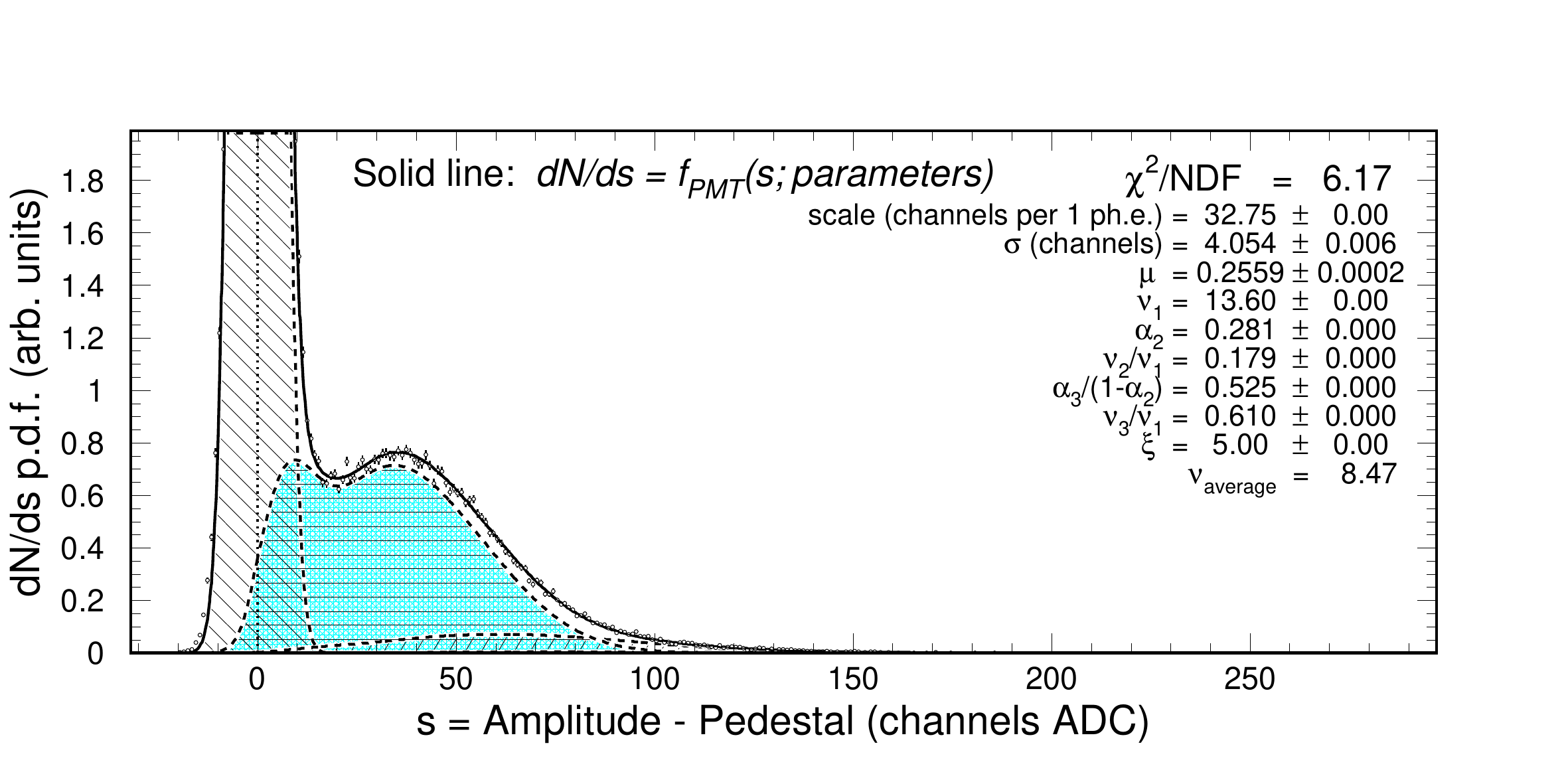}}\hfill
  \subfloat[H8500 MAPMT, anode \#45, test setup at lower-medium light
    conditions corresponding to $\mu~=~0.728$]{%
    \includegraphics[clip=true,trim=0 10 0
      45,width=.48\textwidth,keepaspectratio]
                    {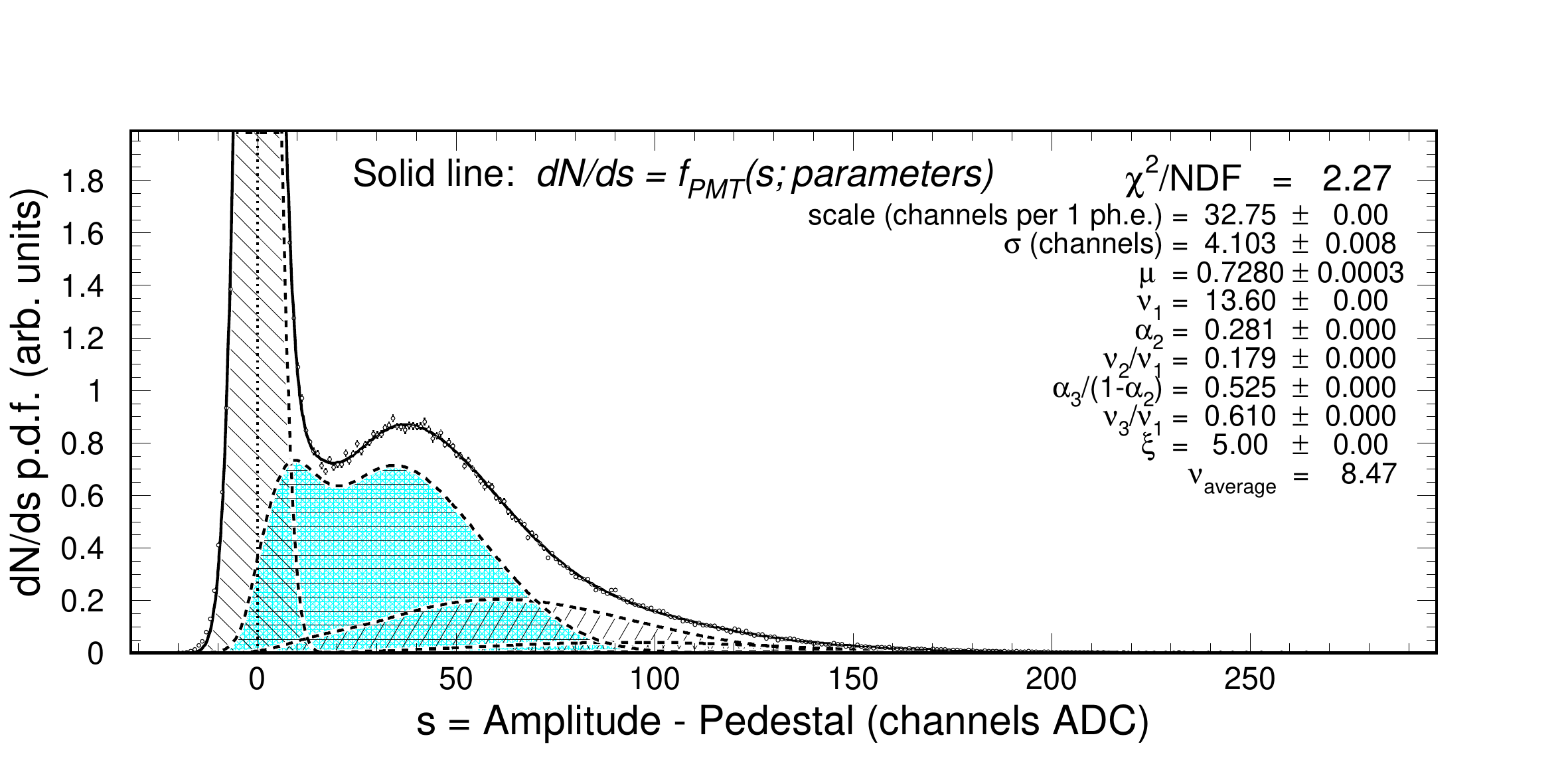}}\\
  \subfloat[H8500 MAPMT, anode \#45, test setup at upper-medium light
    conditions corresponding to $\mu~=~1.383$]{%
    \includegraphics[clip=true,trim=0 10 0
      45,width=.48\textwidth,keepaspectratio]
                    {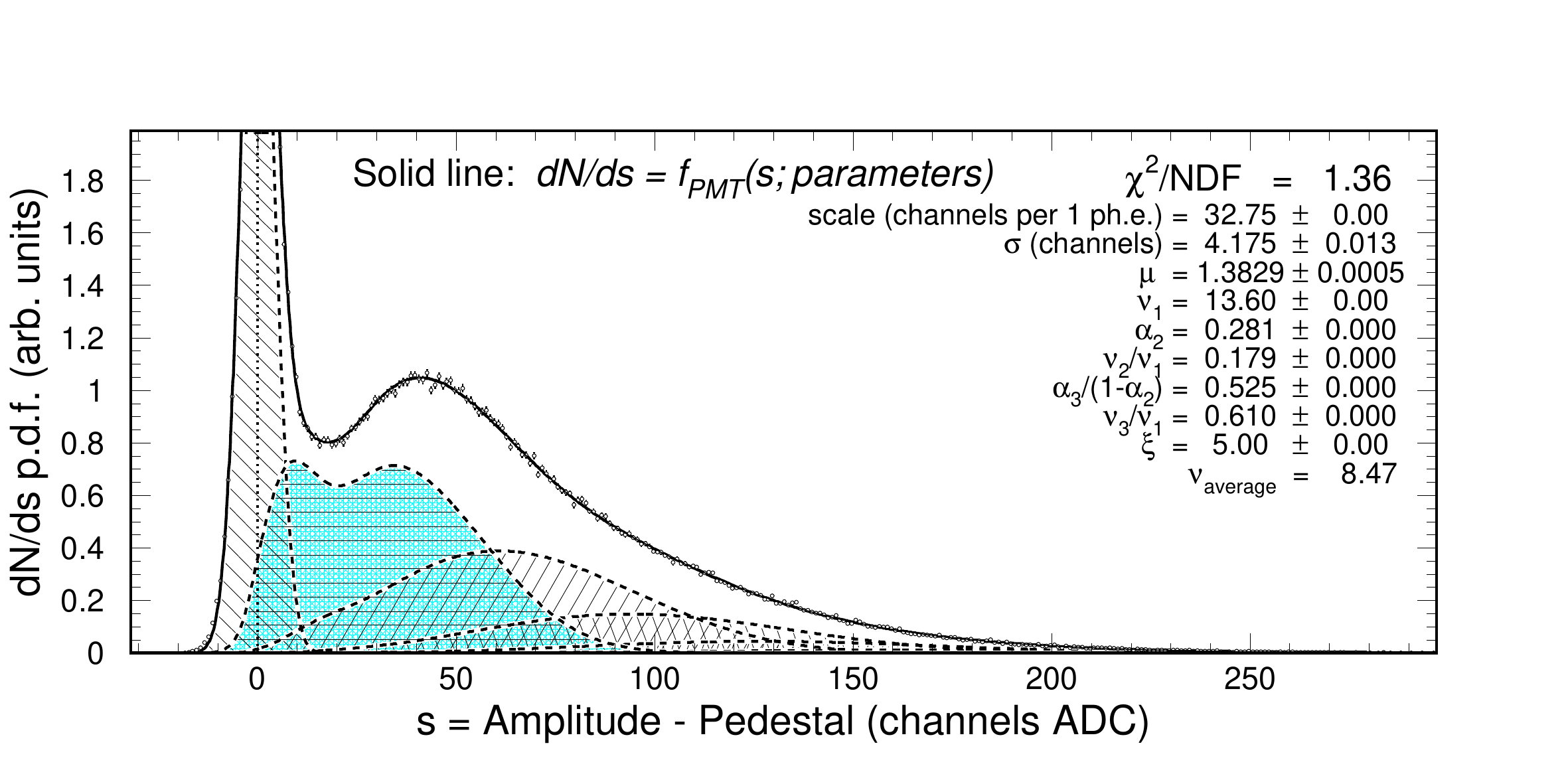}}\hfill
  \subfloat[H8500 MAPMT, anode \#45, test setup at higher light
    conditions corresponding to $\mu~=~2.285$]{%
    \includegraphics[clip=true,trim=0 10 0
      45,width=.48\textwidth,keepaspectratio]
                    {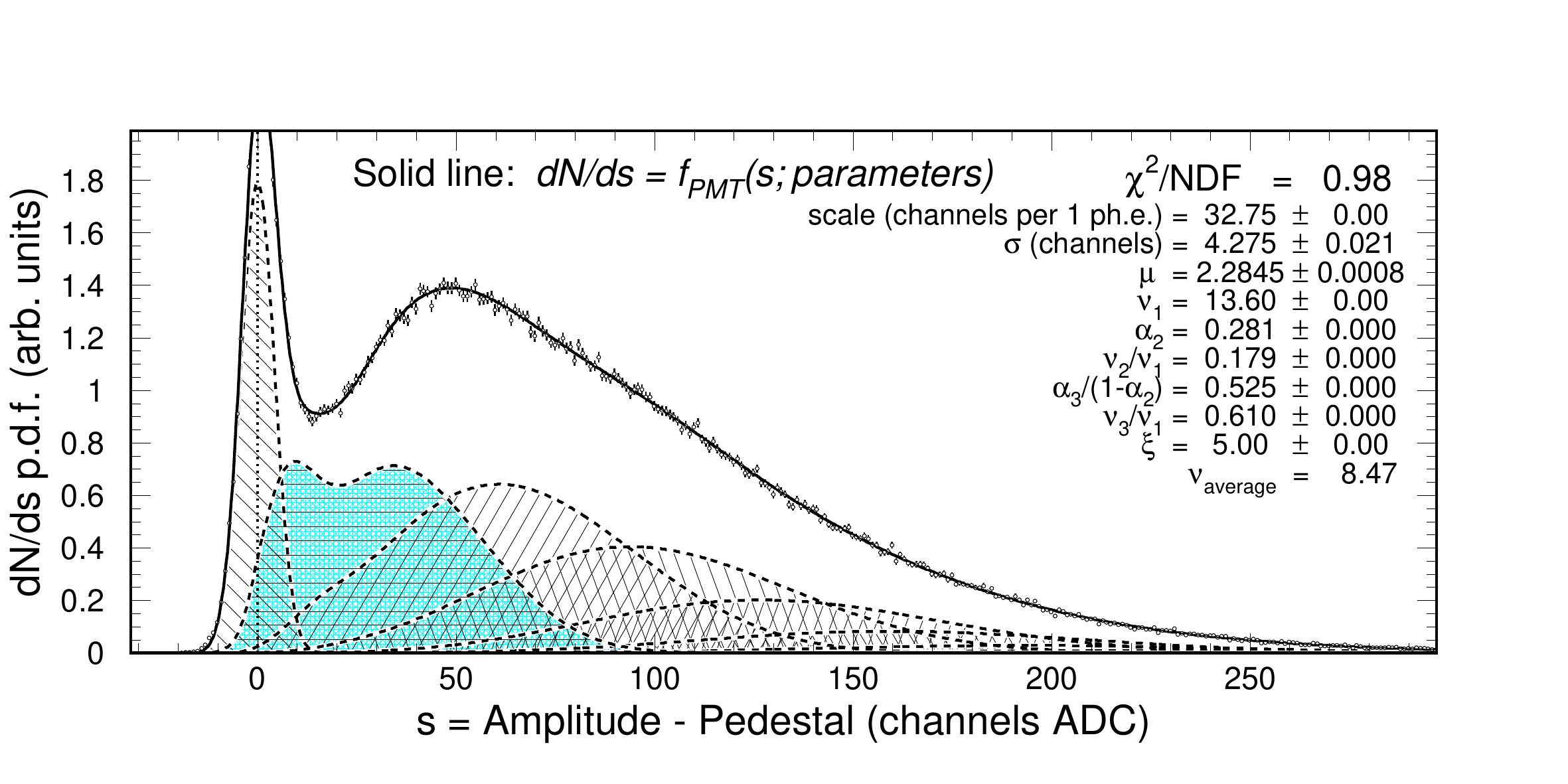}}
  \caption{A set of amplitude distributions measured with a Hamamatsu
    H8500 photomultiplier, similar to the set shown in
    Fig.~\ref{fig:sim_pixel_39_lin}, but on different anode \#45.  Ten
    measured distributions participated in the ``global fit''
    procedure; four of them are shown.}
\label{fig:sim_pixel_45_lin}
\end{figure*}

\begin{figure*}[h!] 
\centering 
  \subfloat[H8500 MAPMT, anode \#61, test setup at low light
    conditions corresponding to $\mu~=~0.388$]{%
    \includegraphics[clip=true,trim=0 10 0
      45,width=.48\textwidth,keepaspectratio]
                    {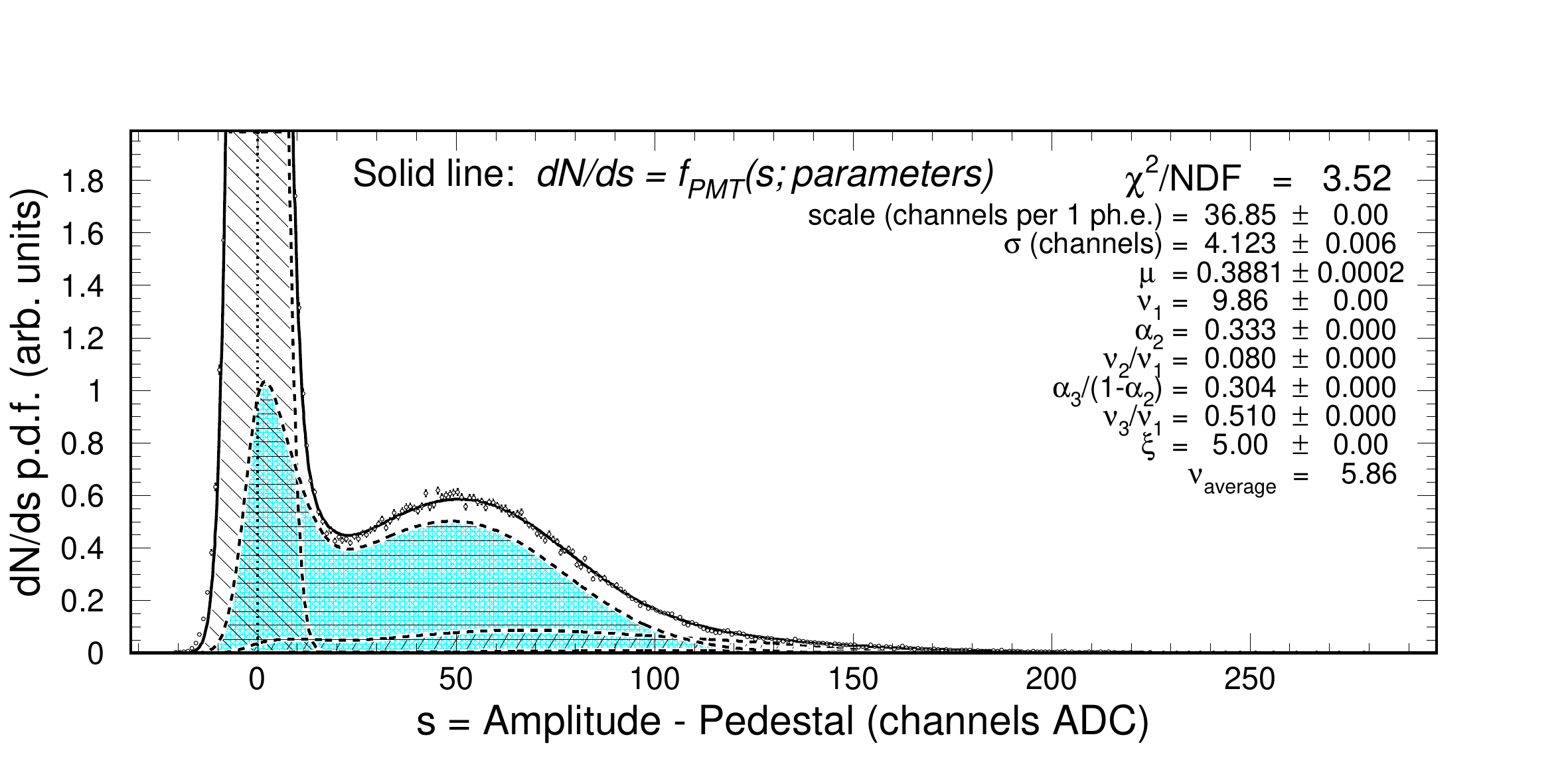}}\hfill
  \subfloat[H8500 MAPMT, anode \#61, test setup at lower-medium light
    conditions corresponding to $\mu~=~0.827$]{%
    \includegraphics[clip=true,trim=0 10 0
      45,width=.48\textwidth,keepaspectratio]
                    {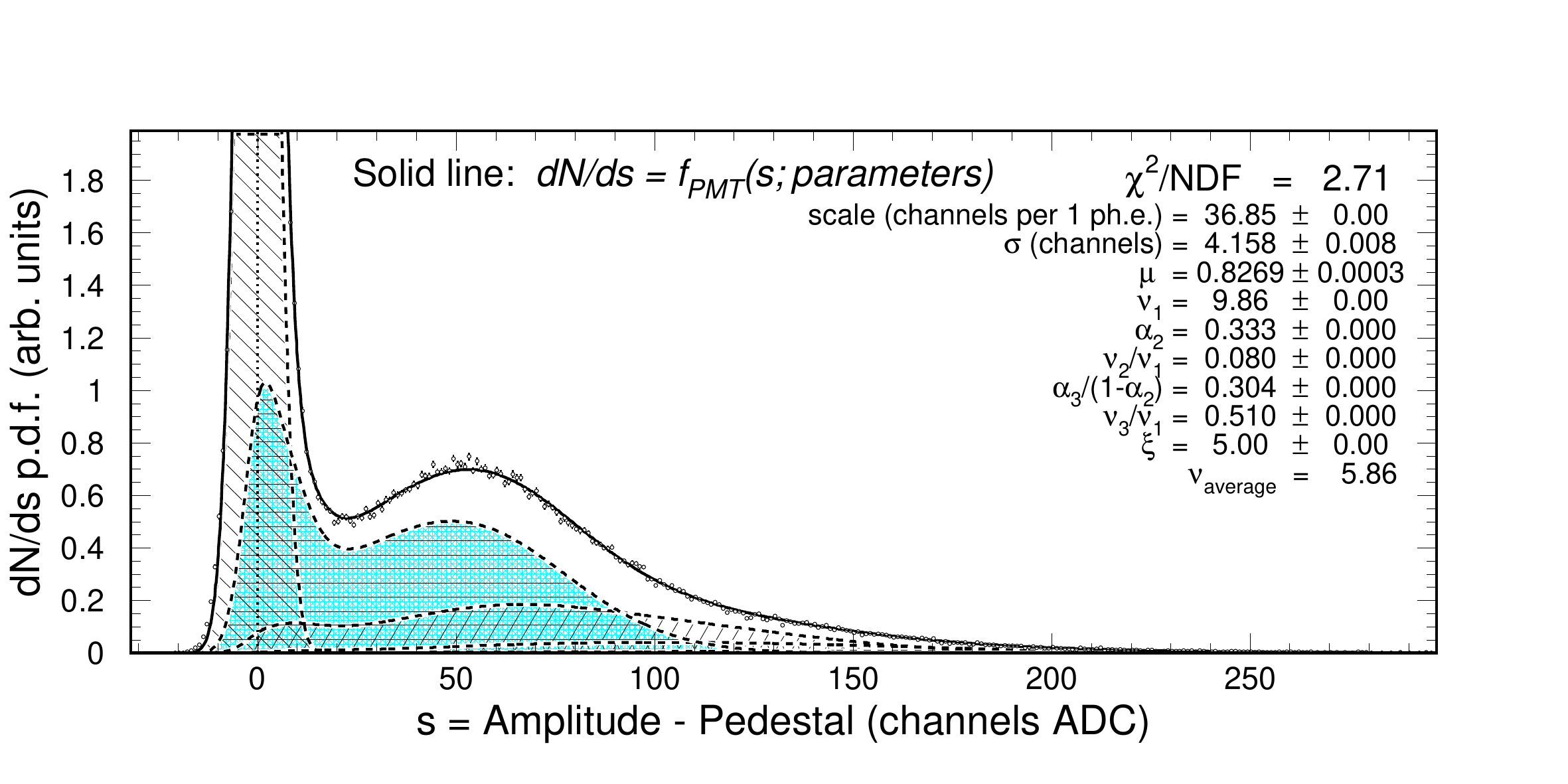}}\\
  \subfloat[H8500 MAPMT, anode \#61, test setup at upper-medium light
    conditions corresponding to $\mu~=~1.636$]{%
    \includegraphics[clip=true,trim=0 10 0
      45,width=.48\textwidth,keepaspectratio]
                    {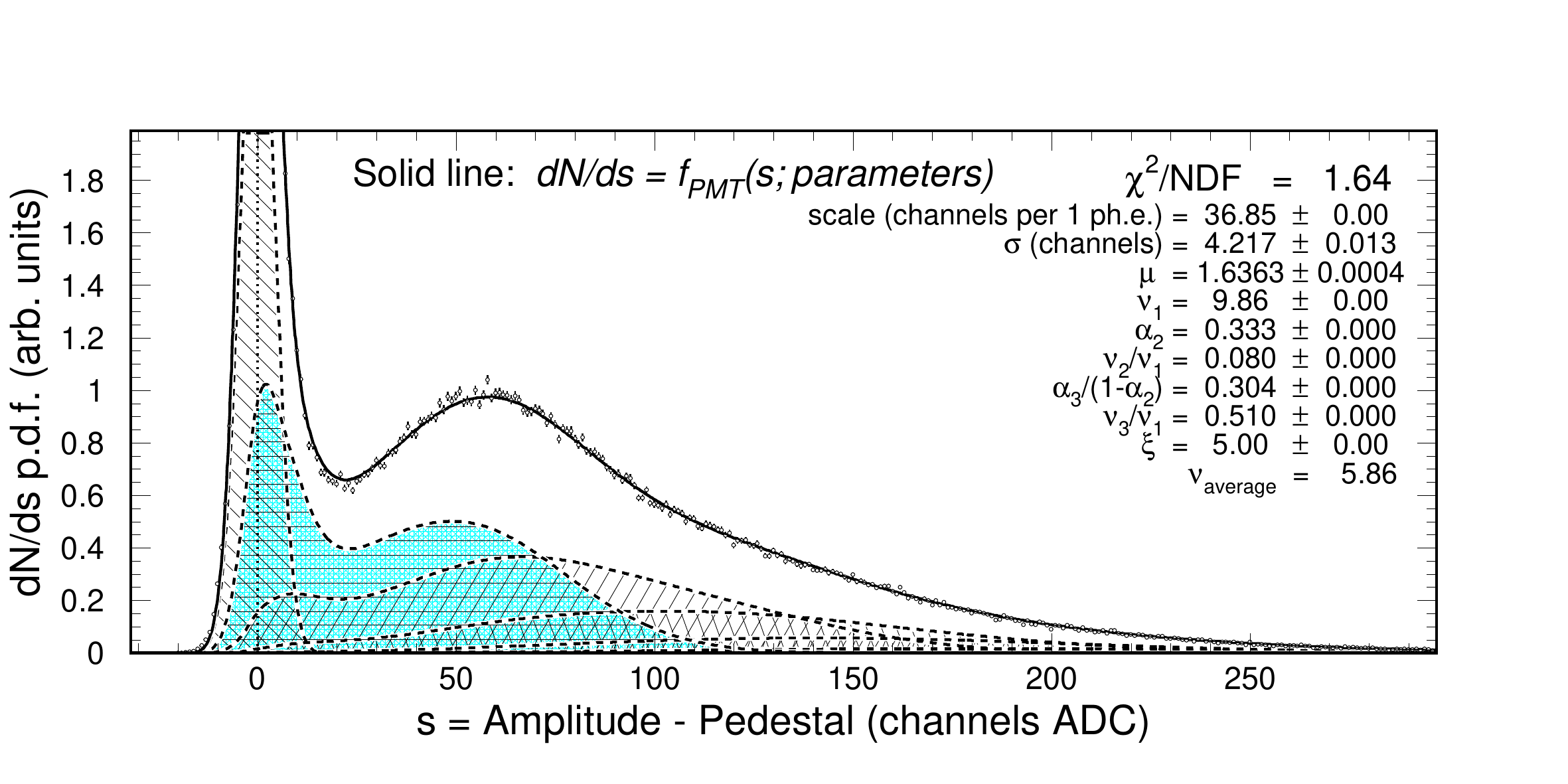}}\hfill
  \subfloat[H8500 MAPMT, anode \#61, test setup at higher light
    conditions corresponding to $\mu~=~2.742$]{%
    \includegraphics[clip=true,trim=0 10 0
      45,width=.48\textwidth,keepaspectratio]
                    {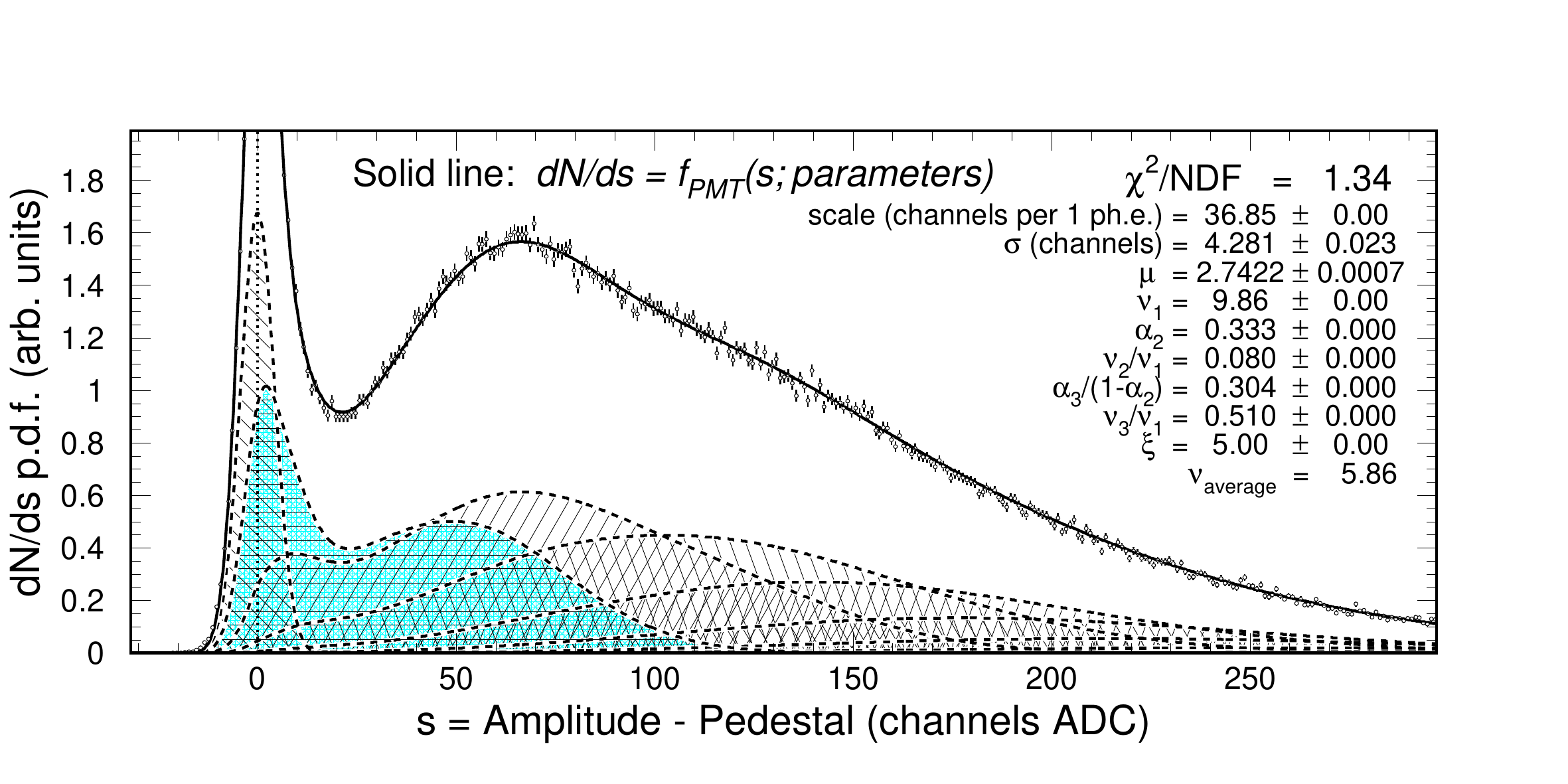}}
  \caption{A set of amplitude distributions measured with a Hamamatsu
    H8500 photomultiplier, similar to the set shown in
    Fig.~\ref{fig:sim_pixel_39_lin}, but on different anode \#61. Nine
    measured distributions participated in the ``global fit''
    procedure; four of them are shown.}
\label{fig:sim_pixel_61_lin}
\end{figure*}

\begin{figure*}[h!] 
\centering 
  \subfloat[H8500 MAPMT, anode \#45, test setup at low light
    conditions corresponding to $\mu~=~0.256$]{%
    \includegraphics[clip=true,trim=0 10 0
      45,width=.48\textwidth,keepaspectratio]
                    {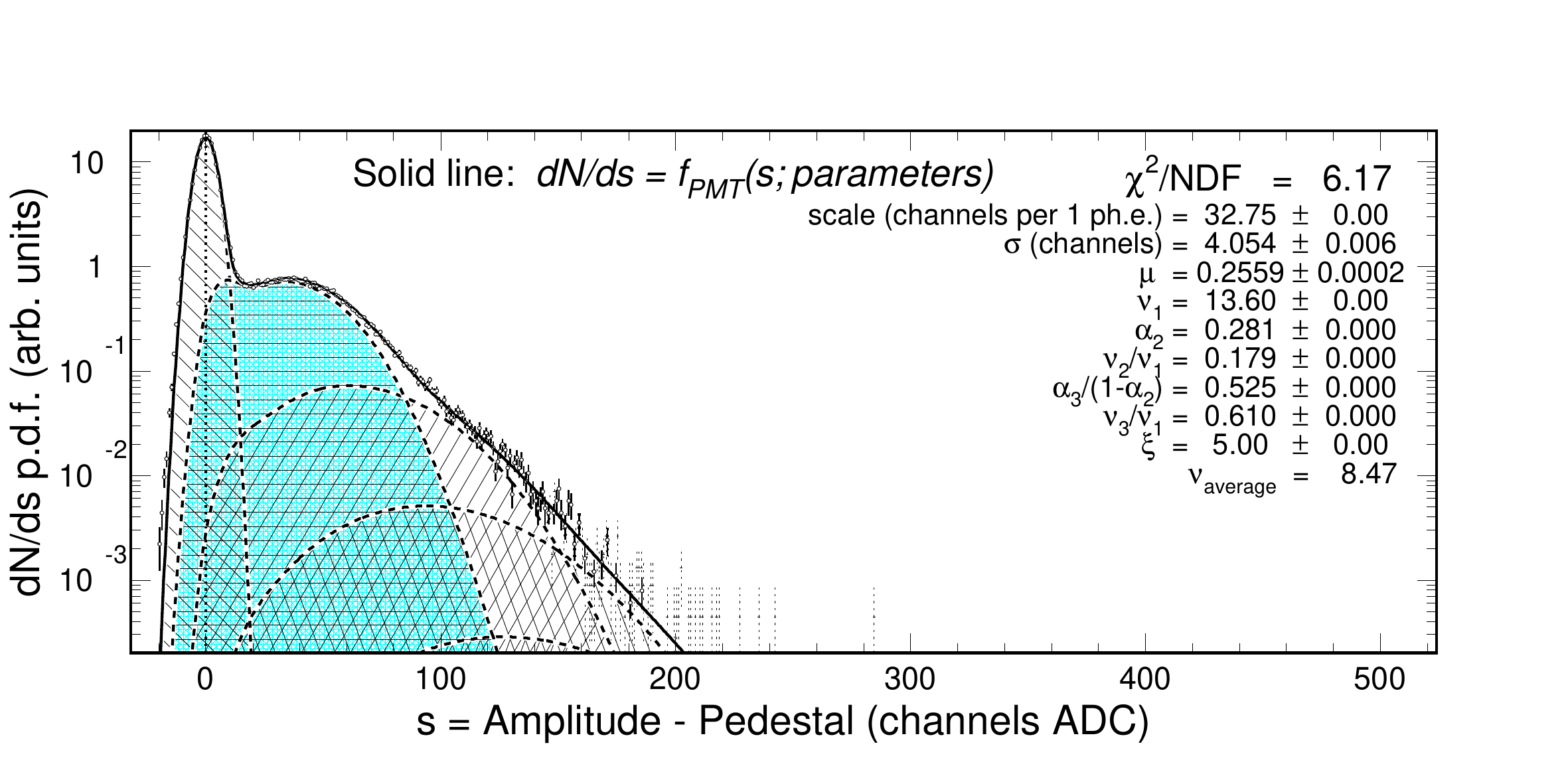}}\hfill
  \subfloat[H8500 MAPMT, anode \#45, test setup at lower-medium light
    conditions corresponding to $\mu~=~0.728$]{%
    \includegraphics[clip=true,trim=0 10 0
      45,width=.48\textwidth,keepaspectratio]
                    {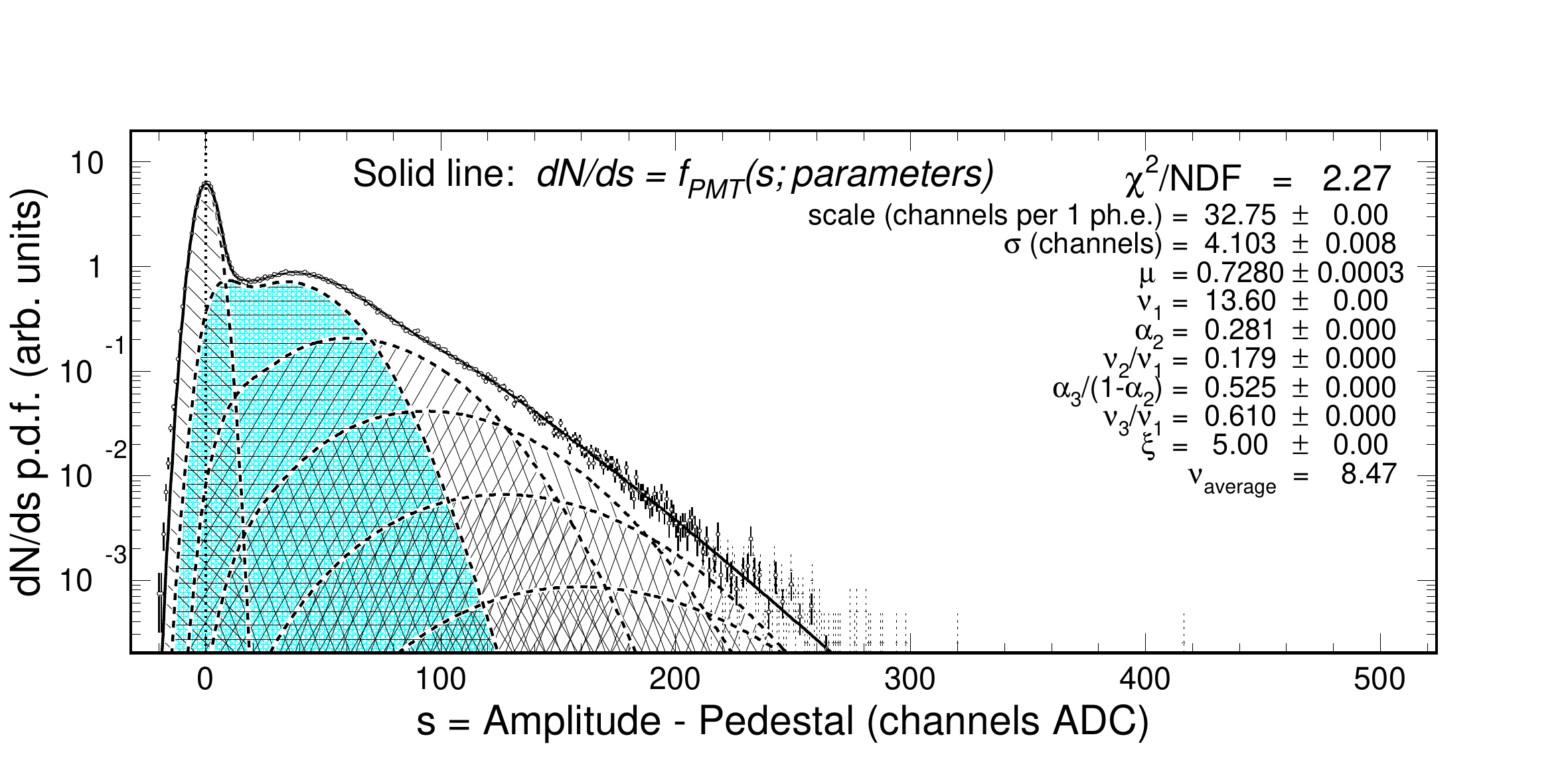}}\\
  \subfloat[H8500 MAPMT, anode \#45, test setup at upper-medium light
    conditions corresponding to $\mu~=~1.383$]{%
    \includegraphics[clip=true,trim=0 10 0
      45,width=.48\textwidth,keepaspectratio]
                    {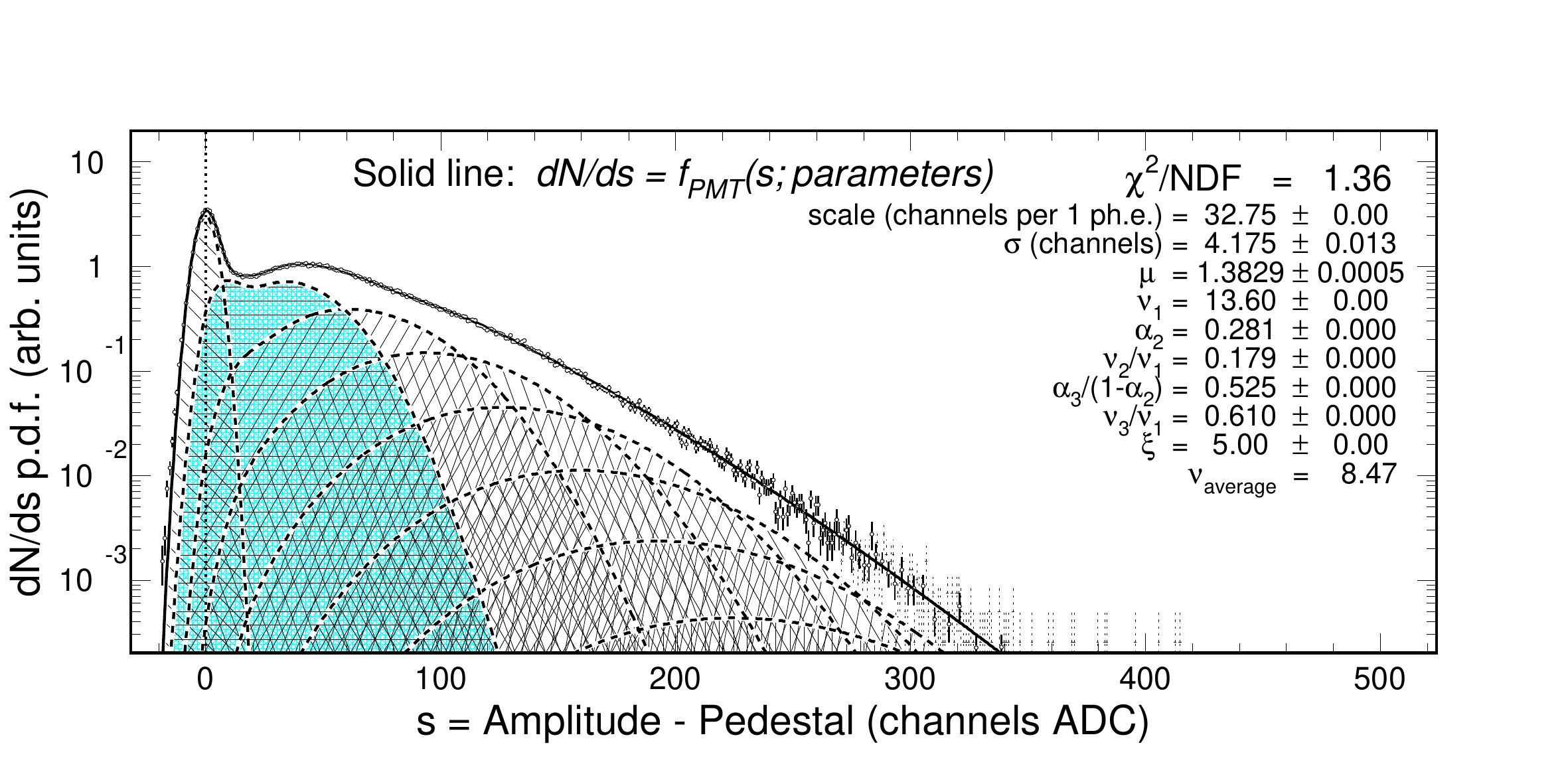}}\hfill
  \subfloat[H8500 MAPMT, anode \#45, test setup at higher light
    conditions corresponding to $\mu~=~2.285$]{%
    \includegraphics[clip=true,trim=0 10 0
      45,width=.48\textwidth,keepaspectratio]
                    {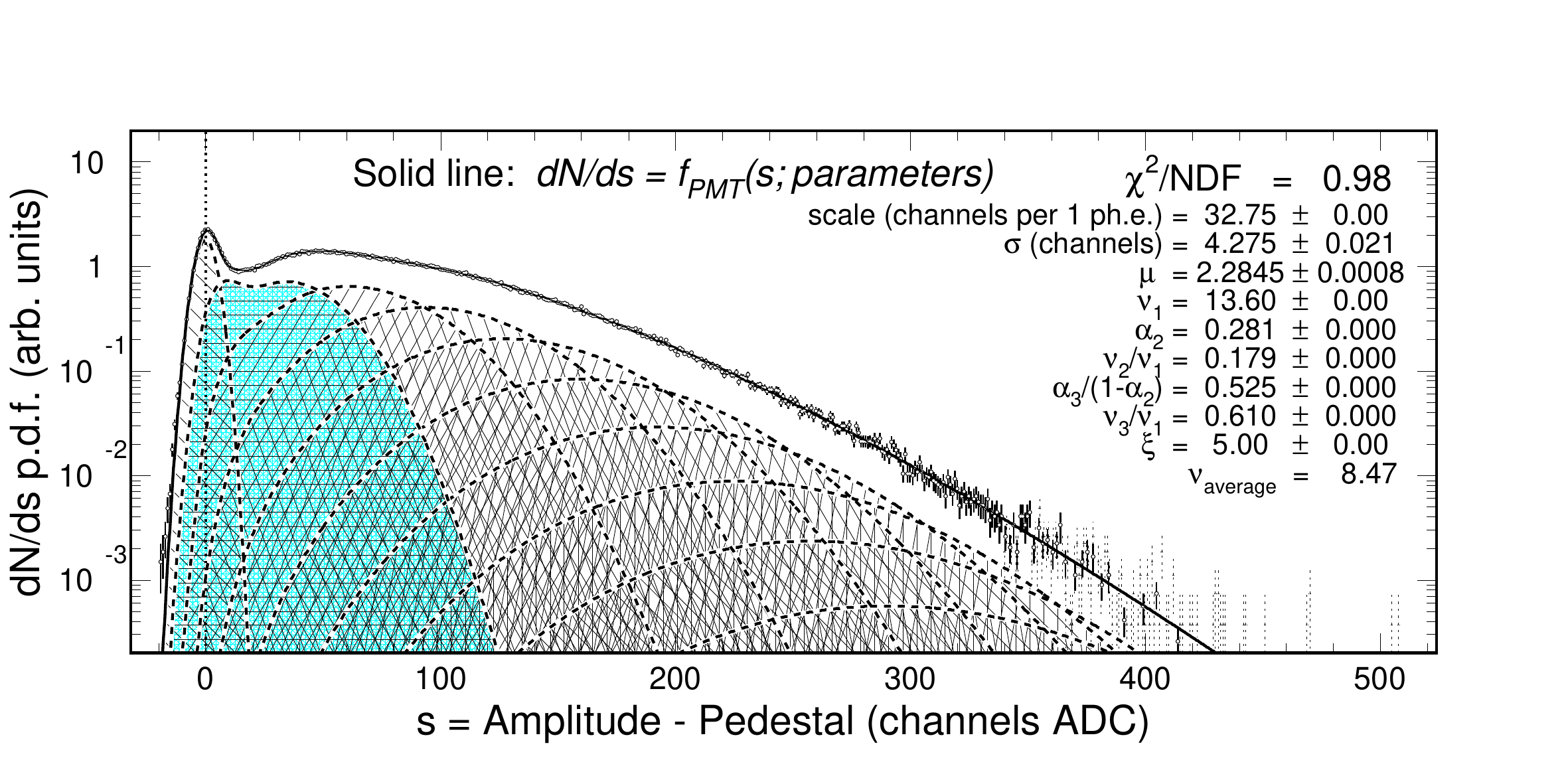}}
  \caption{Same as in Fig.~\ref{fig:sim_pixel_45_lin}, but using
    logarithmic scale in ordinate to illustrate the contribution of
    higher $m$ components in the spectra.}
\label{fig:sim_pixel_45_log}
\end{figure*}

\section{Examples}

This section provides examples of practical applications of the model
used for parameterizations of real signal amplitude spectra 
measured in various conditions and with different types of photomultipliers.

\subsection{Tests of Hamamatsu H8500C-03 Multianode PMT}

Figures \ref{fig:sim_pixel_39_lin}, \ref{fig:sim_pixel_45_lin}, and
\ref{fig:sim_pixel_61_lin} illustrate the general quality of the model
description of the amplitude distributions measured on three different
anodes of the position sensitive Hamamatsu MAPMT
H8500C-03, a 52 mm square 12-stage photomultiplier tube with 64
(matrix 8 by 8) pixels \cite{HamamatsuPMT}.  The measurement was a
part of the dedicated study of the SPE detection capabilities of this
PMT and its performance in a high magnetic field \cite{Malace}. The
spectra from each anode were accumulated in several irradiation
conditions, corresponding to the range of $\mu$ from about 0.3 to 3.
The raw data sets were kindly presented to us by the Authors of
Ref.~\cite{Malace} for the analysis.  Every spectrum was subjected to
the two-level ``global fit'' procedure as described in the previous
section. The fitting function is visibly following the data points
nicely above the pedestal. The values of $\chi^2/NDF$, or
$\chi^2/n_{\mathrm{d}}$ as per \cite{PDG_Stat}, corresponding to the
formal goodness-of-fit statistical evaluators, are mostly determined
by the quality of the Gaussian approximation for the signal
measurement system resolution function in this experiment. At
low-light setups, where the pedestal events dominate, the non-Gaussian
contributions to the shape of the resolution functions increase the
values of $\chi^2/n_{\mathrm{d}}$, but do not disturb significantly
the SPE spectra parameterizations in this example.

We found that the best-fit parameters of the SPE amplitude
distributions, while different for different anodes, are close within
statistical errors for different irradiation conditions of one pixel
(anode) of the PMT. The data are described well in different light
setups with the same fixed set of parameters $\nu_1$, $\alpha_2$,
$\nu_2$, $\alpha_3$, $\nu_3$, $\xi$ of the SPE spectrum $p_1(a)$,
keeping variable only the parameter specifying the light ($\mu$), and
one of the signal measurement parameters, $\sigma$. The values of the
fixed parameters are shown in the plots with zero standard deviations.
The data sets allowed us also to keep the $scale$ parameter fixed in
all fits, indicating to a good stability of the signal measurement
system during the measurements. To illustrate these observations
better, the plots are normalized such that the SPE contribution to the
full spectra is shown (as the dashed line above the highlighted and
horizontally hatched area) visually identical in each plot of the
set. The SPE spectrum approximation extracted from the data in such a
procedure may therefore be considered as a characteristic of the
photon detector (one of the anodes of the MAPMT in this case).

This result demonstrates the predictive functionality of the model,
meaning that the SPE spectrum approximation measured in some
conditions may be used to evaluate the amplitude distributions from
this detector in different light conditions, and with different signal
measurement resolution.

Logarithmic scales in ordinate in Fig.~\ref{fig:sim_pixel_45_log}
illustrate the quality of the model description of the whole spectra
as the sum of the partial terms with $m$ from 0 to about 7-10.

\begin{figure*}[h!] 
\centering 
  \subfloat[ET Enterprises 9823B PMT, test setup at low light
    conditions corresponding to $\mu~=~0.496$]{%
    \includegraphics[clip=true,trim=0 10 0
      45,width=.48\textwidth,keepaspectratio]
                    {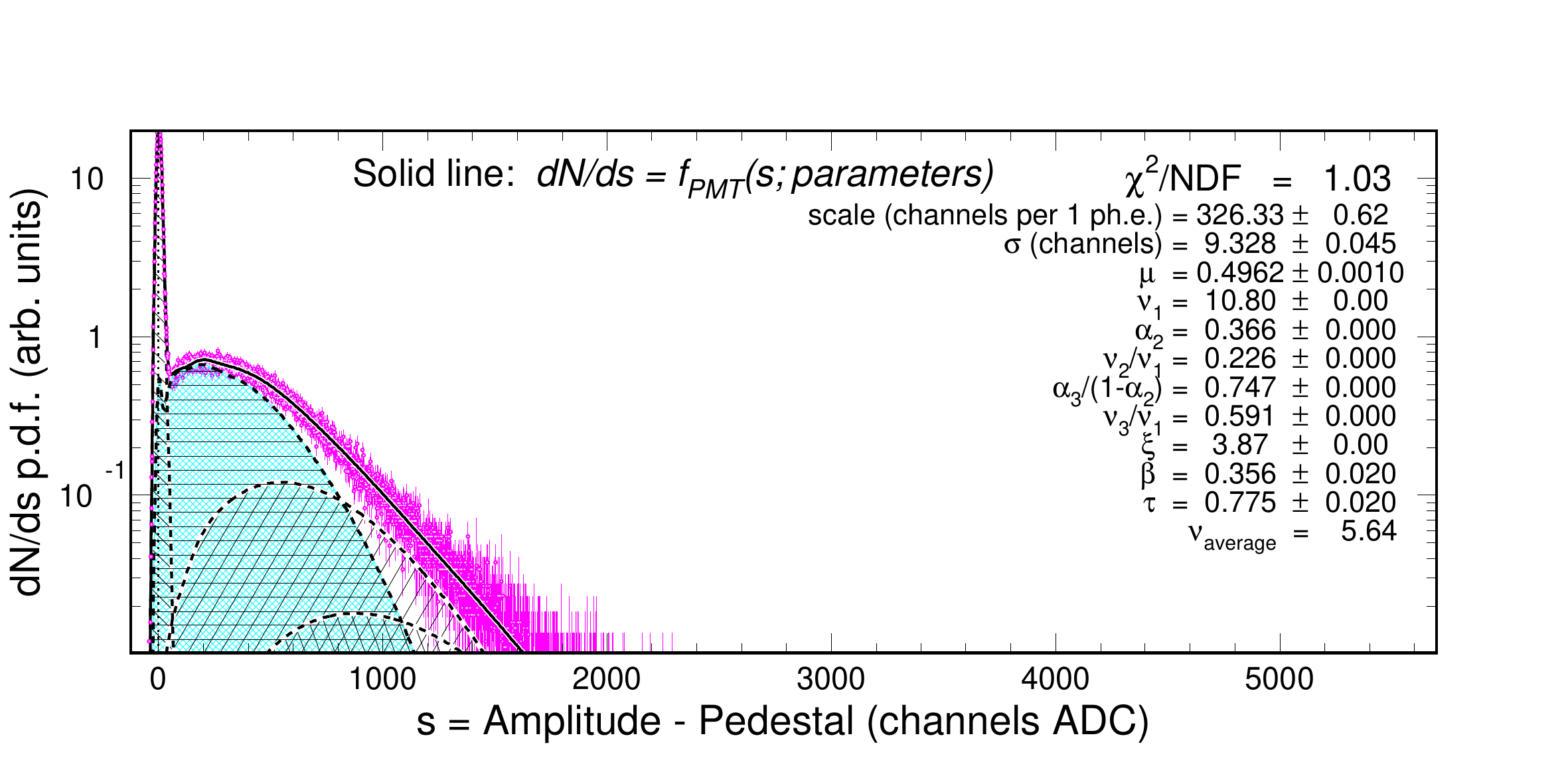}}\hfill
  \subfloat[ET Enterprises 9823B PMT, test setup at lower-medium light
    conditions corresponding to $\mu~=~0.991$]{%
    \includegraphics[clip=true,trim=0 10 0
      45,width=.48\textwidth,keepaspectratio]
                    {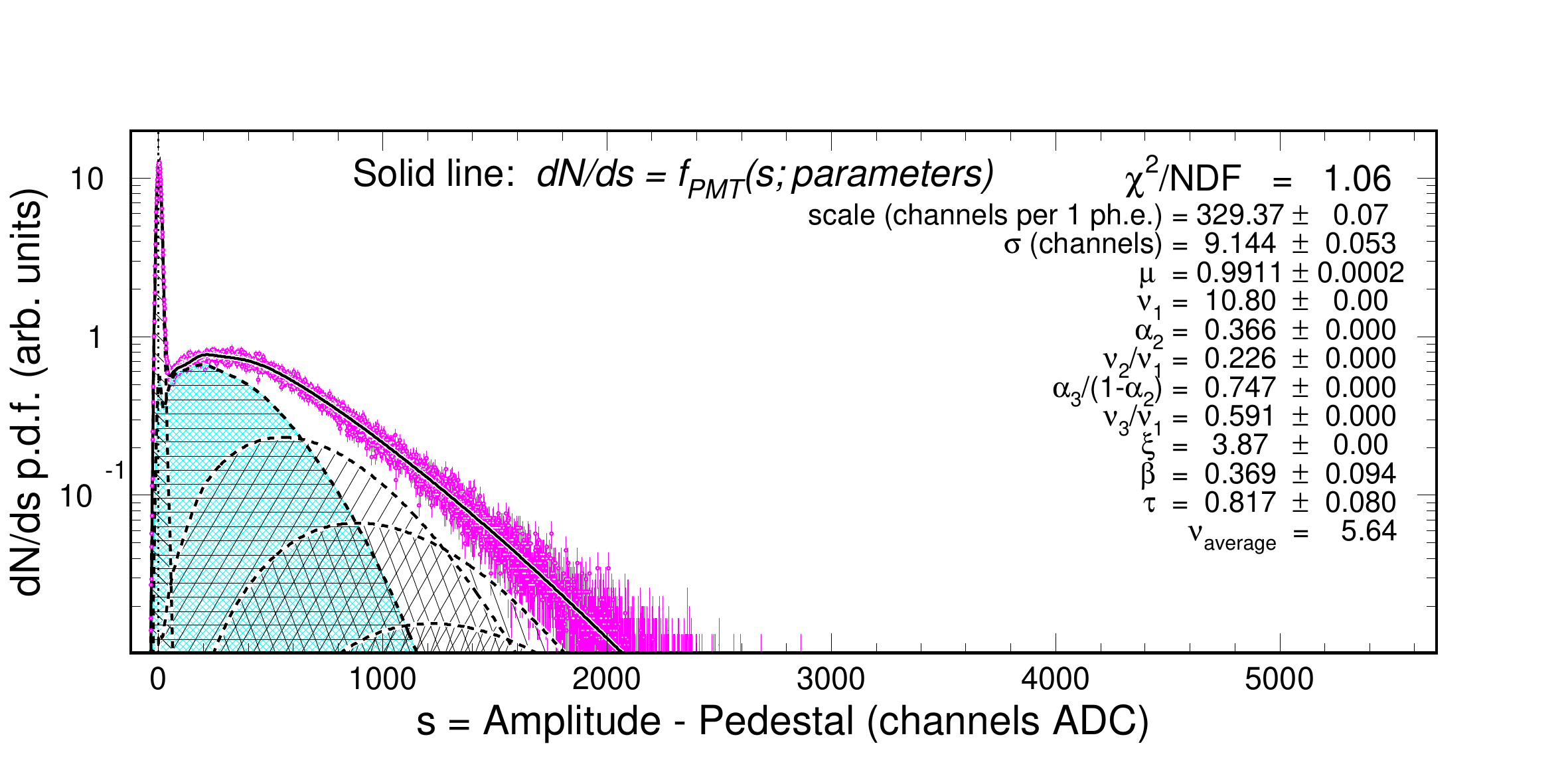}}\\
  \subfloat[ET Enterprises 9823B PMT, test setup at upper-medium light
    conditions corresponding to $\mu~=~1.916$]{%
    \includegraphics[clip=true,trim=0 10 0
      45,width=.48\textwidth,keepaspectratio]
                    {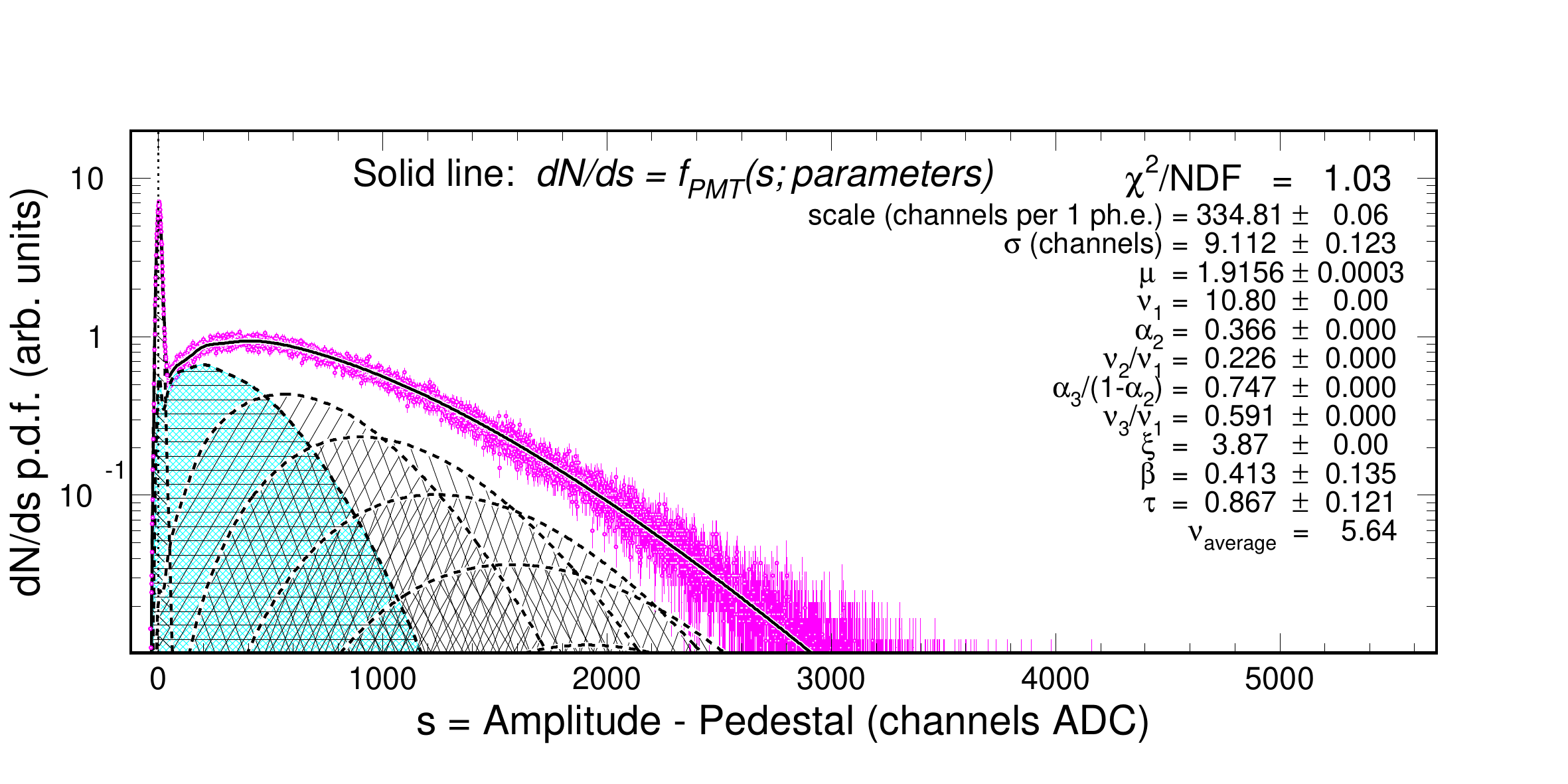}}\hfill
  \subfloat[ET Enterprises 9823B PMT, test setup at higher light
    conditions corresponding to $\mu~=~3.992$]{%
    \includegraphics[clip=true,trim=0 10 0
      45,width=.48\textwidth,keepaspectratio]
                    {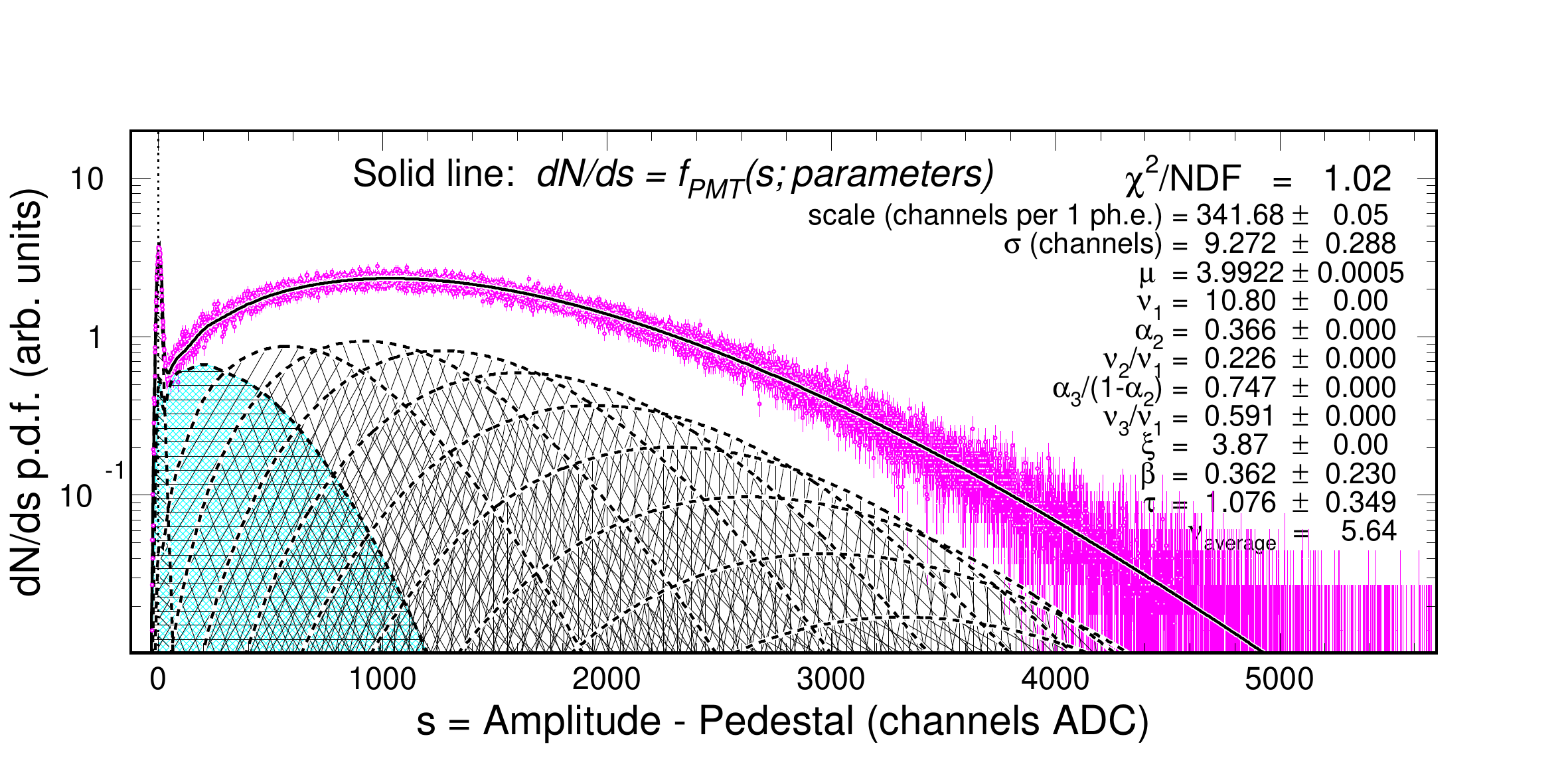}}
  \caption{ A set of amplitude distributions measured with an ET
    Enterprises 9823B photomultiplier at eighteen light conditions,
    four of which are shown.  The experimental data \cite{Hakob} are
    shown as open circles with error bars, other notation and the
    features in the plots are the same as in
    Fig.~\ref{fig:sim_pixel_39_lin}.  The values of $\tau$ parameter
    in the plots are dimensionless and given in the units of $\sigma$.
    Only the parameters related to the performance of the signal
    analysis system ($scale$, $\sigma$, $\beta$, and $\tau$), and also
    the light intensity parameter $\mu$ are left variable in the final
    fits.}
\label{fig:LT9823B_log}
\end{figure*}

\subsection{Tests of ET Enterprises 9823B PMT}

Fig.~\ref{fig:LT9823B_log} shows similar exercise with the amplitude
spectra measured on a very different PMT, ET Enterprises 5-inch 9823B
tube; the data were kindly provided to us by Hakob Voskanyan, Andrey
Kim and Will Phelps \cite{Hakob}. The statistical errors in the data
set are small enough for a stable and accurate multiparametric fitting
procedure. The excellent quality of the data made it possible to
observe and measure the non-Gaussian components in the pedestal
distributions, and adjust the model by introducing and parameterizing
these components of the experimental signal measurement distribution
function, to achieve acceptable model description of the full data set.

We did observe the slight asymmetry in the pedestal function that
could be modeled by introducing an exponential noise component in
addition to the standard Gaussian form.  Such noise may be modeled
{\it ad hoc} as an independent random value $a_{\mathrm{noise}}$ contributing
with a certain probability to the signal value $a$ in any event, and
distributed according to the exponential
\begin{equation}
  \label{noise_exp}
  f_{\mathrm{noise}}(a;\tau) = \frac{1}{\tau} 
  \exp{ \left ( - \frac{a}{\tau} \right ) } ,
\end{equation}
with the parameter $\tau$ describing
the exponential (temperature-like) noise spectrum. 

Adding such a random noise contribution to the model involves the
convolution between the model function (\ref{PMT_9par}) and the
exponential (\ref{noise_exp}).  Using the properties of the
convolution algebra, it can be implemented by the substitution of the
Gaussian form $G(a,n;\sigma_{\mathrm{eff}})$ in Eq.~(\ref{PMT_9par})
with its convolution with the exponential noise in the form
\begin{equation}
  \label{EMGform}
(1 - \beta) G(a,n;\sigma_{\mathrm{eff}}) + \beta G_{em}(a,n;\sigma_{\mathrm{eff}},\tau) ,
\end{equation}
wherein the parameter $\beta$ is the probability for the
noise event to happen in every measurement,
and the function $G_{em}(a,n;\sigma_{\mathrm{eff}},\tau)$ is the
convolution of the Gaussian with the exponential (known also as
exponentially modified Gaussian distribution, see Ref. \cite{EMG}):
\begin{multline}
  \label{EMGdef}
   G_{em}(a,n;\sigma_{\mathrm{eff}},\tau) = \\
 = \frac {1}{2\tau} \exp{\left (\frac{\sigma_{\mathrm{eff}}^2}{2\tau^2} - 
   \frac{a - n/ \nu}{\tau}\right )} \cdot \mathrm{erfc} \left [ 
\frac{\sigma_{\mathrm{eff}}^2/\tau - (a - n/ \nu )}{ \sqrt{2}\sigma_{\mathrm{eff}}} 
\right ] , 
\end{multline}
where 
\begin{equation}
   \mathrm{erfc}(x) \equiv 1 - \mathrm{erf}(x) = 
   \frac{2}{\sqrt{\pi}}\int_x^{\infty}\exp \left ( -t^2 \right ) \mathrm{d}t .
\end{equation}

Adding random exponential noise contributions to every measurement in
this extension of the model eliminates the property of the system
resolution function $R(a)$ to be non-biased, violating the basic
assumption (e) in the model. The
correspondingly modified relation of Eq.~(\ref{scalemu}) between the
values of the $scale$, $\langle s \rangle $, and $\mu$ parameters that
should be applied in the minimization procedure in this case is as
follows:
\begin{equation}
\label{scalebt}
 scale_{\beta \tau} = (\langle s \rangle - \beta \tau)/\mu .
\end{equation}

The results of application of such approach to the data are
illustrated in Fig.~\ref{fig:LT9823B_log}.  The set of 18 measurements
at different light intensities in the range of $\mu$ values from about
0.5 to about 4.0 was approximated using the identical SPE spectrum
defined by the parameters $\nu_1$, $\alpha_2$, $\nu_2$, $\alpha_3$,
$\nu_3$ and $\xi$. The signal measurement system parameters $scale$,
$\sigma$, $\beta$ and $\tau$ were left variable in the global fit
procedure to allow for their slight modification between different
measurements, but their variations are quite small, showing the
stability of the test setup. The only major variable parameter in the
fitting procedure is $\mu$, characterizing average number of
photoelectrons in each test. The goodness-of-fit evaluator
$\chi^2/n_{\mathrm{d}}$ is in the range between 1.0 and 1.1 in all 18
approximations, indicating to a model description of the data close to
a theoretically perfect.

\begin{figure*}[hbt] 
\centering 
  \subfloat[H8500 MAPMT, anode \#39]{%
    \includegraphics[clip=true,trim=0 10 0
      45,width=.48\textwidth,keepaspectratio]
                    {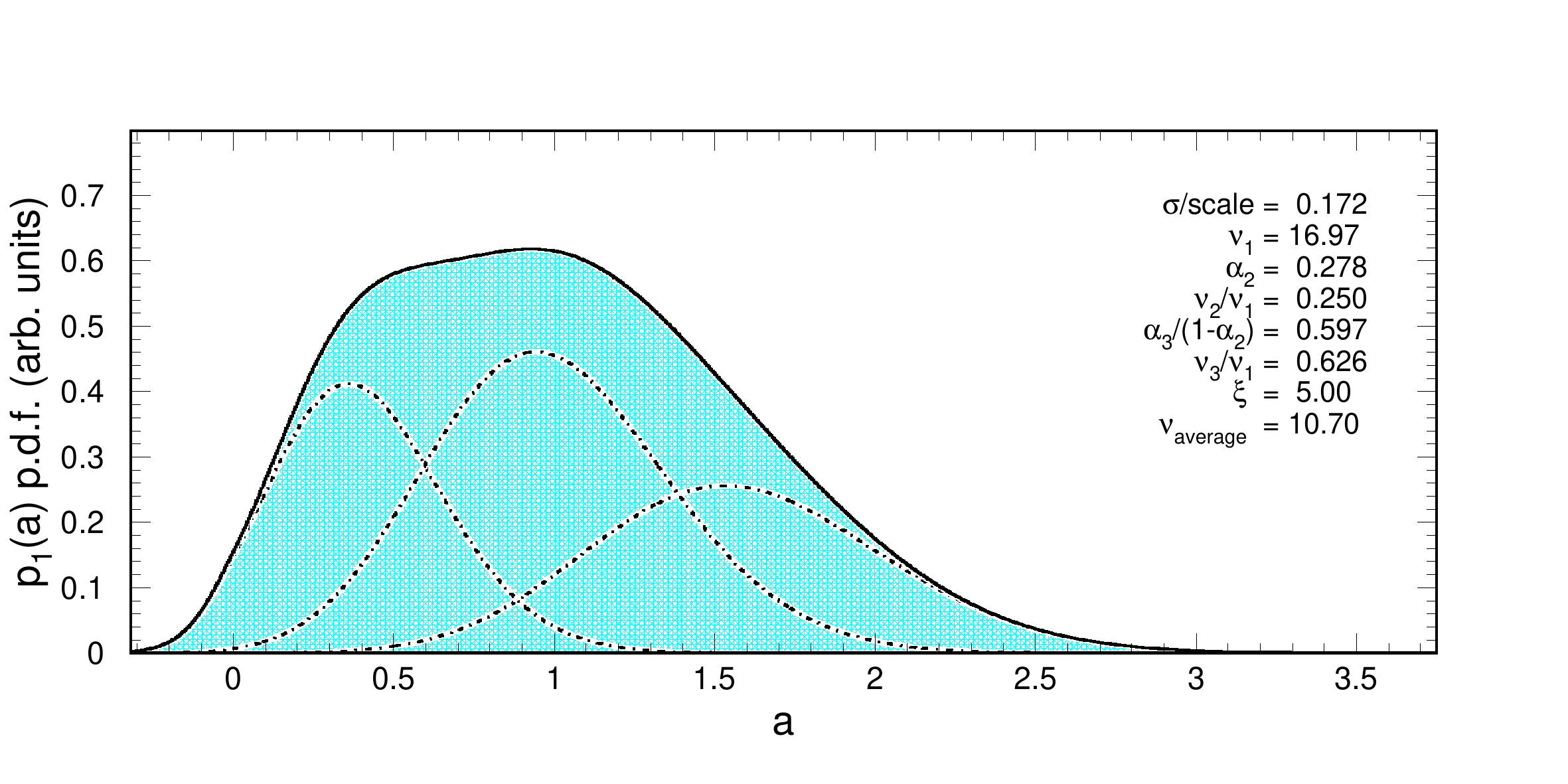}}\hfill
  \subfloat[H8500 MAPMT, anode \#45]{%
    \includegraphics[clip=true,trim=0 10 0
      45,width=.48\textwidth,keepaspectratio]
                    {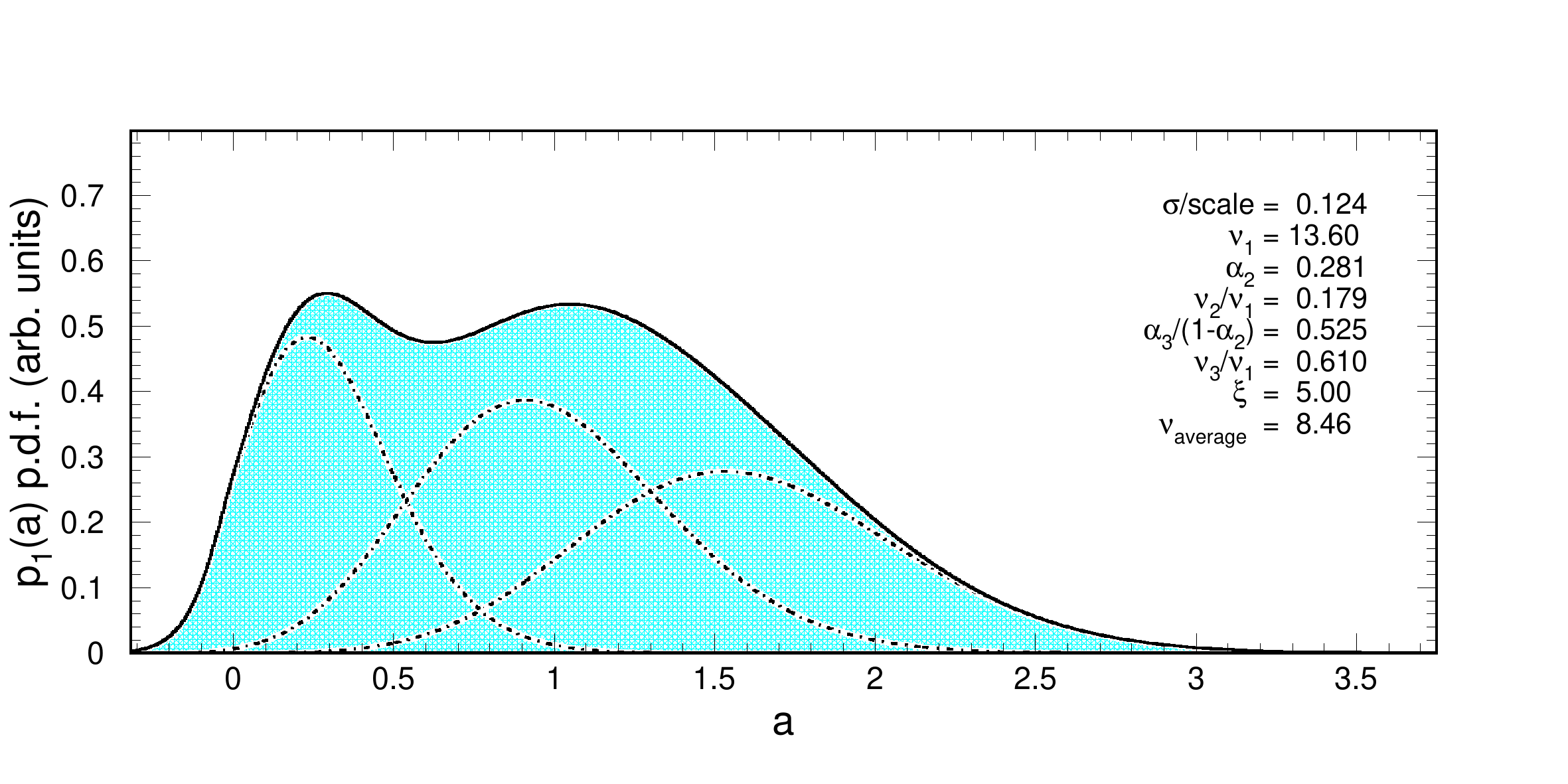}} \\
  \subfloat[H8500 MAPMT, anode \#61]{%
    \includegraphics[clip=true,trim=0 10 0
      45,width=.48\textwidth,keepaspectratio]
                    {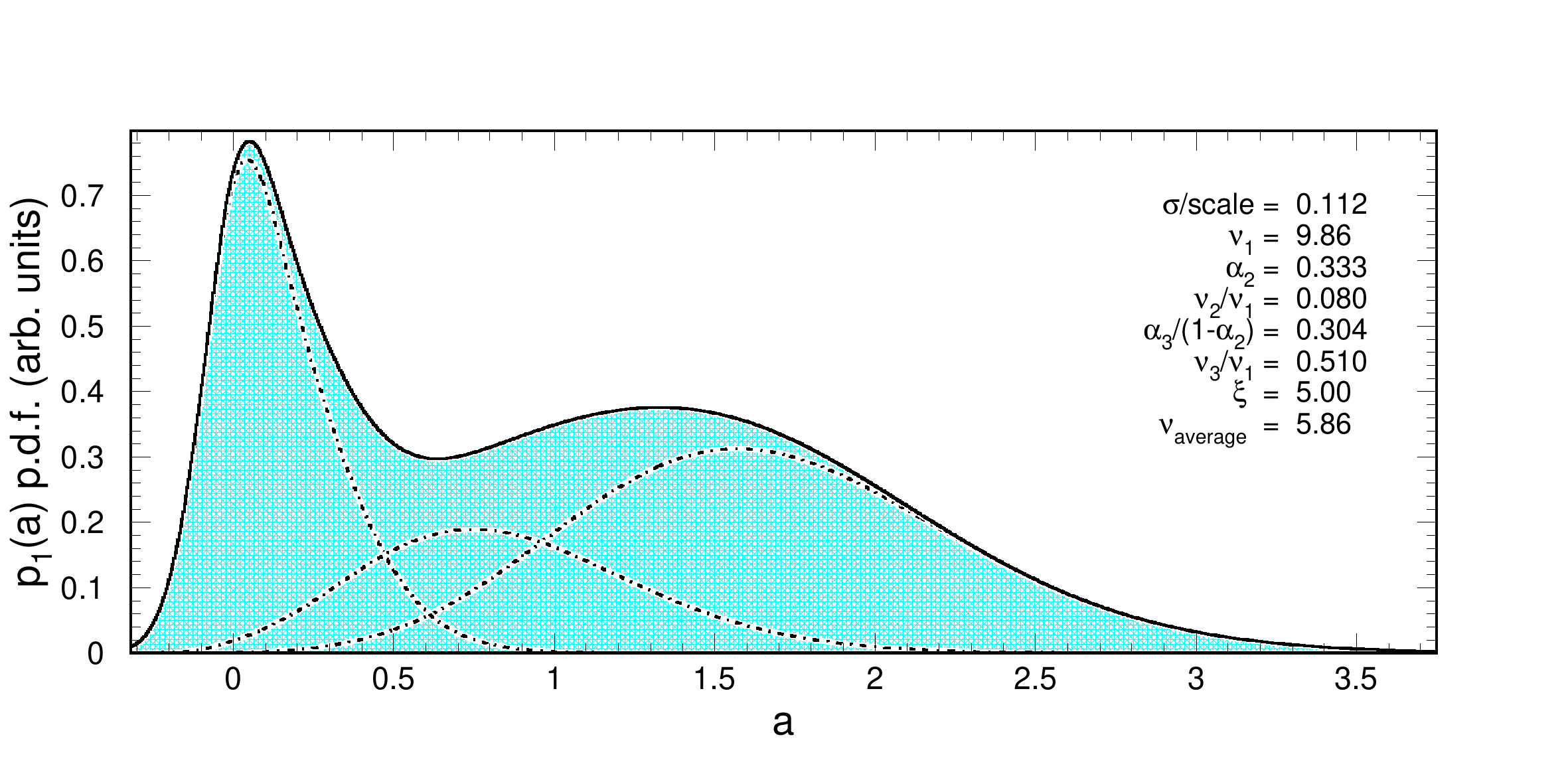}}\hfill
  \subfloat[ET Enterprises 9823B PMT]{%
    \includegraphics[clip=true,trim=0 10 0
      45,width=.48\textwidth,keepaspectratio]
                    {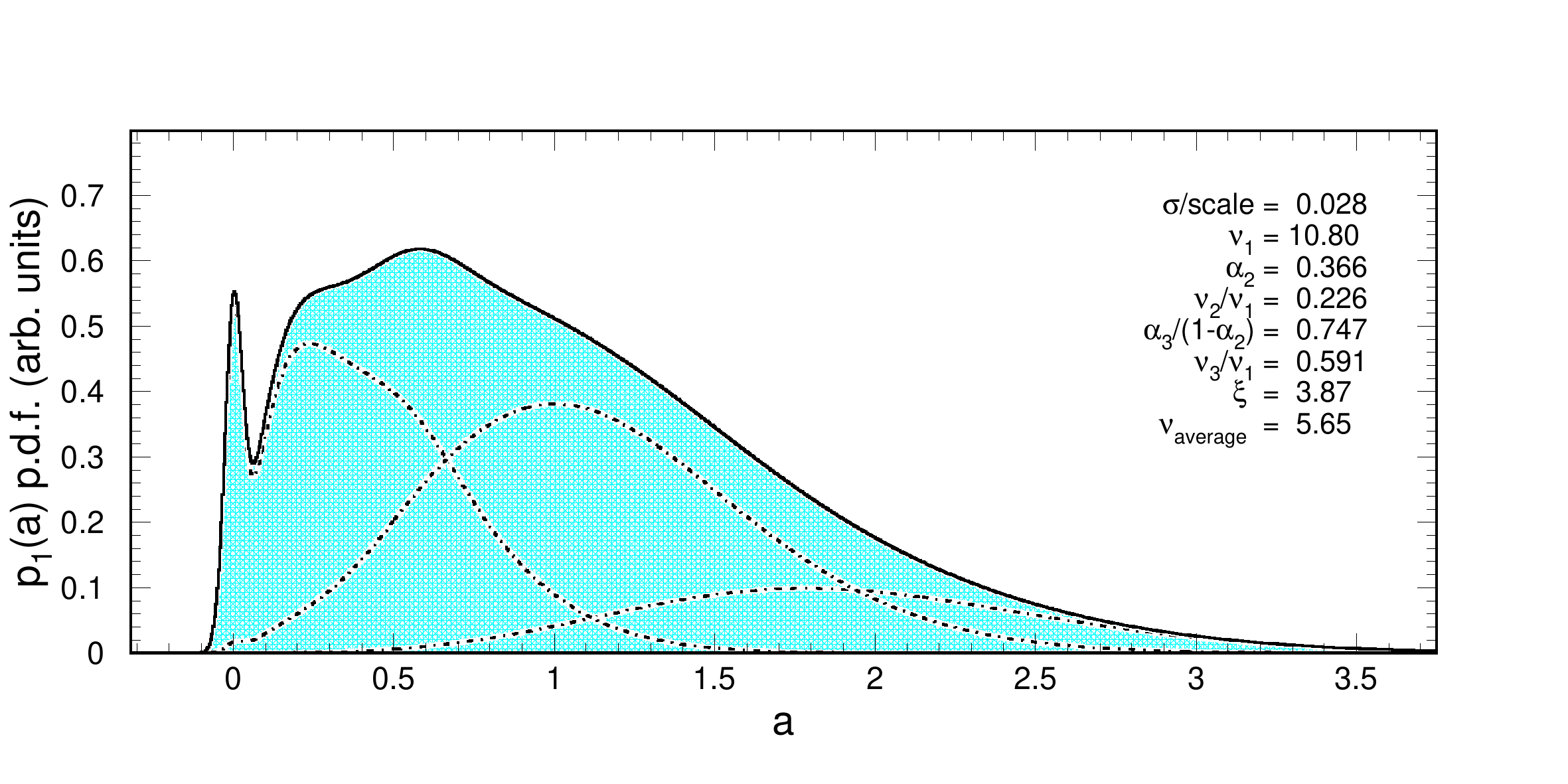}}
  \caption{Solid lines show the $p_1(a)$ p.d.f. corresponding to the
    amplitude spectra of a single photoelectron as determined in the
    plots shown in
    Figs.~\ref{fig:sim_pixel_39_lin}-\ref{fig:sim_pixel_61_lin}
    (panels a-c), and Fig.~\ref{fig:LT9823B_log} (panel d). The
    trinomial components of the functions are shown by the dash-dotted
    lines.}
\label{fig:single_e_spec_ls}
\end{figure*}

\subsection{Extracted SPE spectra}

Fig.~\ref{fig:single_e_spec_ls} further illustrates the inner
structure of the SPE spectra extracted from the data sets shown in
Figs.~\ref{fig:sim_pixel_39_lin}-\ref{fig:sim_pixel_61_lin} and
Fig.~\ref{fig:LT9823B_log}.  The $p_1(a)$ p.d.f. are drawn as
functions of the normalized signal amplitude $a$, together with their
three Poissonian components defined by the vectors of parameters ${\bf
  t}$. The $p_1(a)$ functions are shown convoluted with the
corresponding effective signal measurement Gaussians
$G(a,n;\sigma_{\mathrm{eff}})$ in
Figs.~\ref{fig:single_e_spec_ls}a-\ref{fig:single_e_spec_ls}c, and
convoluted with the modified signal measurement function of
Eq.~(\ref{EMGform}) in Fig.~\ref{fig:single_e_spec_ls}d. The
parameters for the signal measurement functions used were averaged
over the test light conditions.

The three components of the $p_1(a)$ functions originate from the
three elementary Poissonian constituents of the discrete $q_1(n)$
distributions, as defined in Eq.~(\ref{SPE_L}), and are converted to
the continuous $a$ scale by the same convolutions applied to each
component separately, similar to how it's done in Eq.~(\ref{pmqm}).
The components add up to fit the complicated SPE amplitude
distribution functions that would be difficult to approximate using a
smaller number of parameters.

Good normalized signal amplitude resolution of the measurement system
for the data shown in Fig.~\ref{fig:single_e_spec_ls}d allowed us to
clearly distinguish between the events with $n = 0$ and events with $n
> 0$ in the SPE spectrum, that is, to evaluate according to the model
the portion of events when a photoelectron fails to generate any
response from the PMT. Portion of such events in the $p_1(a)$ function
may be linked to the ``collection efficiency'' characteristic reported
by the PMT manufacturers, see, for example, Ref. \cite{Hamamatsu},
page 45.

\begin{figure*}[h!] 
\centering 
  \subfloat[Hamamatsu H12700 MAPMT GA0133, $\rm{HV}=1000$~V]{%
    \includegraphics[clip=true,trim=0 10 0
      45,width=.48\textwidth,keepaspectratio]
                    {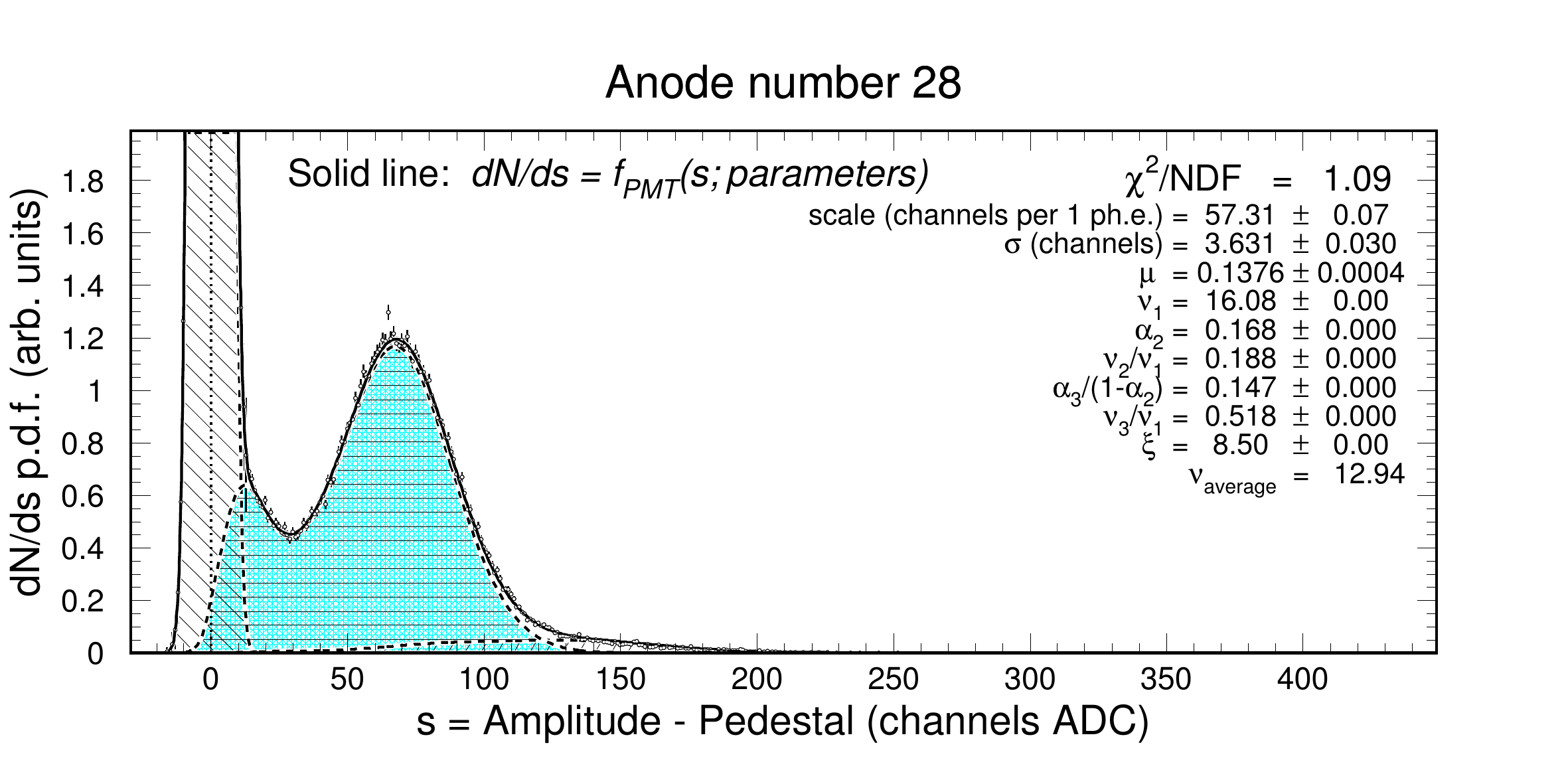}}\hfill
  \subfloat[Hamamatsu H8500 MAPMT CA7782, $\rm{HV}=1000$~V]{%
    \includegraphics[clip=true,trim=0 10 0
      45,width=.48\textwidth,keepaspectratio]
                    {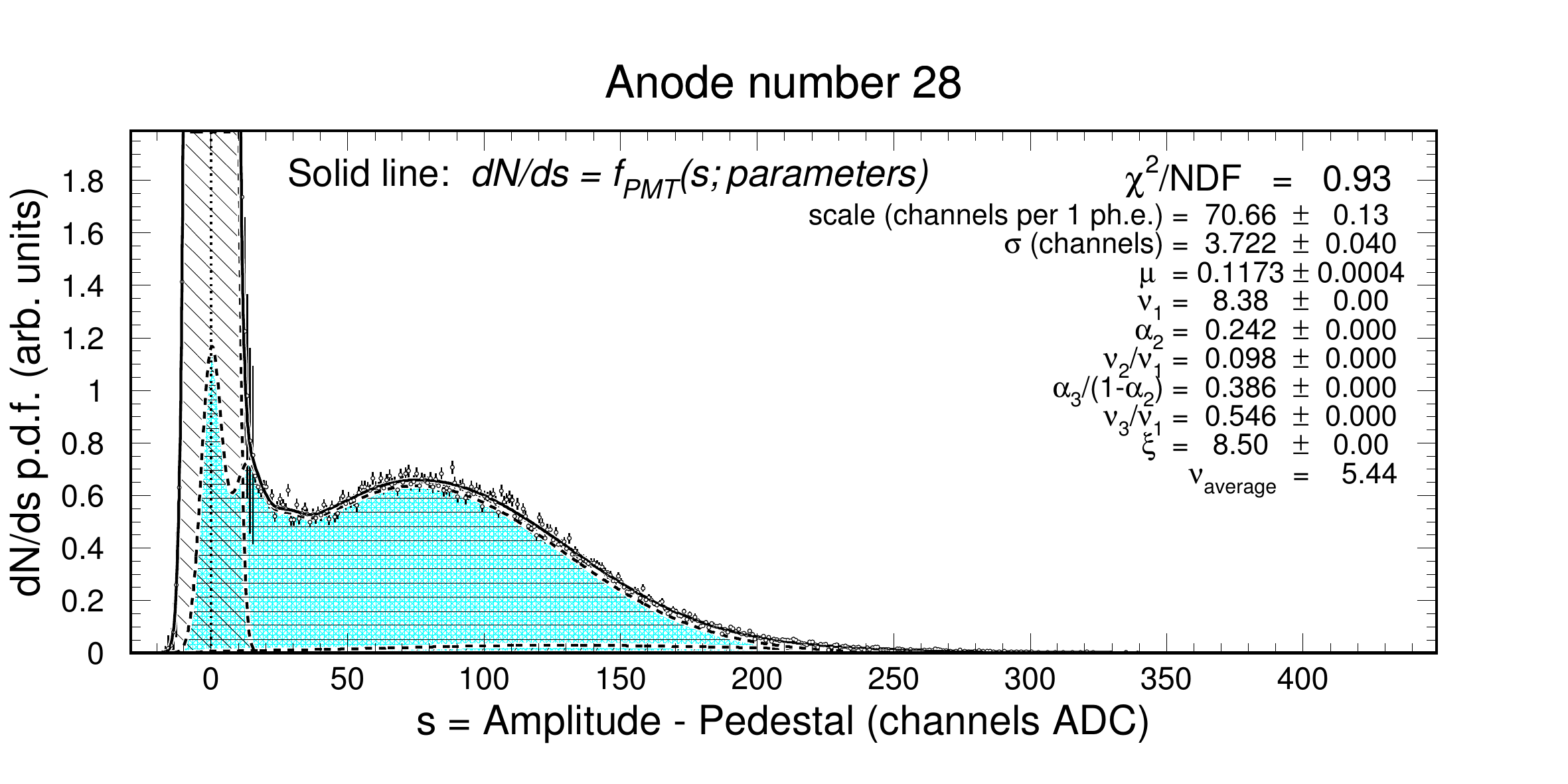}}\\
  \subfloat[Hamamatsu H12700 MAPMT GA0133, $\rm{HV}=1050$~V]{%
    \includegraphics[clip=true,trim=0 10 0
      45,width=.48\textwidth,keepaspectratio]
                    {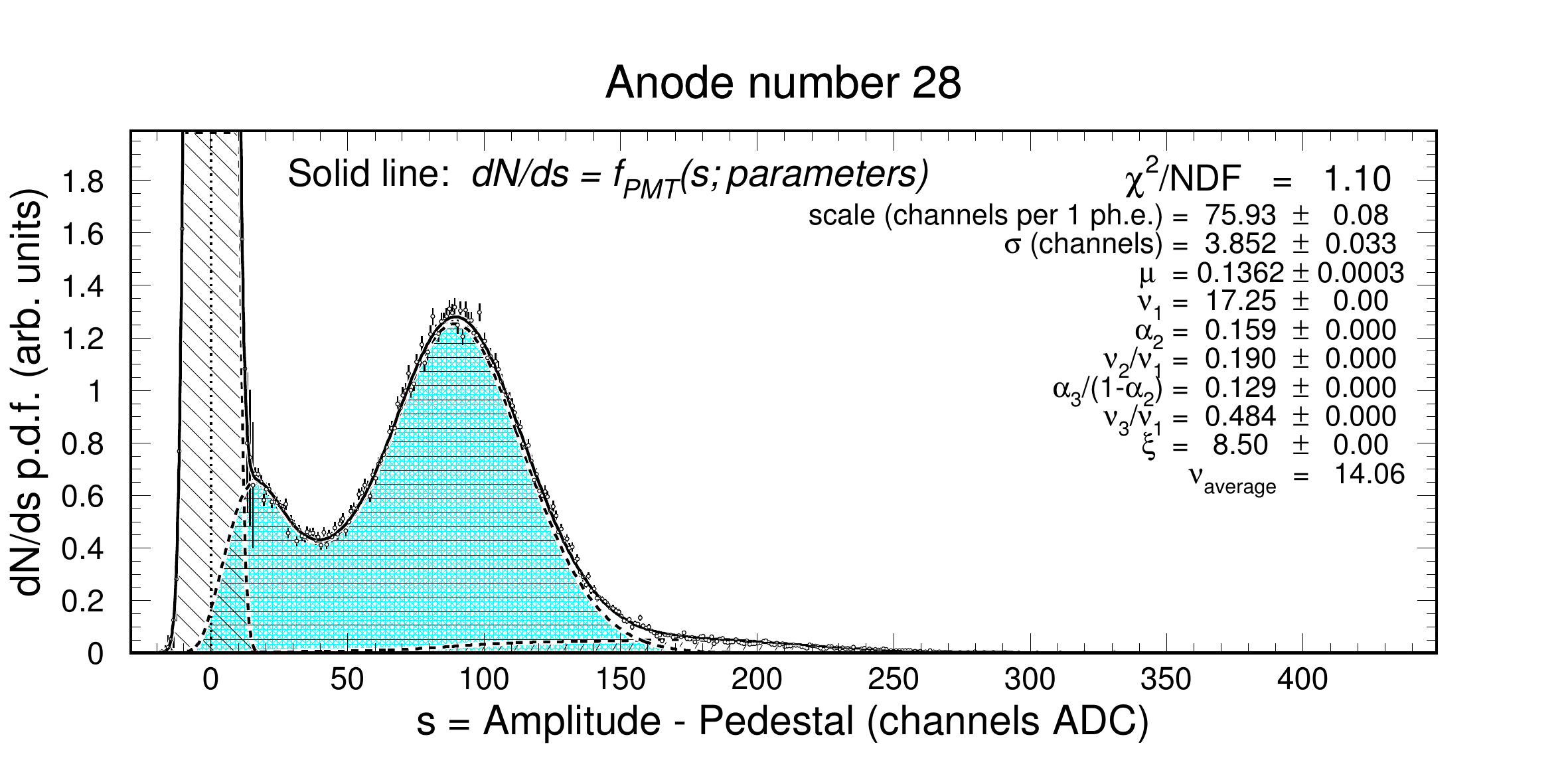}}\hfill
  \subfloat[Hamamatsu H8500 MAPMT CA7782, $\rm{HV}=1050$~V]{%
    \includegraphics[clip=true,trim=0 10 0
      45,width=.48\textwidth,keepaspectratio]
                    {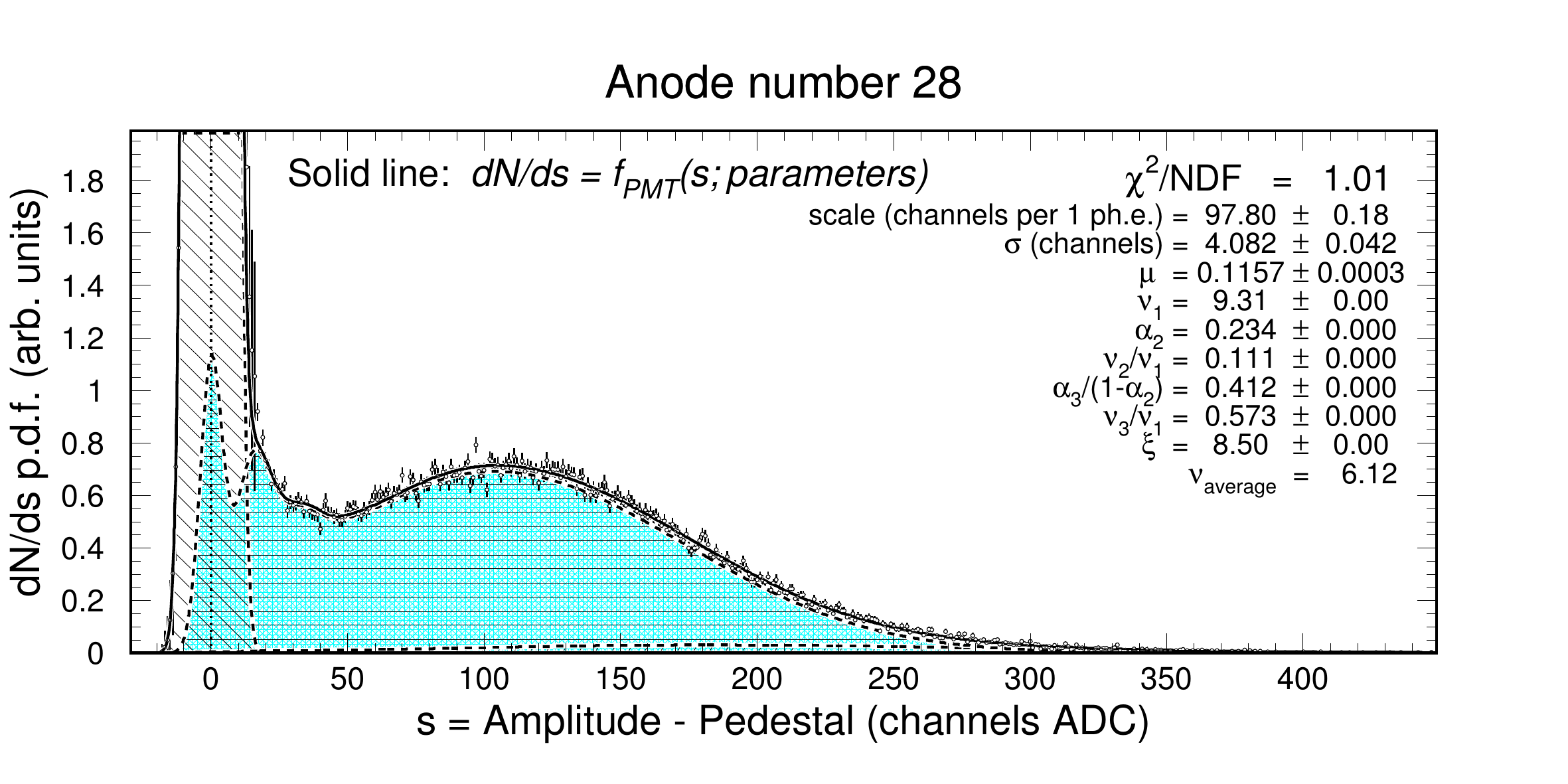}}\\
  \subfloat[Hamamatsu H12700 MAPMT GA0133, $\rm{HV}=1075$~V]{%
    \includegraphics[clip=true,trim=0 10 0
      45,width=.48\textwidth,keepaspectratio]
                    {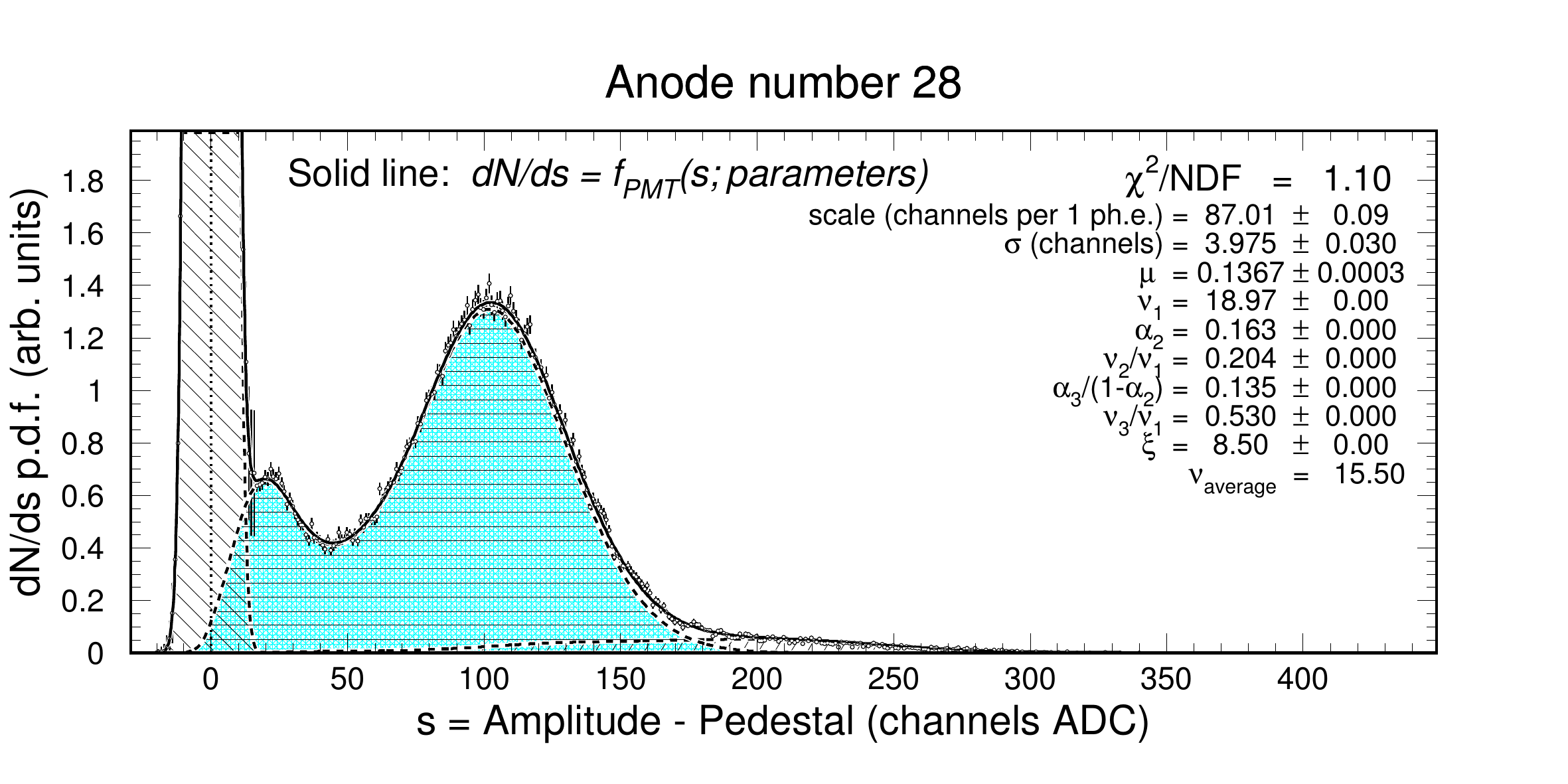}}\hfill
  \subfloat[Hamamatsu H8500 MAPMT CA7782, $\rm{HV}=1075$~V]{%
    \includegraphics[clip=true,trim=0 10 0
      45,width=.48\textwidth,keepaspectratio]
                    {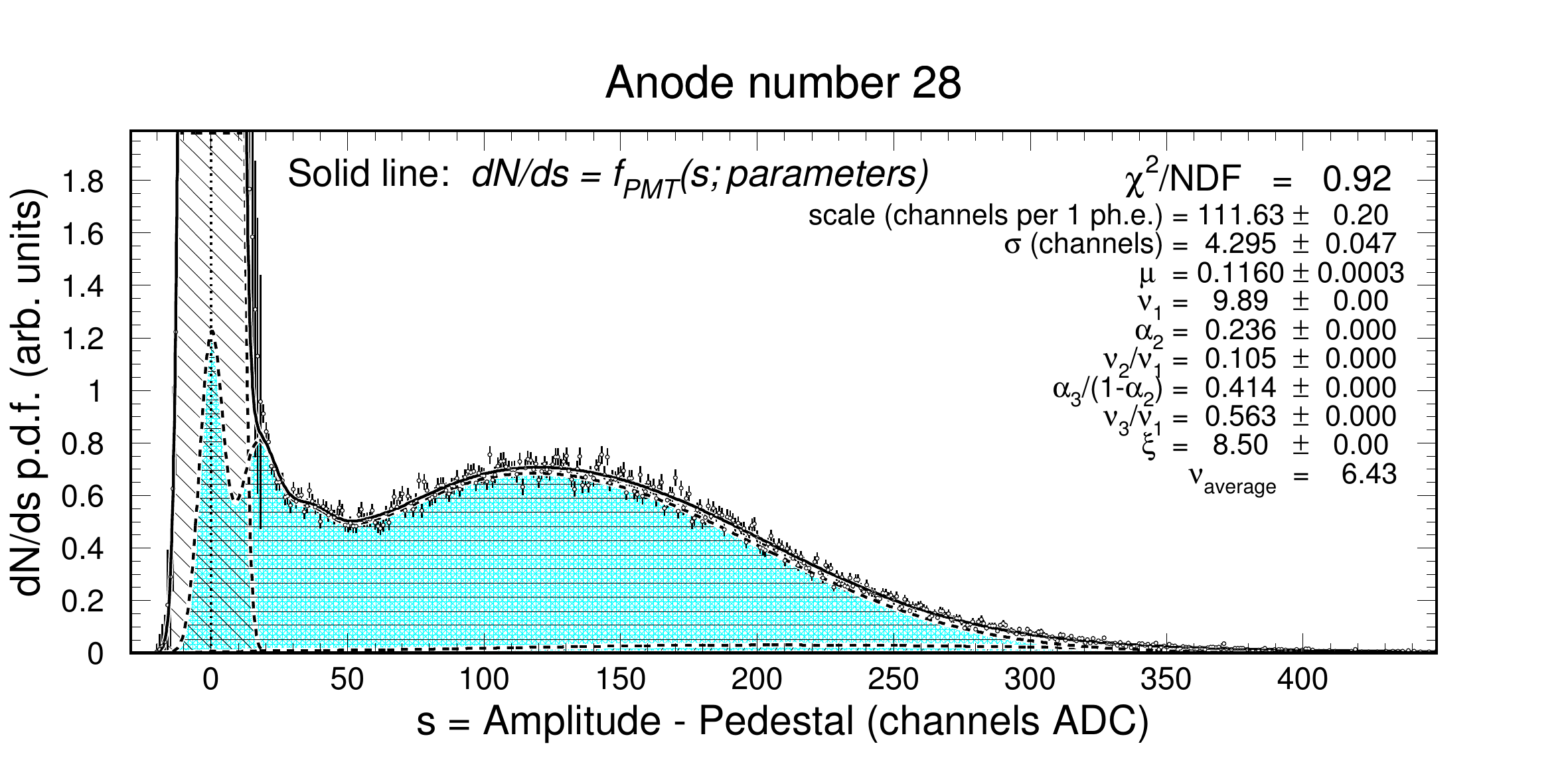}}\\
  \subfloat[Hamamatsu H12700 MAPMT GA0133, $\rm{HV}=1100$~V]{%
    \includegraphics[clip=true,trim=0 10 0
      45,width=.48\textwidth,keepaspectratio]
                    {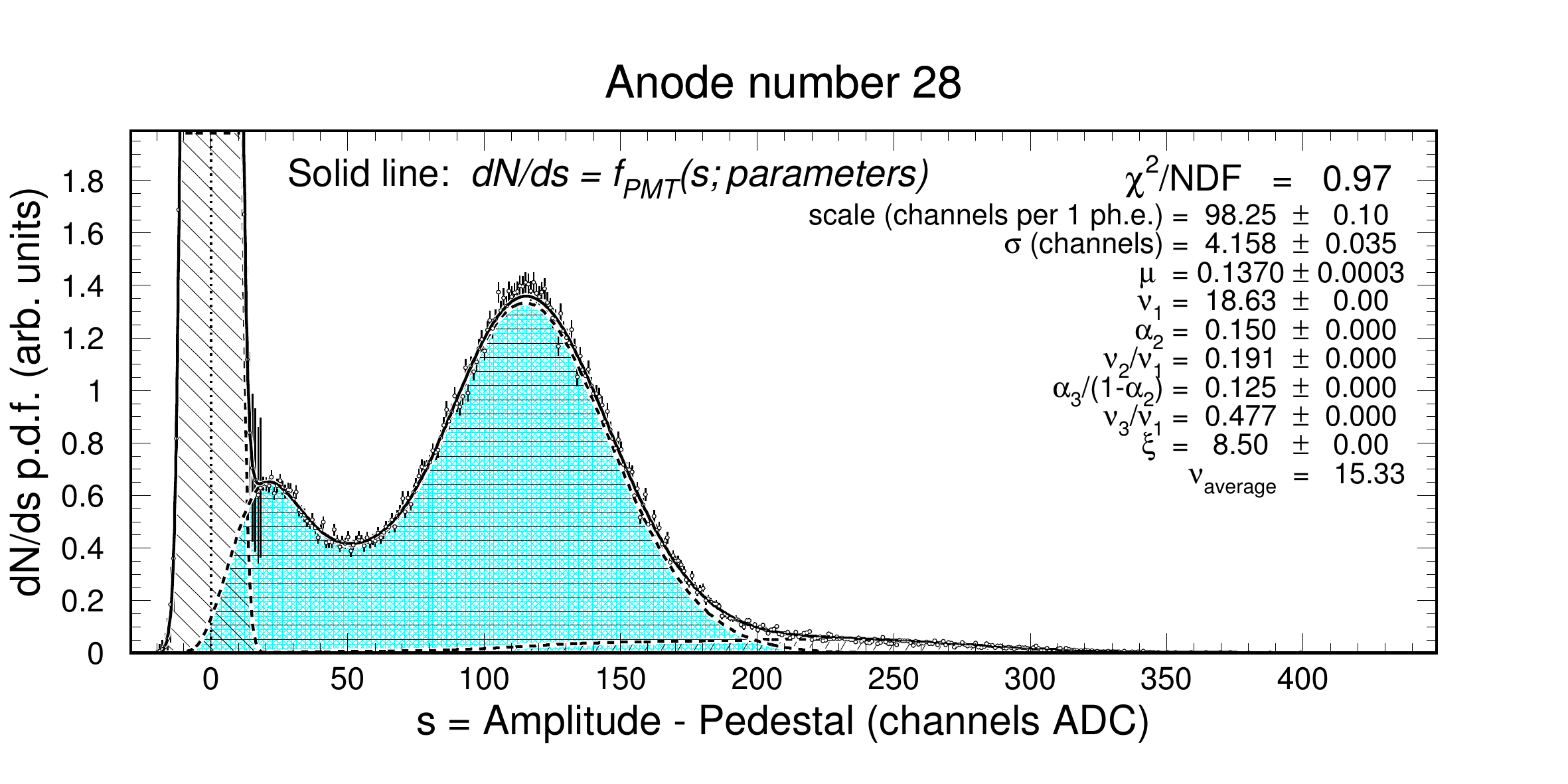}}\hfill
  \subfloat[Hamamatsu H8500 MAPMT CA7782, $\rm{HV}=1100$~V]{%
    \includegraphics[clip=true,trim=0 10 0
      45,width=.48\textwidth,keepaspectratio]
                    {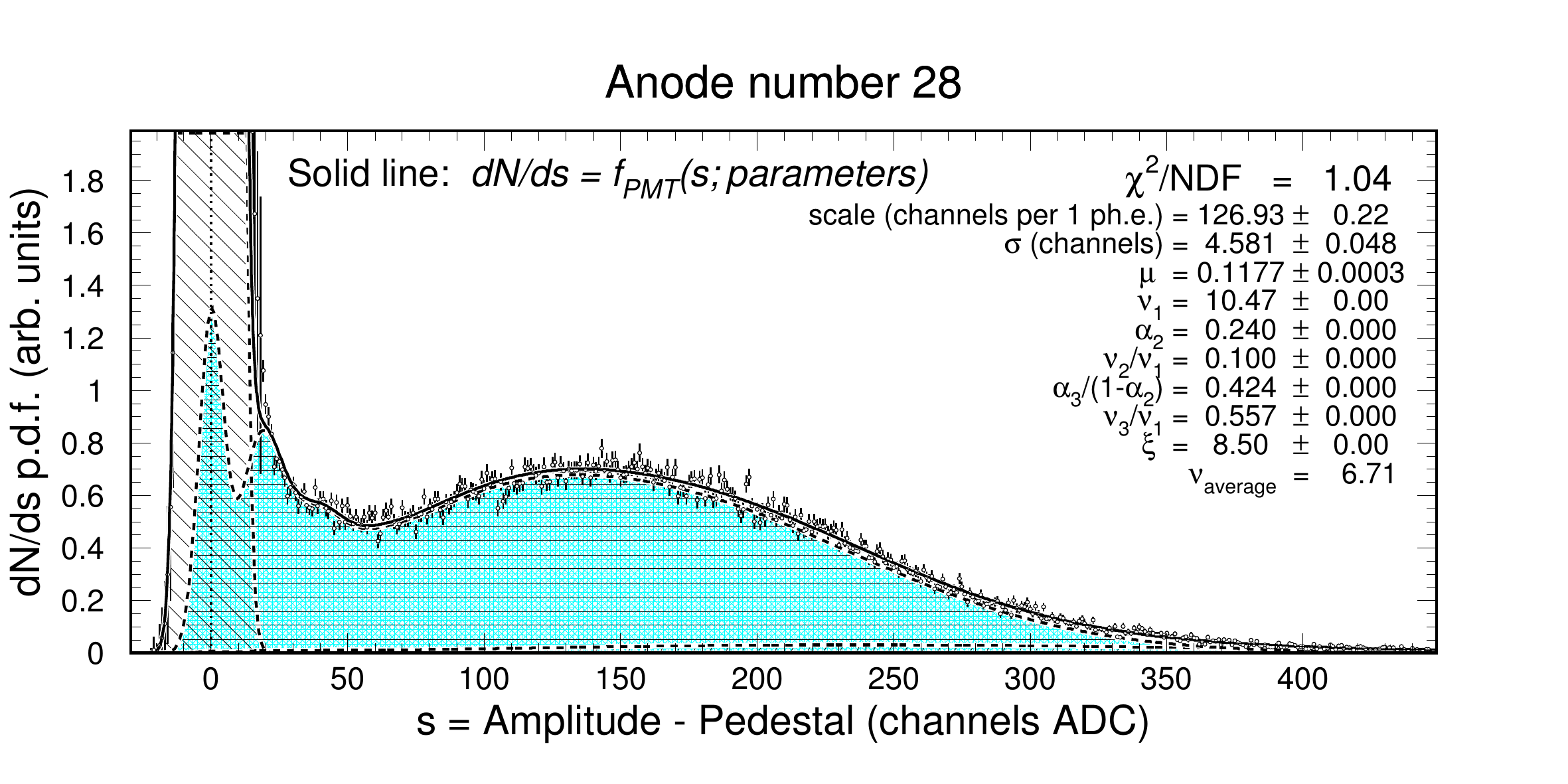}}
\caption{A set of amplitude distributions measured at four high
  voltages on one Hamamatsu H12700 MAPMT (left panels), and one
  Hamamatsu H8500 MAPMT (right panels), shown for one of the central
  anodes (\#28) in each MAPMT at the meduim ``OD50'' light
  condition. The notation and other features in the plots are the same
  as in Fig.~\ref{fig:sim_pixel_39_lin}. Only the set of parameters
  related to the performance of the signal analysis system ($scale$
  and $\sigma$), and also the light intensity parameter $\mu$ are left
  variable in the final fits.}
\label{fig:rich_lin}
\end{figure*}

\begin{figure*}[h!] 
\centering 
  \subfloat[Hamamatsu H12700 MAPMT GA0133, anode \#28]{%
    \includegraphics[clip=true,trim=0 10 0
      48,width=.48\textwidth,keepaspectratio]
                    {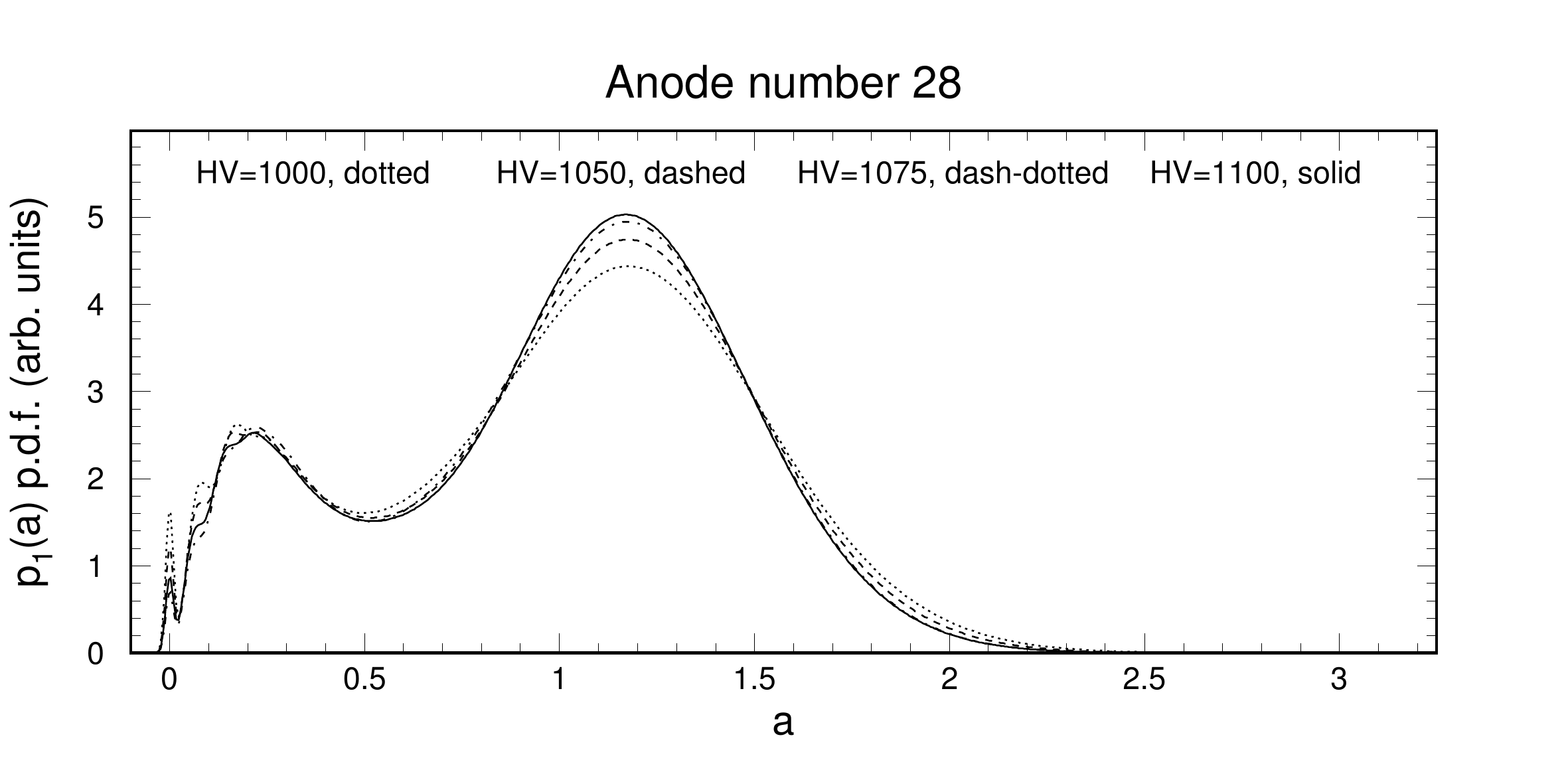}}\hfill
  \subfloat[Hamamatsu H8500 MAPMT CA7782, anode \#28]{%
    \includegraphics[clip=true,trim=0 10 0
      48,width=.48\textwidth,keepaspectratio]
                    {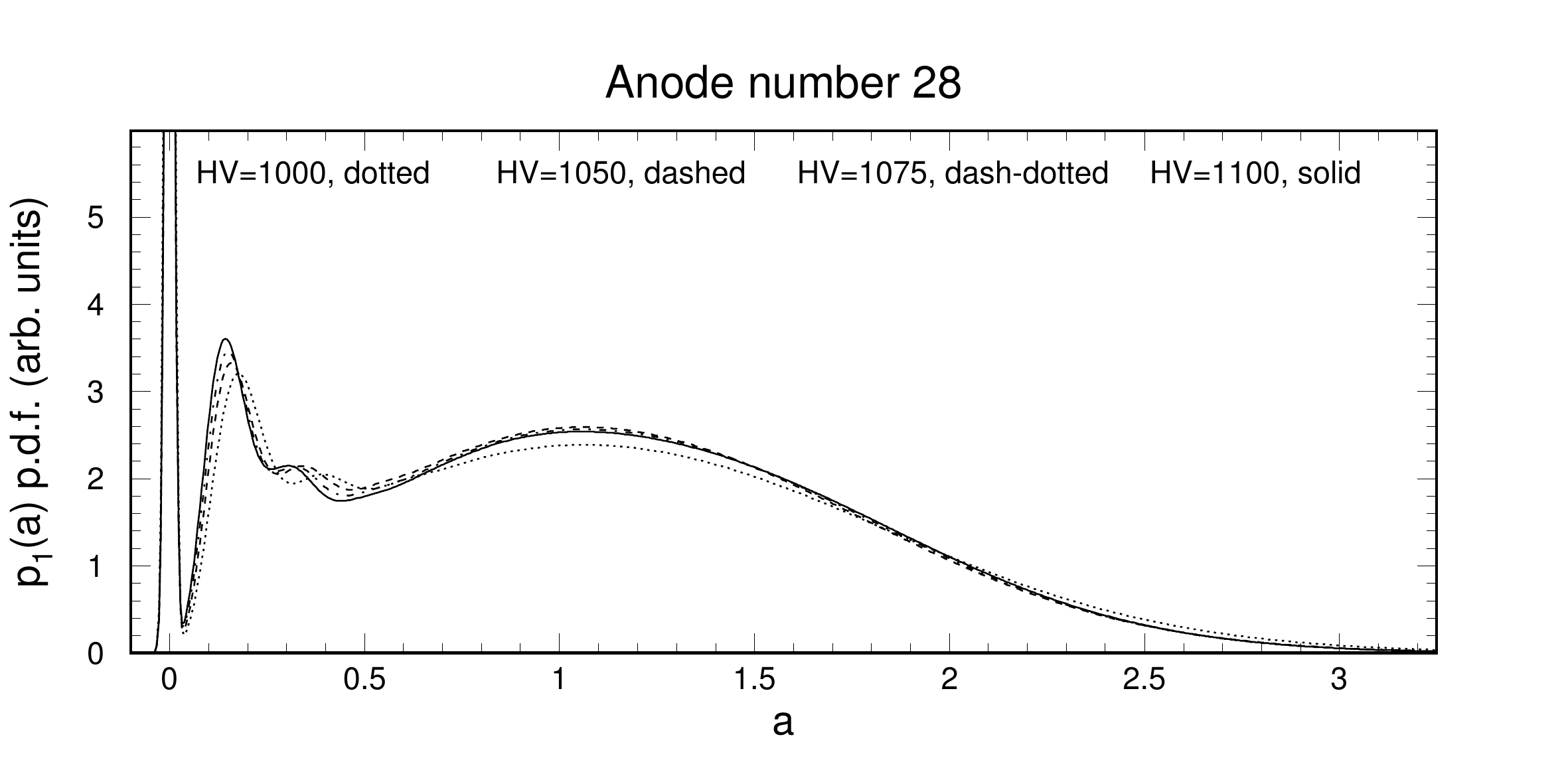}}\\
  \subfloat[Hamamatsu H12700 MAPMT GA0133, anode \#32]{%
    \includegraphics[clip=true,trim=0 10 0
      48,width=.48\textwidth,keepaspectratio]
                    {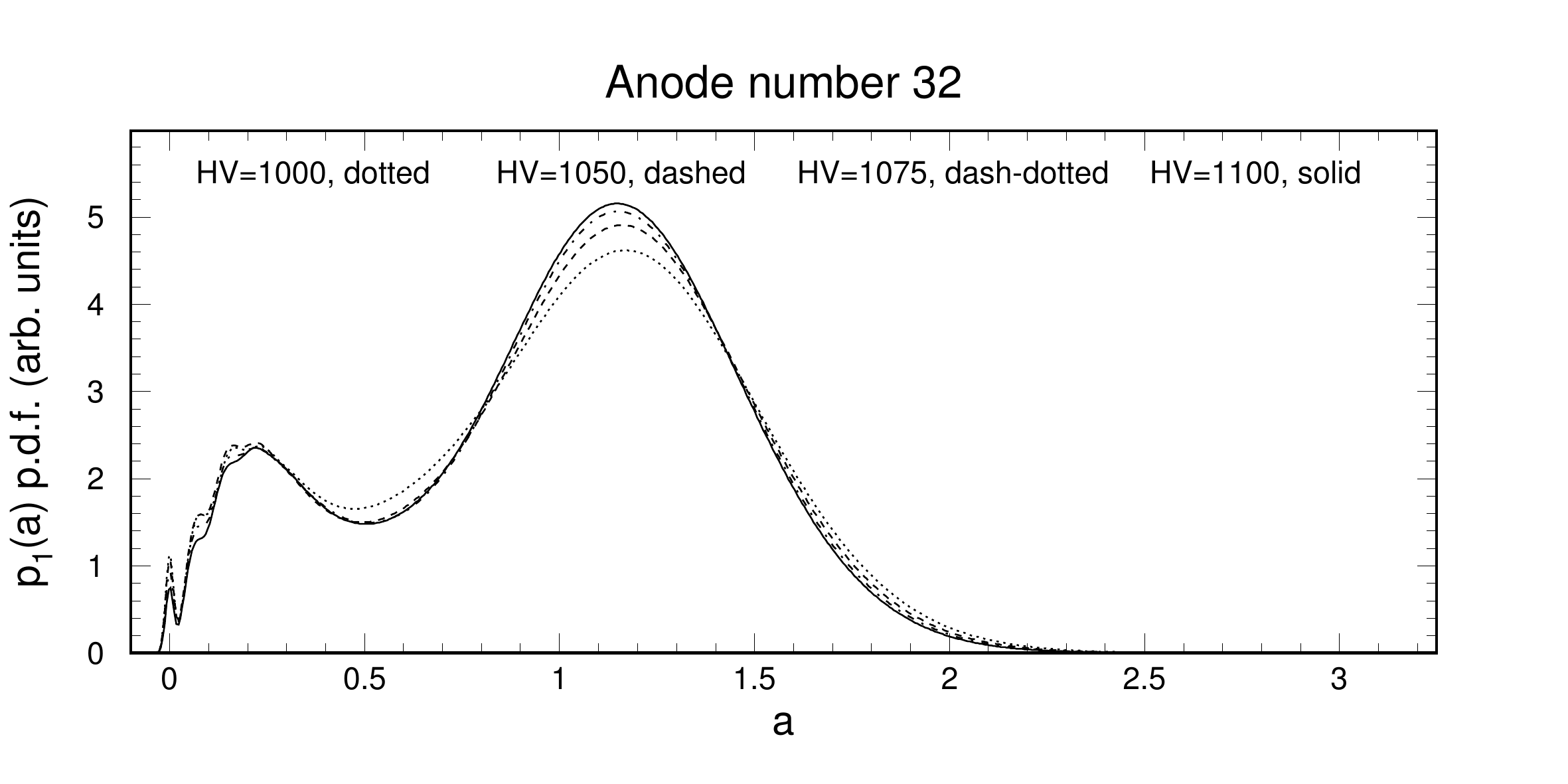}}\hfill
  \subfloat[Hamamatsu H8500 MAPMT CA7782, anode \#32]{%
    \includegraphics[clip=true,trim=0 10 0
      48,width=.48\textwidth,keepaspectratio]
                    {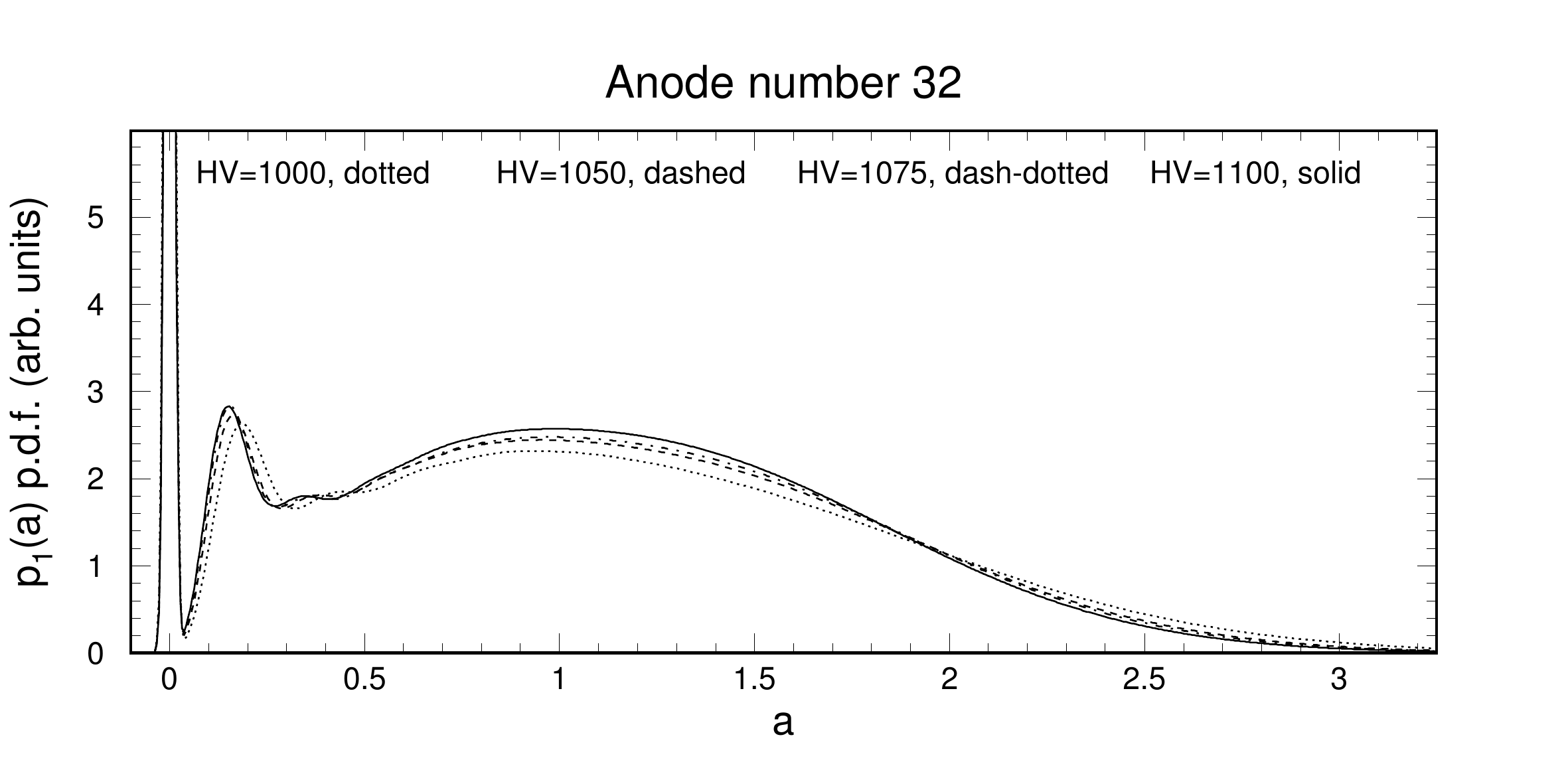}}\\
  \subfloat[Hamamatsu H12700 MAPMT GA0133, anode \#57]{%
    \includegraphics[clip=true,trim=0 10 0
      48,width=.48\textwidth,keepaspectratio]
                    {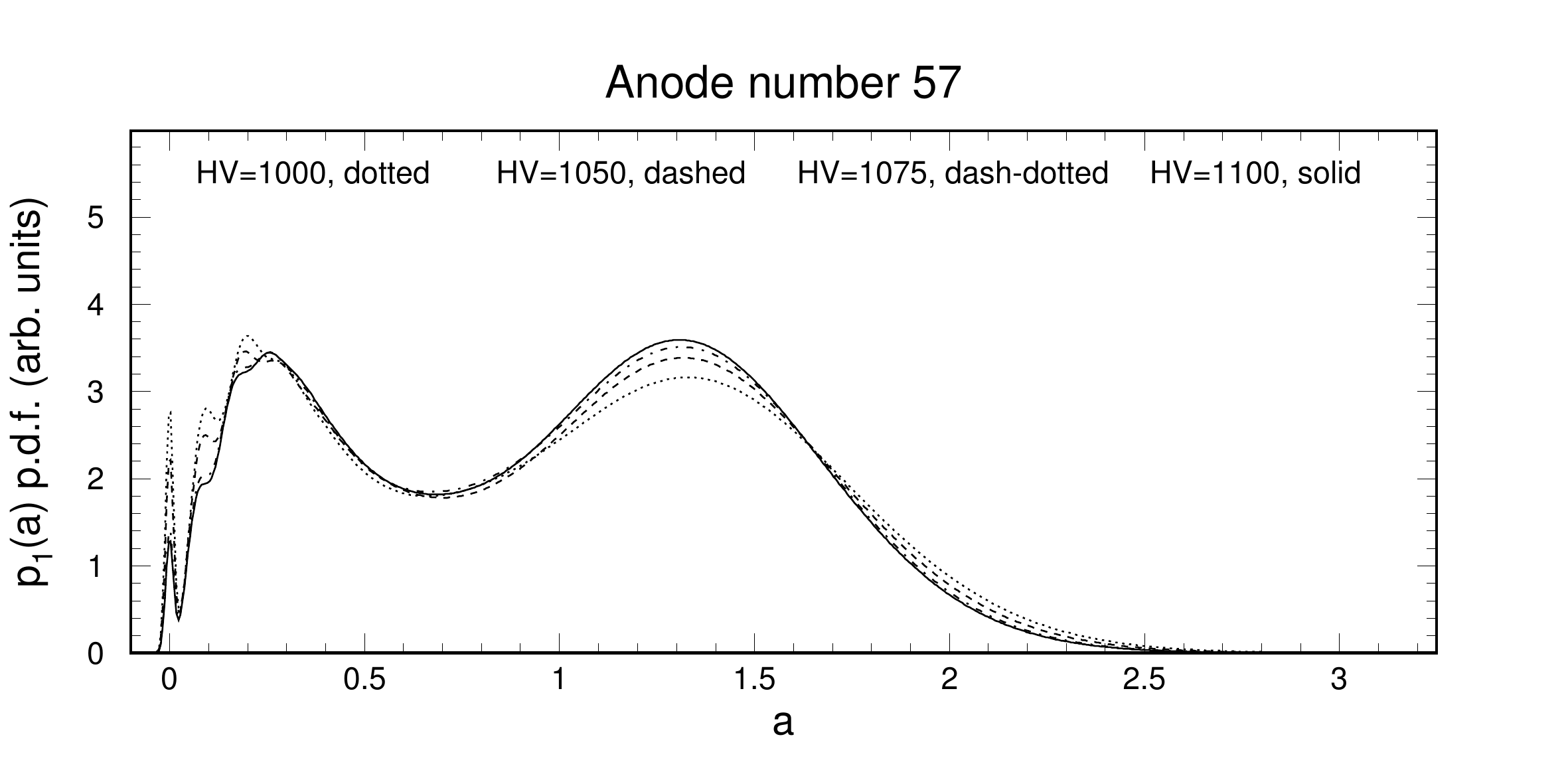}}\hfill
  \subfloat[Hamamatsu H8500 MAPMT CA7782, anode \#57]{%
    \includegraphics[clip=true,trim=0 10 0
      48,width=.48\textwidth,keepaspectratio]
                    {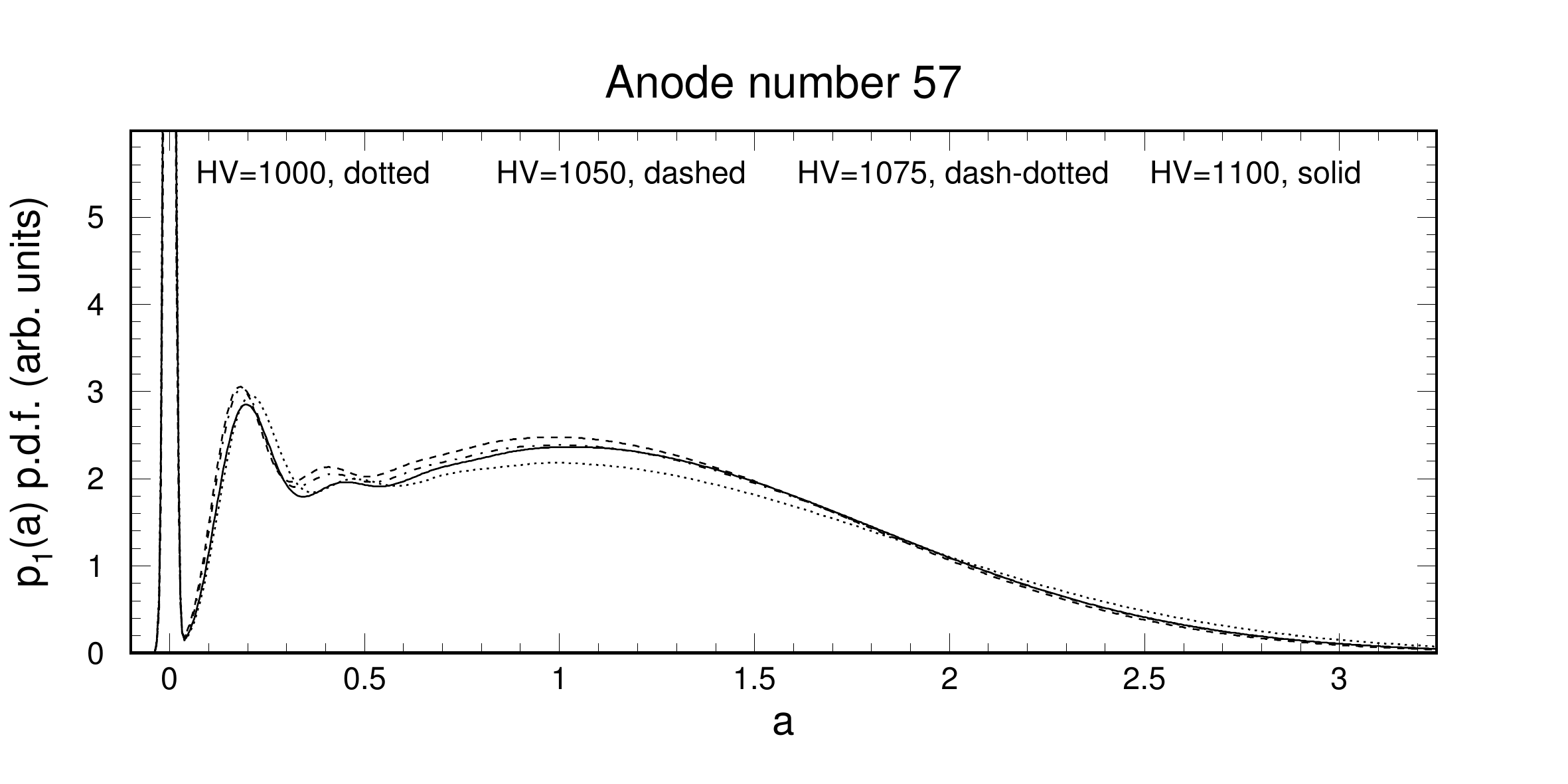}}\\
  \subfloat[Hamamatsu H12700 MAPMT GA0133, anode \#64]{%
    \includegraphics[clip=true,trim=0 10 0
      48,width=.48\textwidth,keepaspectratio]
                    {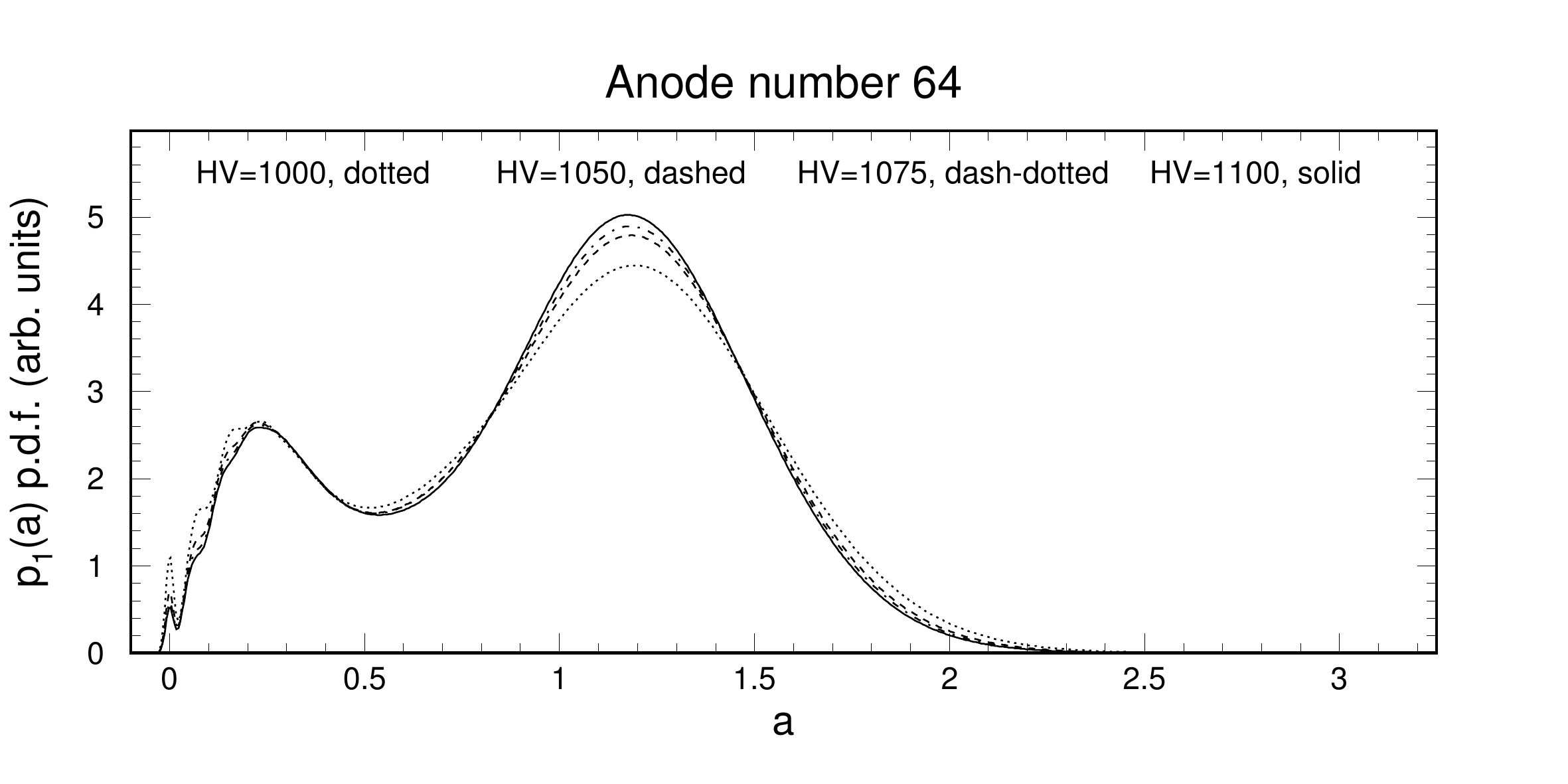}}\hfill
  \subfloat[Hamamatsu H8500 MAPMT CA7782, anode \#64]{%
    \includegraphics[clip=true,trim=0 10 0
      48,width=.48\textwidth,keepaspectratio]
                    {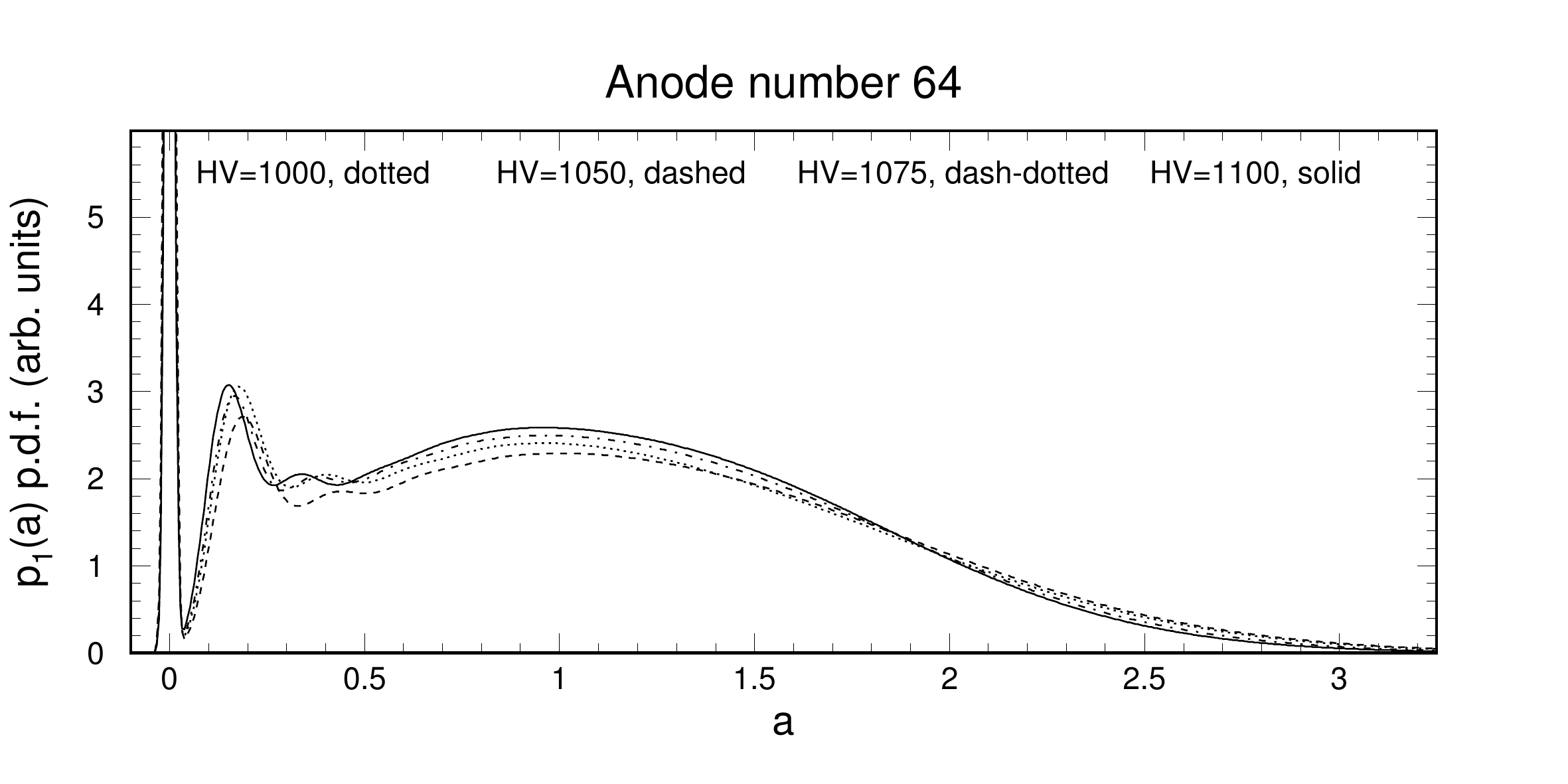}}
\caption{ Each panel shows the SPE p.d.f., measured using the ``global
  fit'' procedure at four high voltage values. Four anodes are shown
  for both types of MAPMTs, corresponding to the anode positions at
  the center, at the center of the edge, and at the two corners of the
  MAPMT's face. All plots show the $p_1(a)$ p.d.f. assuming
  artificially small normalized experimental measurement resolution
  $\sigma / scale = 0.02$.}
\label{fig:SPE}
\end{figure*}

\subsection{Tests of Hamamatsu H8500 and H12700 MAPMTs}

The following example illustrates some of the results of the study of
large number (430) of the Hamamatsu H8500 and H12700 MAPMTs, obtained
in the process of PMT selection for the new RICH detector, which is
presently underway at JLab \cite{RICH,Andrey}.  As opposed to H8500,
the new 10-stage H12700 series of MAPMTs from Hamamatsu
\cite{HamamatsuPMT} is designed specifically to suit better for the
applications requiring reliable single photoelectron detection, such
as RICH detectors.

All the MAPMTs were tested in the conditions of a relatively low light
(three illumination conditions identified as ``OD54'', ``OD50'' and
``OD46'', corresponding to the parameters of average $\mu$ of about
0.06, 0.13, and 0.20, and at four values (1000, 1050, 1075 and 1100
volts) of the operational high voltage (HV) applied.

The total number of measured amplitude distributions recorded and
analyzed is about 340 thousand. The signal measurement system did not
provide perfect Gaussian pedestal amplitude distributions during these
tests. While the pedestal shapes were very close to Gaussian form, the
small statistical errors in the peak made the fitting procedure very
sensitive to the small distortions, and thus unstable.  To avoid
parameterization instabilities caused by the discrepancies between the
ideal Gaussian pedestal shape and the measured pedestal peaks, in
every spectrum the statistical errors in the data points constituting
the pedestal peaks were increased and re-normalized such that the
peaks could be approximated by the Gaussian functions with the new
modified $(\chi^2/n_{\mathrm{d}})_{\mathrm{Gaussian}}$ equal to one.
That way during the multiparameter fitting procedure the disturbed
pedestal peak shapes did not influence the main $\chi^2$ of the full
spectrum minimization. Essentially only pedestal position and
effective Gaussian width were used in the main fitting procedure, not
details of the shape.

The parameters of the SPE spectrum for each of the 27,520 anodes were
obtained using the ``global fit'' procedure. The SPE parameters were
averaged over the runs with different illumination conditions and
fixed in final fits.

Fig.~\ref{fig:rich_lin} shows the characteristic examples of the
spectra measured on one of the central anodes belonging to a MAPMT
H12700 (left panels) and to a MAPMT H8500 (right panels) at four
different applied high voltages from 1000~V to 1100~V, together with
the model approximations.

Each plot shows the quality of the overall fit of the data by the
model function, mostly defined by the SPE contribution at such
low-light test conditions.  The significant increase of the $scale$
parameter with the increase of the applied high voltage may be
seen clearly, corresponding to the well-known dependence of PMT gains
on the applied high voltage. Notice that the extracted values of $\mu$
parameter are quite stable and practically do not depend on HV. The
H12700 MAPMTs generally exhibit a more prominent high-$\nu$ component
of the SPE spectrum compared with the H8500 tubes.

The model-approximated SPE spectra measured for several anodes of the
sample MAPMTs, including those corresponding to the set of plots from
Fig.~\ref{fig:rich_lin}, are shown in Fig.~\ref{fig:SPE}, function of
the normalized amplitude $a$. Despite the strong dependence of the
$scale$ parameter on the applied high voltage observed earlier, the
shapes of the SPE spectra function of $a$ are stable and only slightly
depend on the HV, possibly due to the changes in the average
multiplicity $\nu$ of the second-stage electrons knocked from the
first dynode. Qualitatively this result may be understood such that as
energy of the photoelectron acceleration from the photocathode to the
first dynode increases at higher voltages, the average number of the
knocked-out electrons increases slightly. Such pattern is observed in
all anodes and all photomultipliers in the study.

\begin{figure*}[h!] 
\centering 
  \subfloat[Parameter $scale$, proportional to the overall PMT gain at
    each anode. H12700 MAPMT ``GA0133'']{
    \includegraphics[clip=true,trim=0 10 0
      27,width=.48\textwidth,keepaspectratio]
                    {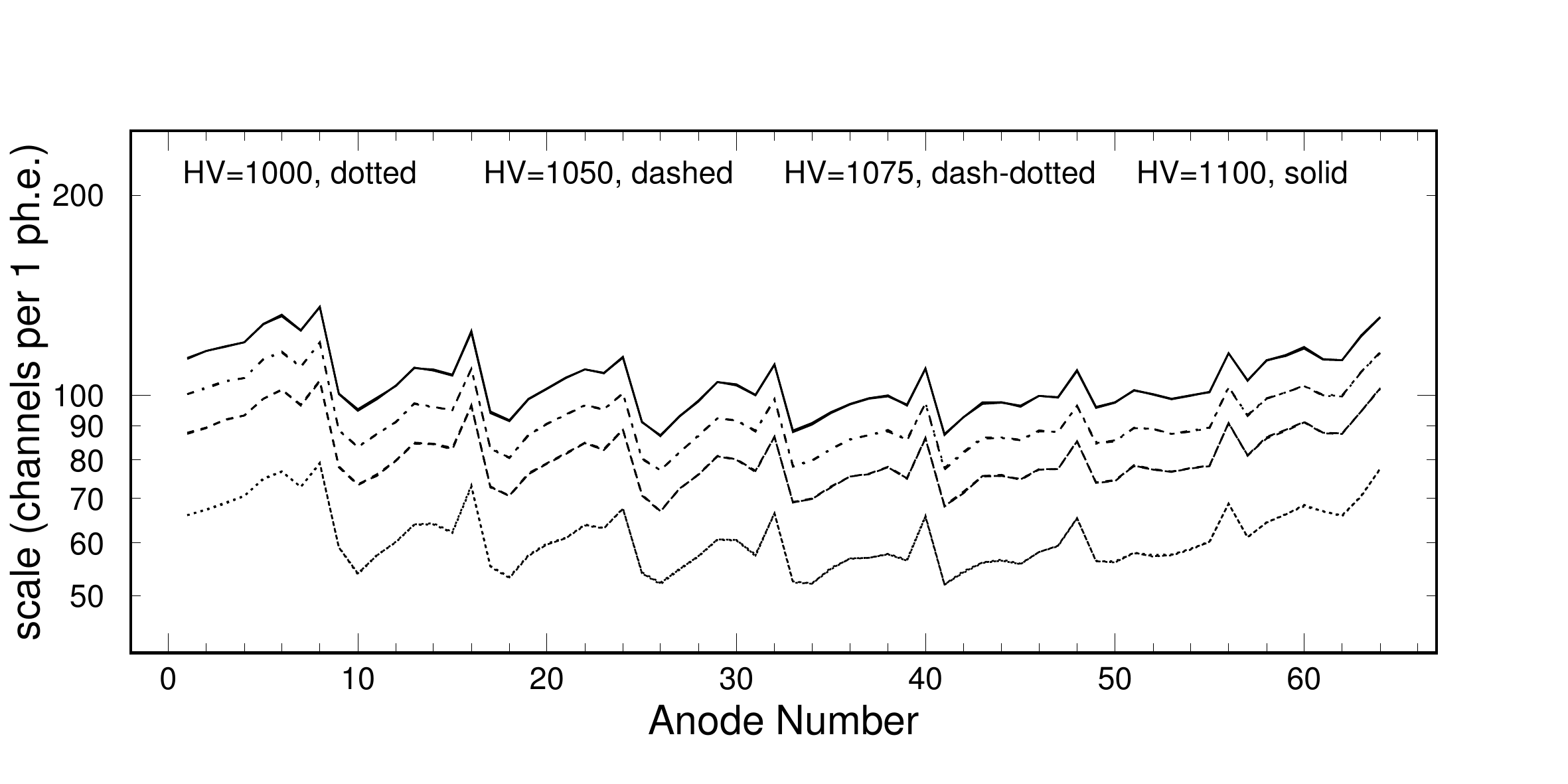}}\hfill
  \subfloat[Same as in (a), but for the H8500 MAPMT ``CA7782'']{
    \includegraphics[clip=true,trim=0 10 0
      27,width=.48\textwidth,keepaspectratio]
                    {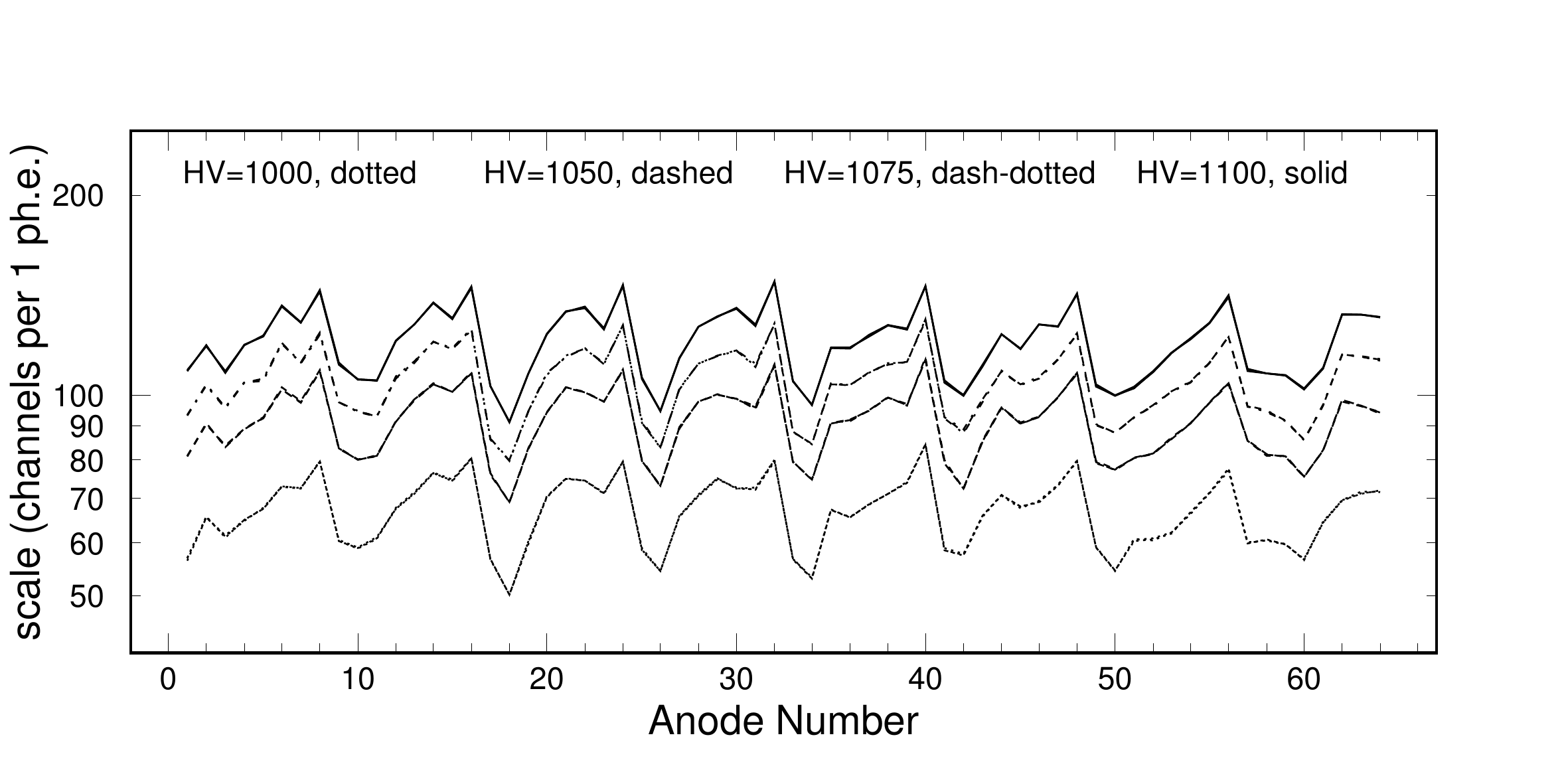}}\\
  \subfloat[Parameter $\mu$, proportional to the light intensity at
    each anode. H12700 MAPMT ``GA0133'']{
    \includegraphics[clip=true,trim=0 10 0
      27,width=.48\textwidth,keepaspectratio]
                    {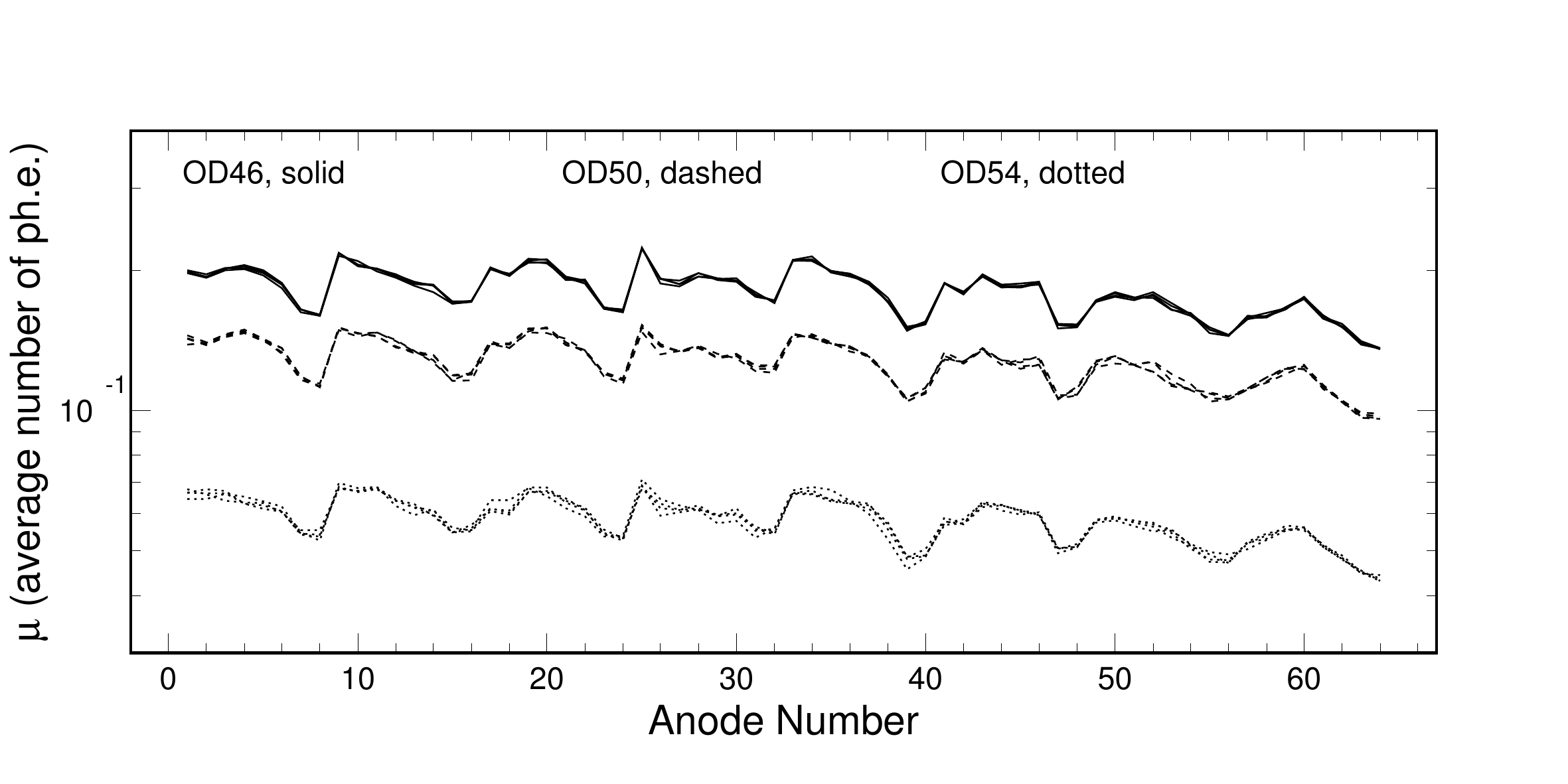}}\hfill
  \subfloat[Same as in (c), but for the H8500 MAPMT ``CA7782'']{
    \includegraphics[clip=true,trim=0 10 0
      27,width=.48\textwidth,keepaspectratio]
                    {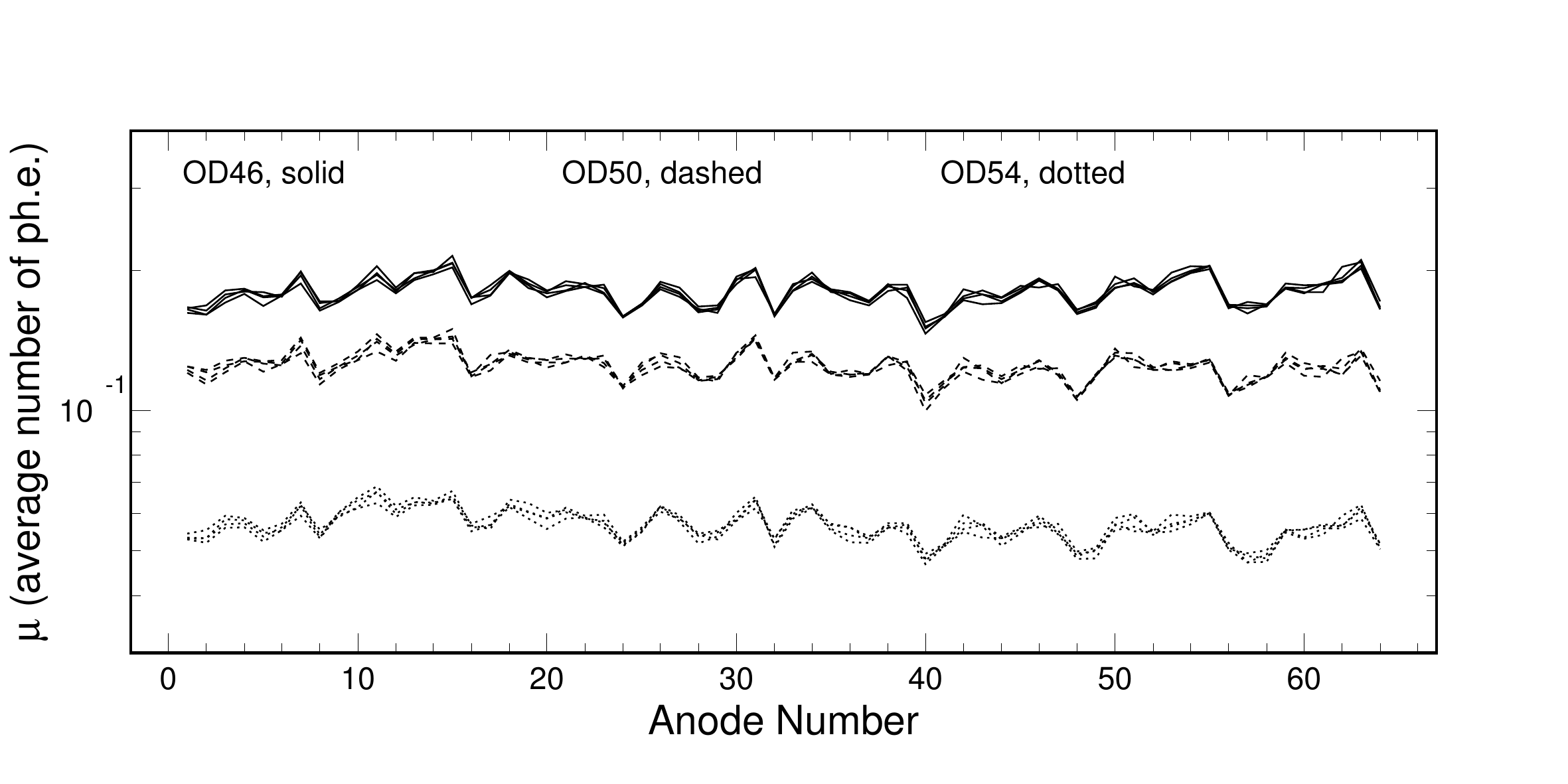}}\\
  \subfloat[Average number of the second-stage electrons knocked out by one
  photoelectron, for each anode. H12700 MAPMT ``GA0133'']{
    \includegraphics[clip=true,trim=0 10 0
      27,width=.48\textwidth,keepaspectratio]
                    {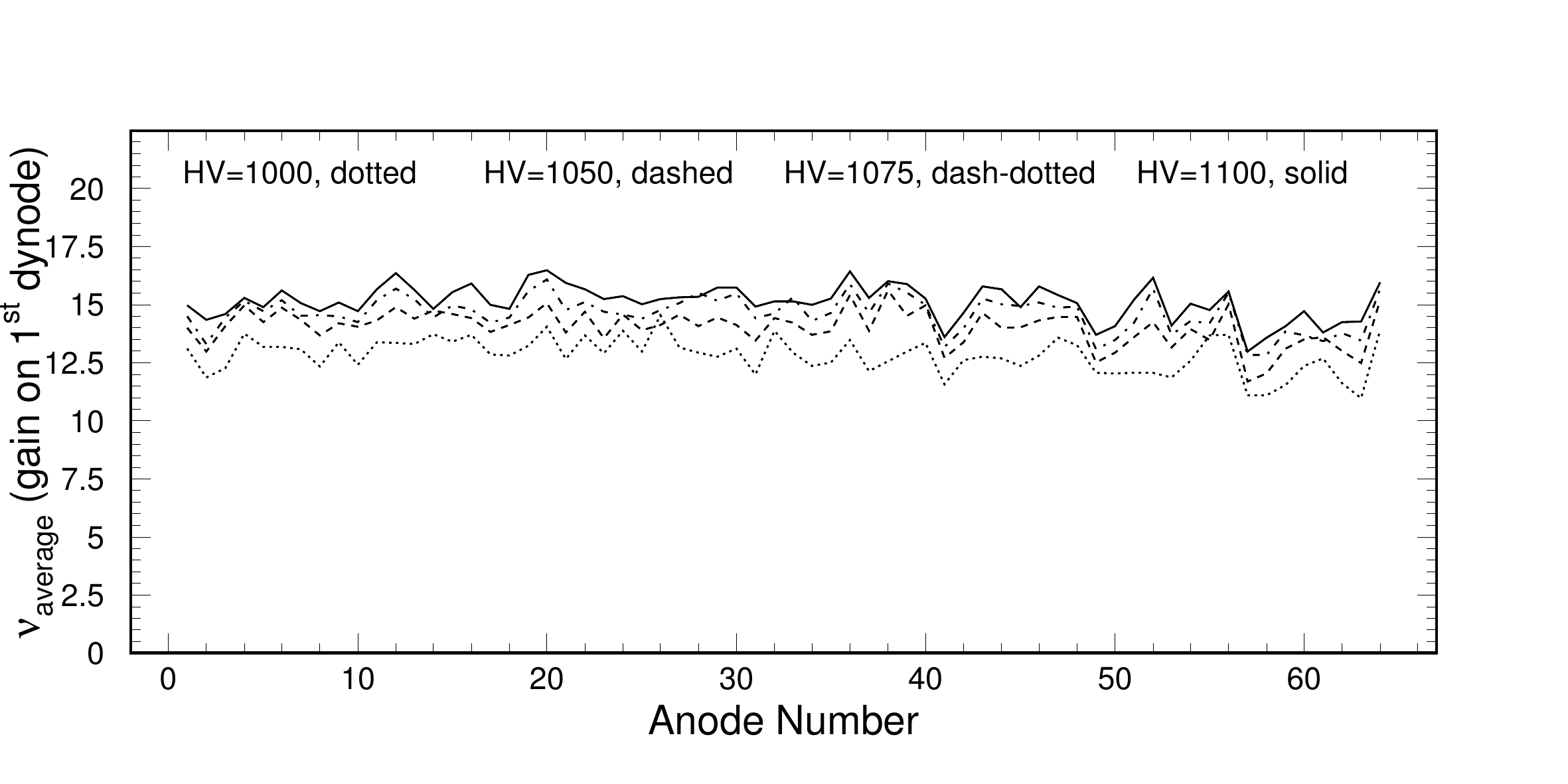}}\hfill
  \subfloat[Same as in (e), but for the H8500 MAPMT ``CA7782'']{
    \includegraphics[clip=true,trim=0 10 0
      27,width=.48\textwidth,keepaspectratio]
                    {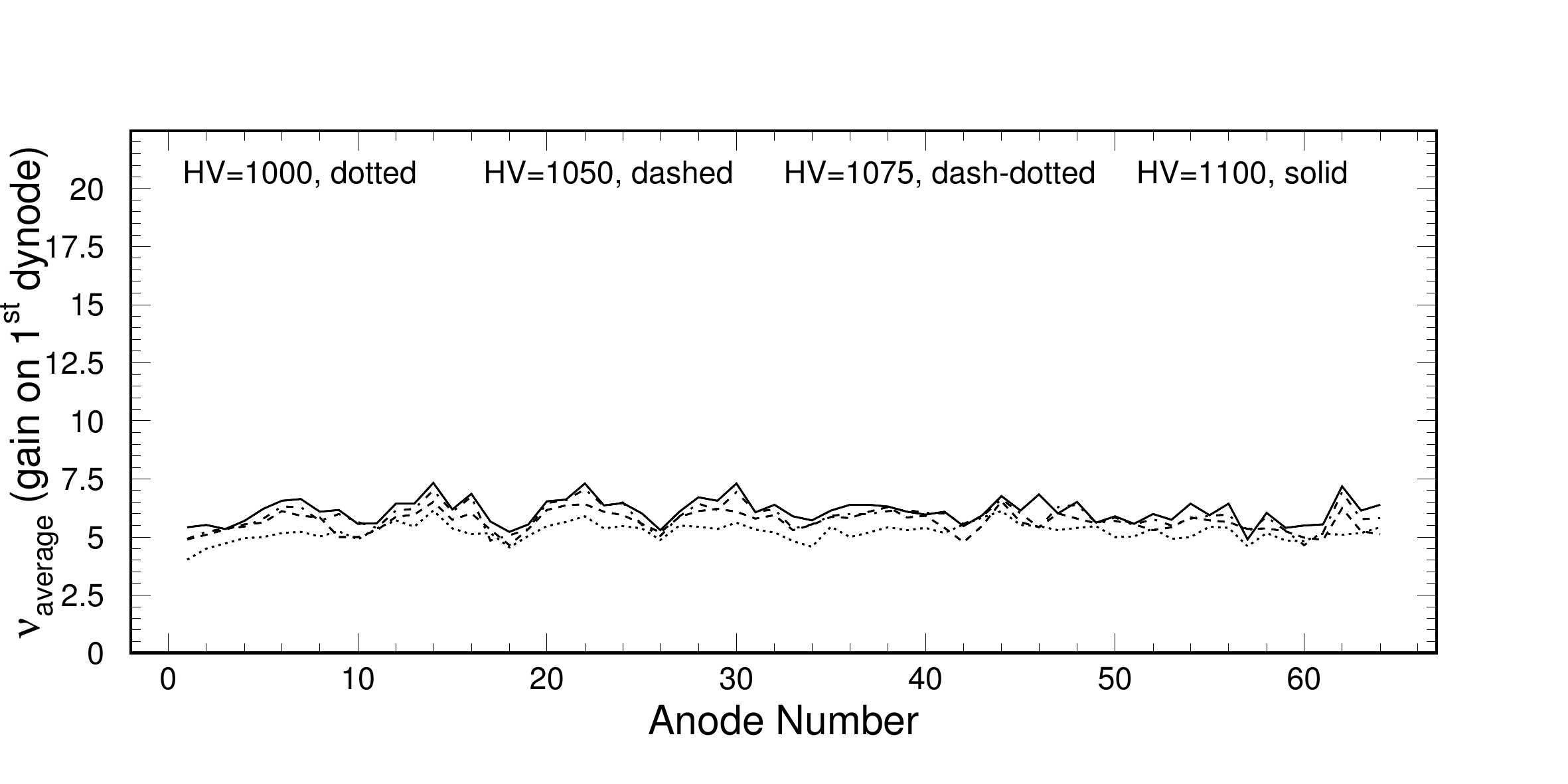}}\\
  \subfloat[Efficiency $\varepsilon$ of one photoelectron
    detection at each anode. H12700 MAPMT ``GA0133'']{
    \includegraphics[clip=true,trim=0 10 0
      27,width=.48\textwidth,keepaspectratio]
                    {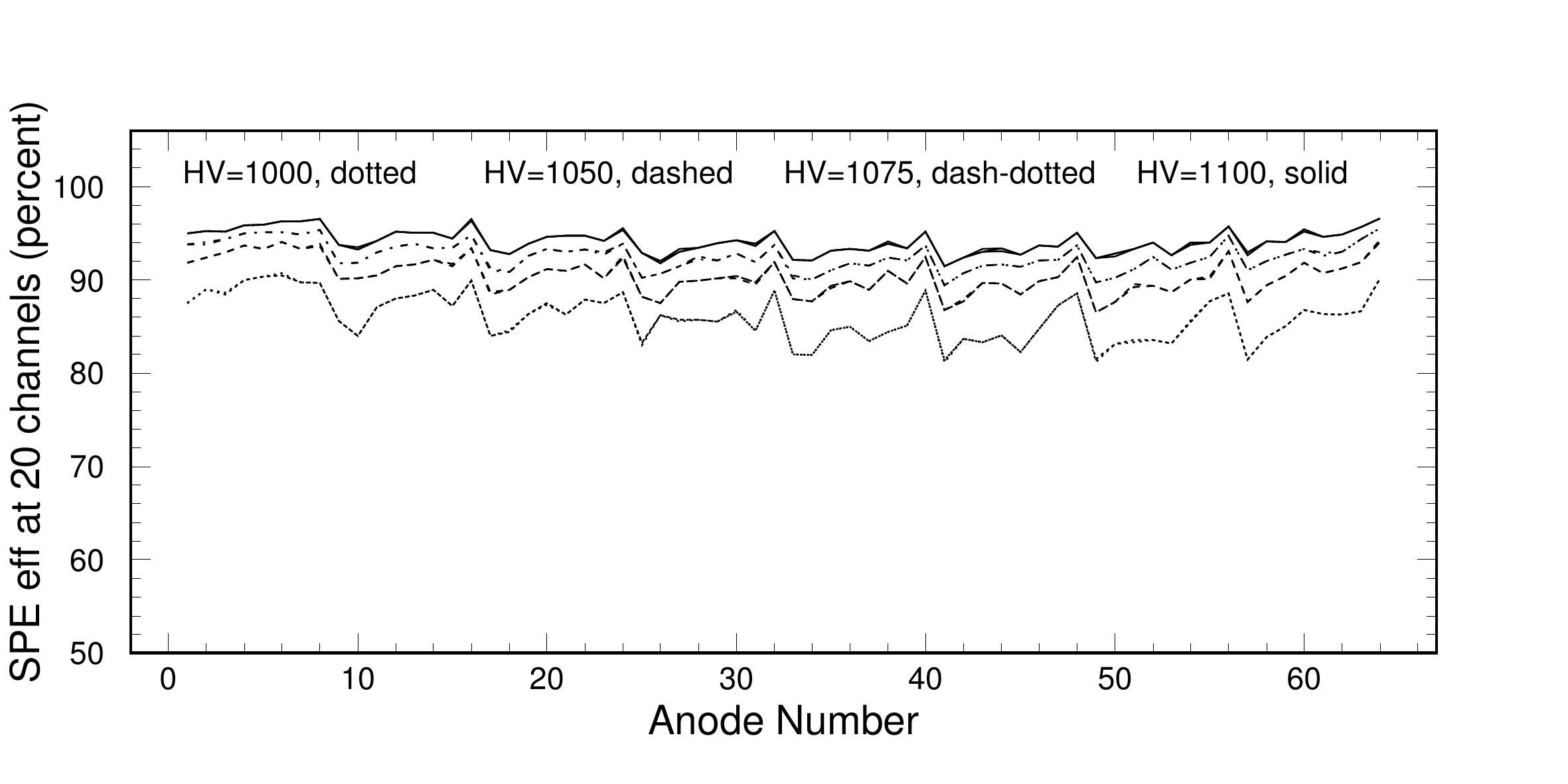}}\hfill
  \subfloat[Same as in (g), but for the H8500 MAPMT ``CA7782'']{
    \includegraphics[clip=true,trim=0 10 0
      27,width=.48\textwidth,keepaspectratio]
                    {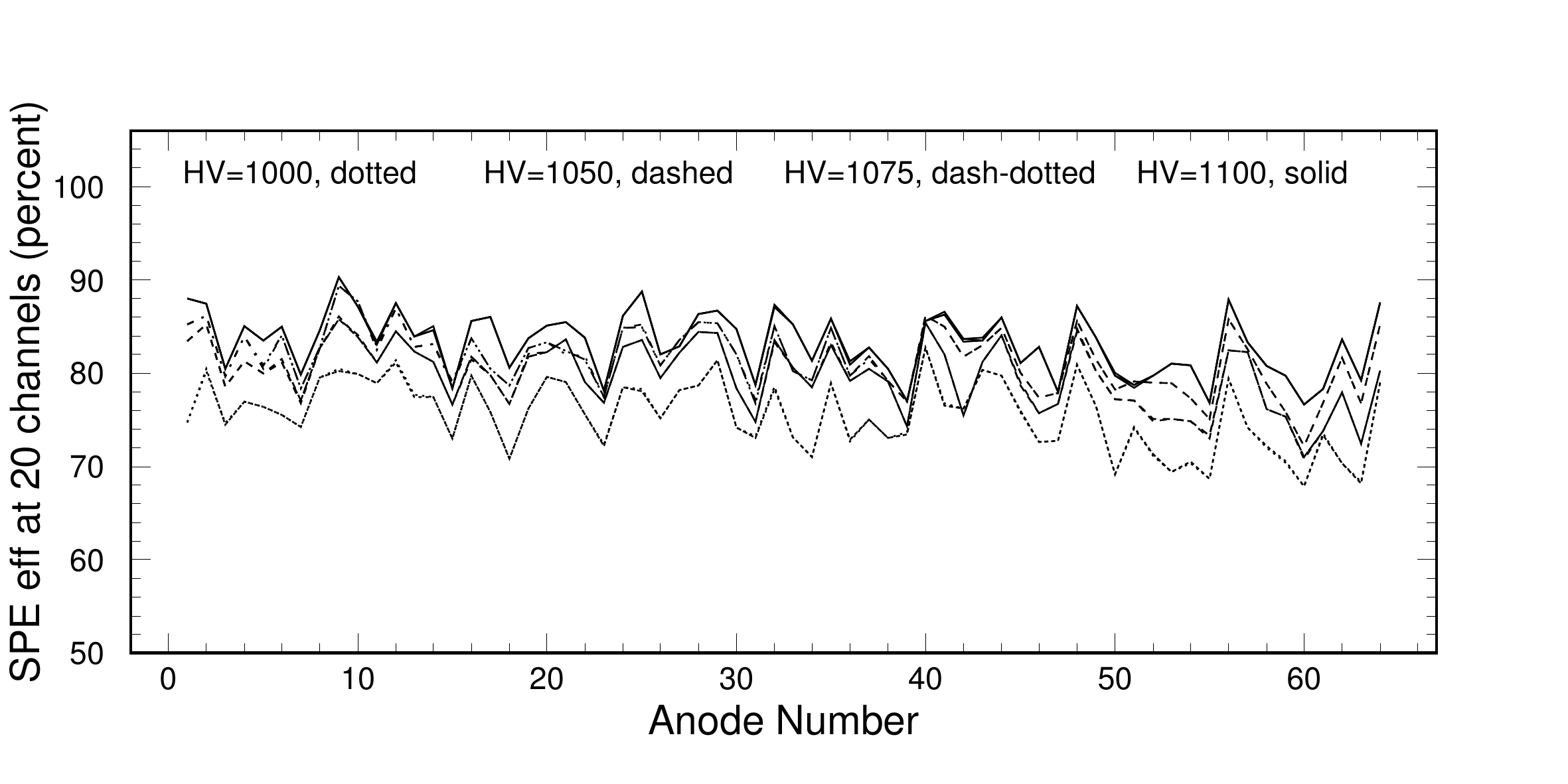}}
\caption{MAPMT passport plots: a selection of the model parameters
  $scale$, $\mu$, and the derived values of $\nu$ and $\varepsilon$,
  evaluated using the ``global fit'' procedure for the two sample
  devices, Hamamatsu H12700 MAPMT ``GA0133'' (left panels), and H8500
  MAPMT ``CA7782'' (right panels), plotted as functions of the anode
  numbers of these MAPMTs.  All twelve data sets are shown in each
  plot, corresponding to the three illumination conditions measured at
  each of four different applied high voltages.}
\label{fig:Passport}
\end{figure*}

The variability of the SPE parameters between different anodes in each
MAPMT is found to be quite significant. Also significant is the
difference between average SPE parameters for H8500 and H12700 MAPMT
types. Fig.~\ref{fig:Passport} illustrates this statement by showing
some of the ``PMT Passport'' plots for the above two MAPMT example
devices.  Model approximation parameters were obtained for every anode
independently using the ``global fit'' procedure, and plotted as a
function of the anode number for every photomultiplier. Model
parameters $scale$, $\mu$, and the derived values of
$\nu_{\mathrm{average}}$ and $\varepsilon$ on the left panels in
Fig.~\ref{fig:Passport} are obtained for the H12700 example MAPMT, and
corresponding right panels show the results for the H8500 MAPMT. The
SPE efficiency evaluation parameter $\varepsilon$ will be discussed
further in the text.

The top panels (a) and (b) in Fig.~\ref{fig:Passport} illustrate
typical variable patterns of the $scale$ parameters as a function of
anode number.  The plots show all twelve test conditions that the
MAPMTs were subjected to in this study, namely, four HV values times
three light conditions. Quite striking feature of the model
approximation is that the extracted $scale$ parameters do not depend
on the light conditions to a very high degree of accuracy, such that
those differences practically cannot be resolved on the plots. The
dependence of the $scale$ parameter on HV is on the other hand quite
clear and corresponds to the well-known characteristic exponential
dependence of output amplitudes (PMT gain) on high voltage. The data
sets, measured at different high voltages and plotted as a function of
the anode number, look essentially parallel in logarithmic scale in
the plots, meaning that their difference may be approximately
attributed to multiplication by a factor.

Panels (c) and (d) in Fig.~\ref{fig:Passport} are complementary to the
previous two in a sense that they show the stability of the model in
determining the model parameter $\mu$ during the varying test
conditions. Naturally $\mu$ must be proportional to the average light
delivered during the test, and ideally it wouldn't depend on the HV
applied.  These regularities are generally observed in the data. As
the irradiation of the MAPMT face was uniform, $\mu$ measured in each
of the 64 channels change in sync with changing light conditions. The
dependence of the $\mu$ parameter on the high voltage applied is very
minimal, and possibly could be explained by such effects as the tiny
increase in the probability of photoelectron emission in higher
gradients of electric fields in the photocathode region, or by better
focusing of the photoelectrons at higher voltages. However, these
hypotheses weren't further investigated in this work.

Panels (e) and (f) in Fig.~\ref{fig:Passport} show the $\nu$ derived
value as defined in Eq.~(\ref{nu_average}), function of the anode
number for the sample MAPMTs. The values of $\nu$ are averaged over
the three light conditions at each of the HV settings using the
``global fit'' procedure. The difference in $\nu$ values between the
H12700 and H8500 MAPMT models is quite significant and is observed in
other MAPMTs through the whole data set. Most likely explanation of
this observation is the difference in the design of these
MAPMTs. Other typical feature that could be seen in these two panels
is the relatively weak, but noticeable, dependence of $\nu$ on the
high voltage applied. Such dependence of $\nu$ on HV may be
qualitatively understood as increasing probability of knocking out
electrons from the first dynode at higher voltages due to higher
energy that a photoelectron acquires when accelerating from the
photocathode to the first dynode.

Panels (g) and (h) in Fig.~\ref{fig:Passport} illustrate one of the
possible final goals of such studies: evaluate efficiency
$\varepsilon$ of the photoelectron detection by the
photodetectors. Here $\varepsilon$ is defined as the probability of
events distributed according to the evaluated SPE amplitude
distributions $p_1(s)$ to have their signal amplitude $s$ above 20
channels ADC or QDC as recorded by the signal measurement system
during these tests. The value of $\varepsilon$ generally varies from
anode to anode, as shown in the plots in correlation with the anode
gain, which in turn depends on the high voltage applied. The
efficiency is systematically higher for the H12700 MAPMT series,
despite generally higher $scale$ parameters observed for the H8500
MAPMTs.

The overall features of the massive analyzed MAPMT data set are
presented in the following plots.

\begin{figure}[h!] 
 \centering 
  \includegraphics[clip=true,trim=0 10 0
    27,width=.50\textwidth,keepaspectratio] {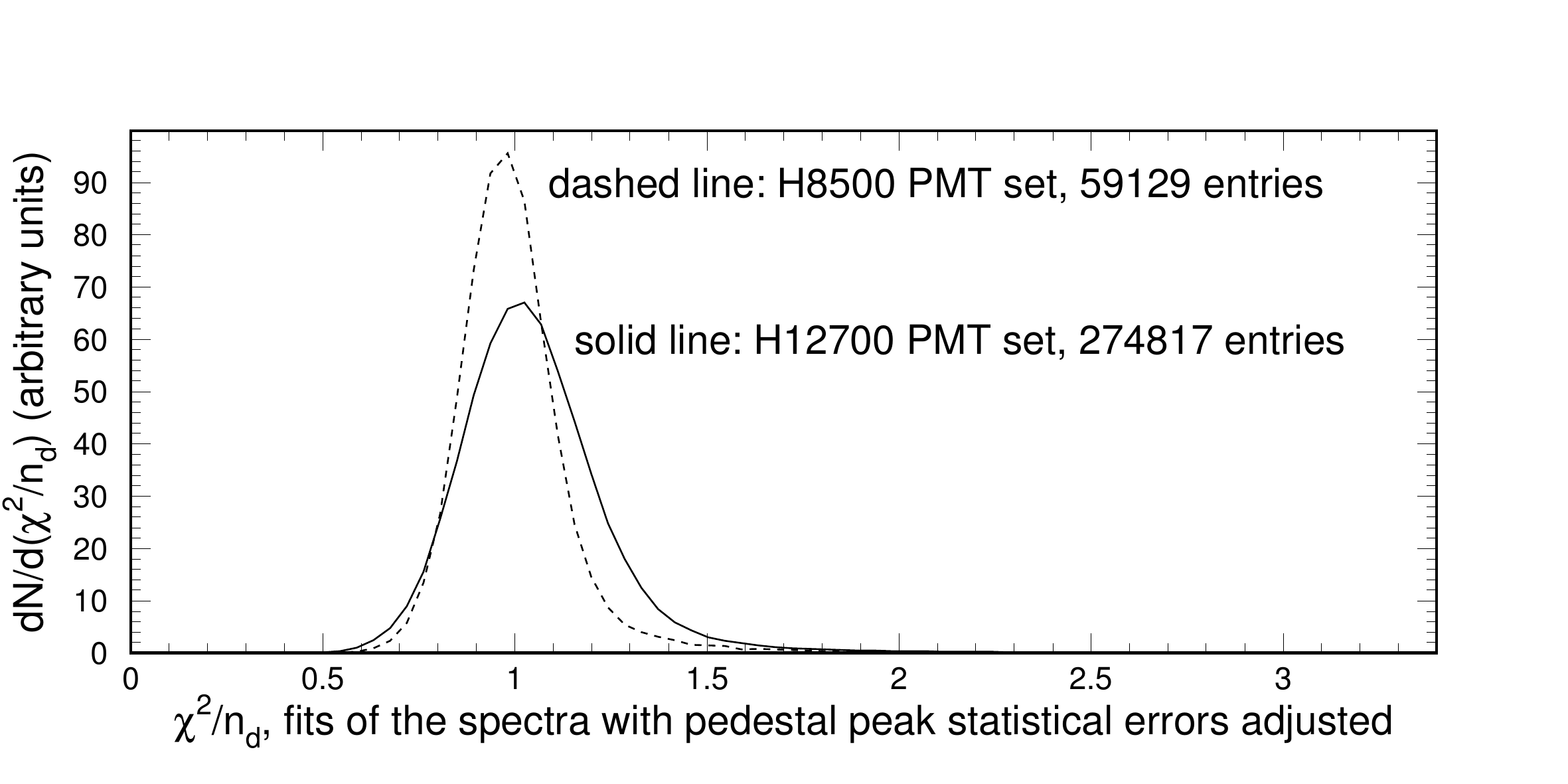}
\caption{Distribution of the goodness-of-the-fit evaluator
  $\chi^2/n_{\mathrm{d}}$ on the number of model
  parameterizations. Dashed line shows the H8500 set, solid line - the
  set of fits for the H12700 MAPMTs. The distributions are normalized
  to equal areas in the plot.}
\label{fig:chi2}
\end{figure}

Fig.~\ref{fig:chi2} shows the distributions of the goodness-of-fit
evaluator $\chi^2/n_{\mathrm{d}}$ for about 60000 parameterizations
for the H8500 MAPMTs in this study, and about 275000 parameterizations
for the H12700 MAPMT series. The $\chi^2/n_{\mathrm{d}}$ value is
taken from the last stage of the ``global fit'' procedure in which the
six parameters characterizing the SPE spectra were averaged and fixed
for the 3 setups at different light conditions, and other variables
were optimized to minimize the $\chi^2$. As it was explained above,
the values of statistical errors in the bins around the pedestal in
the raw spectra were artificially adjusted to make the fit insensitive
to the slightly non-Gaussian shape of the measured peak and avoid fit
instabilities.  The $\chi^2/n_{\mathrm{d}}$ distributions are
normalized to equal areas under the curves. While both distributions
indicate to a reasonably good quality of the fits, the H8500 series is
closer to being ``theoretically perfect'', and the H12700 series
distribution has more instances of the fits with a somewhat less than
perfect quality.

\begin{figure*}[h!] 
 \centering 
  \subfloat[Distributions of $scale$ parameter for the two MAPMT data
    sets]{
    \includegraphics[clip=true,trim=0 10 0
      27,width=.48\textwidth,keepaspectratio]
                    {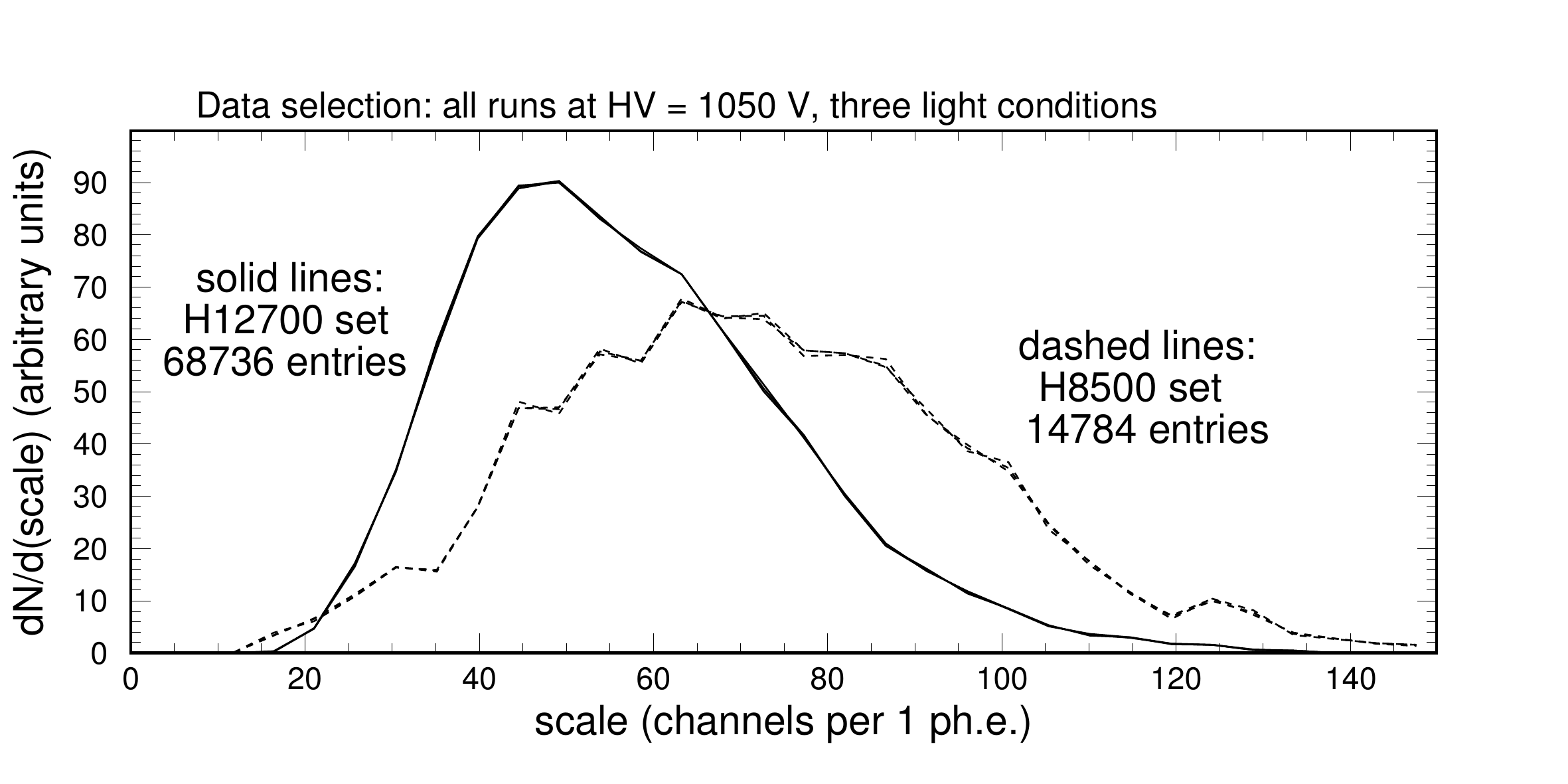}}\hfill
  \subfloat[Light-averaged distributions of $scale$ parameter for the
    two MAPMT data sets, in the two HV settings]{
    \includegraphics[clip=true,trim=0 10 0
      27,width=.48\textwidth,keepaspectratio]
                    {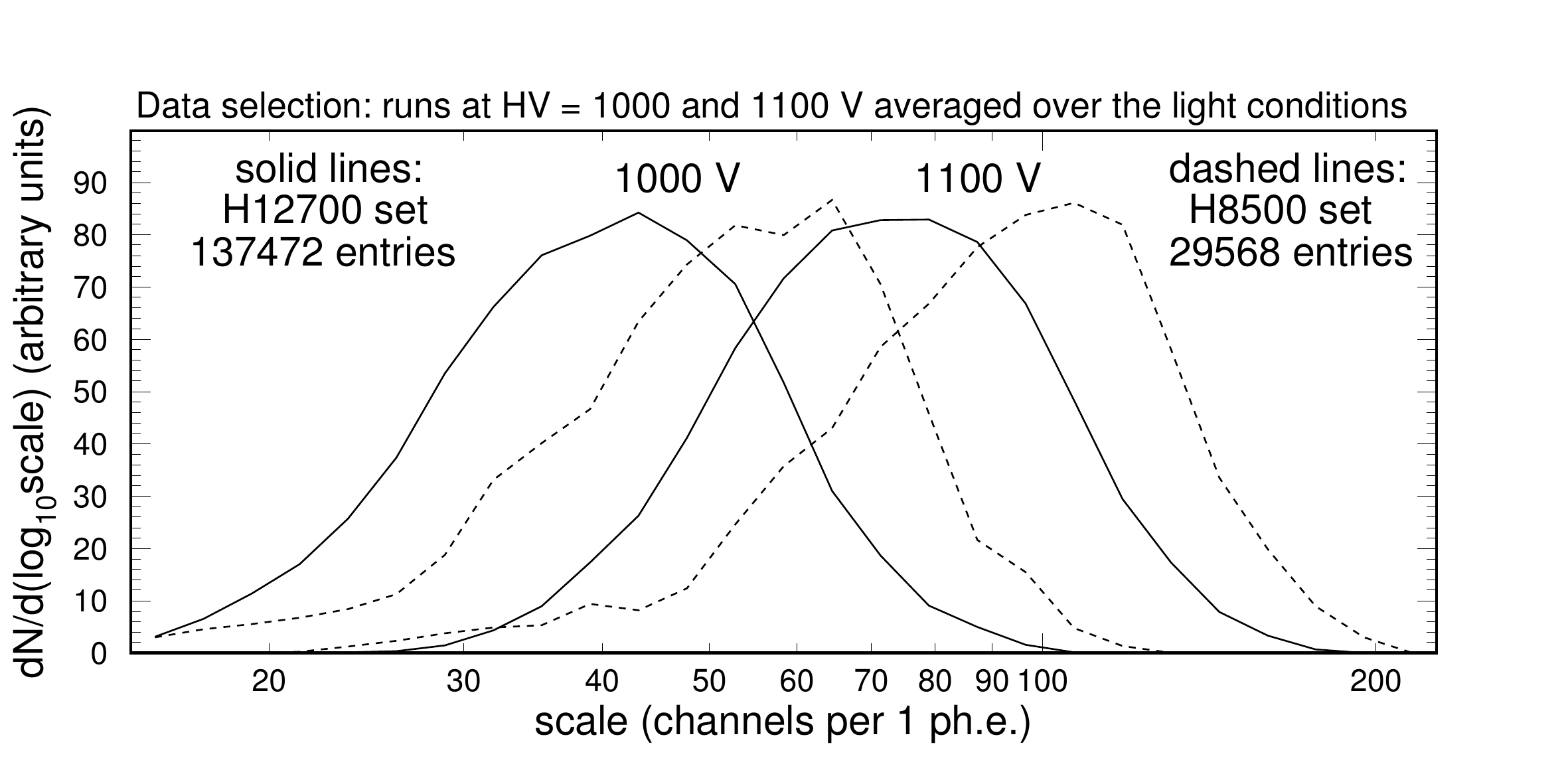}}\\
  \subfloat[Distributions of $\mu$ parameter for the two MAPMT data
    sets]{
    \includegraphics[clip=true,trim=0 10 0
      27,width=.48\textwidth,keepaspectratio]
                    {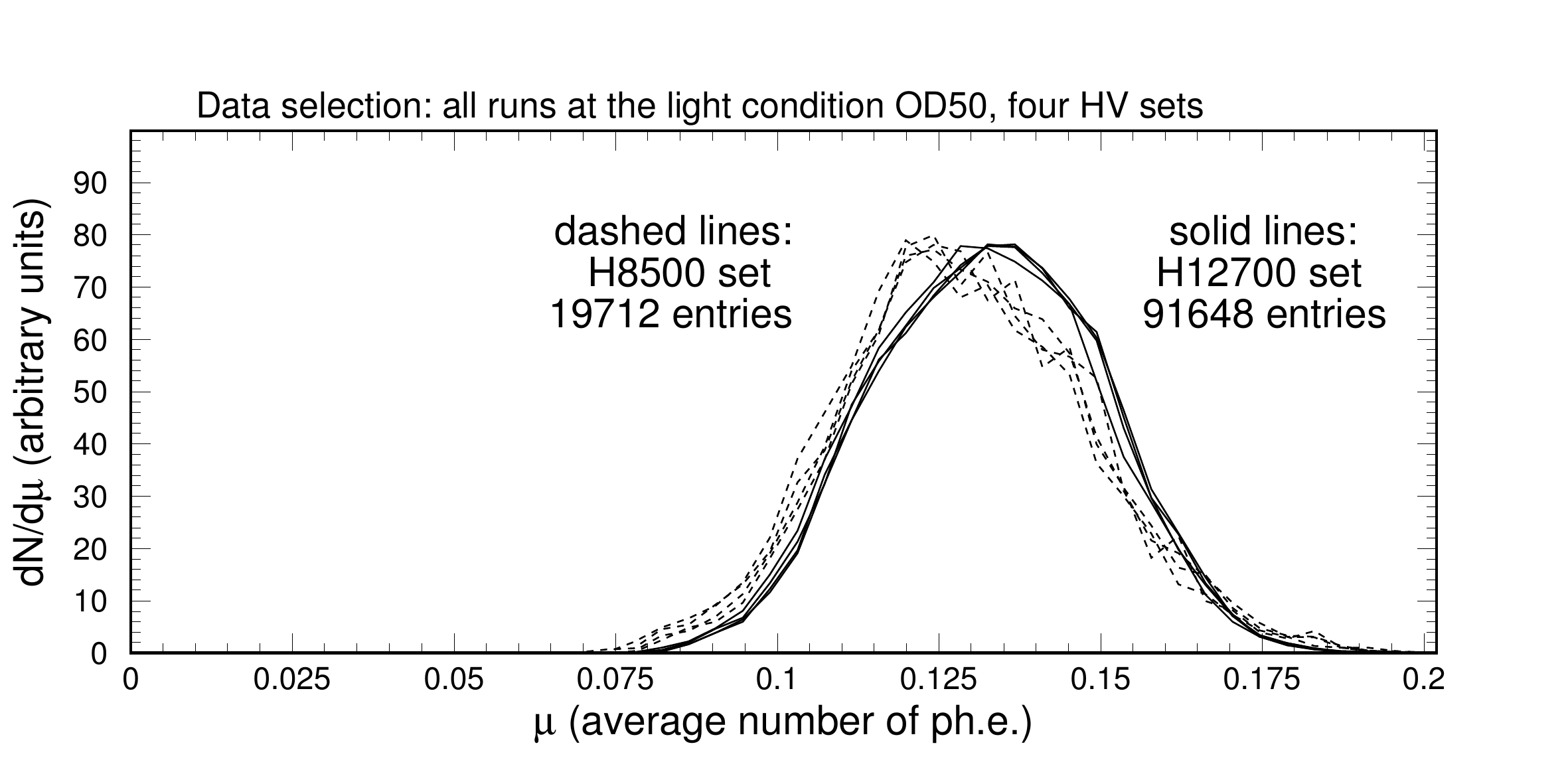}}\hfill
  \subfloat[HV-averaged distributions of $\mu$ parameter for the two
    MAPMT data sets, in the two light conditions]{
    \includegraphics[clip=true,trim=0 10 0
      27,width=.48\textwidth,keepaspectratio]
                    {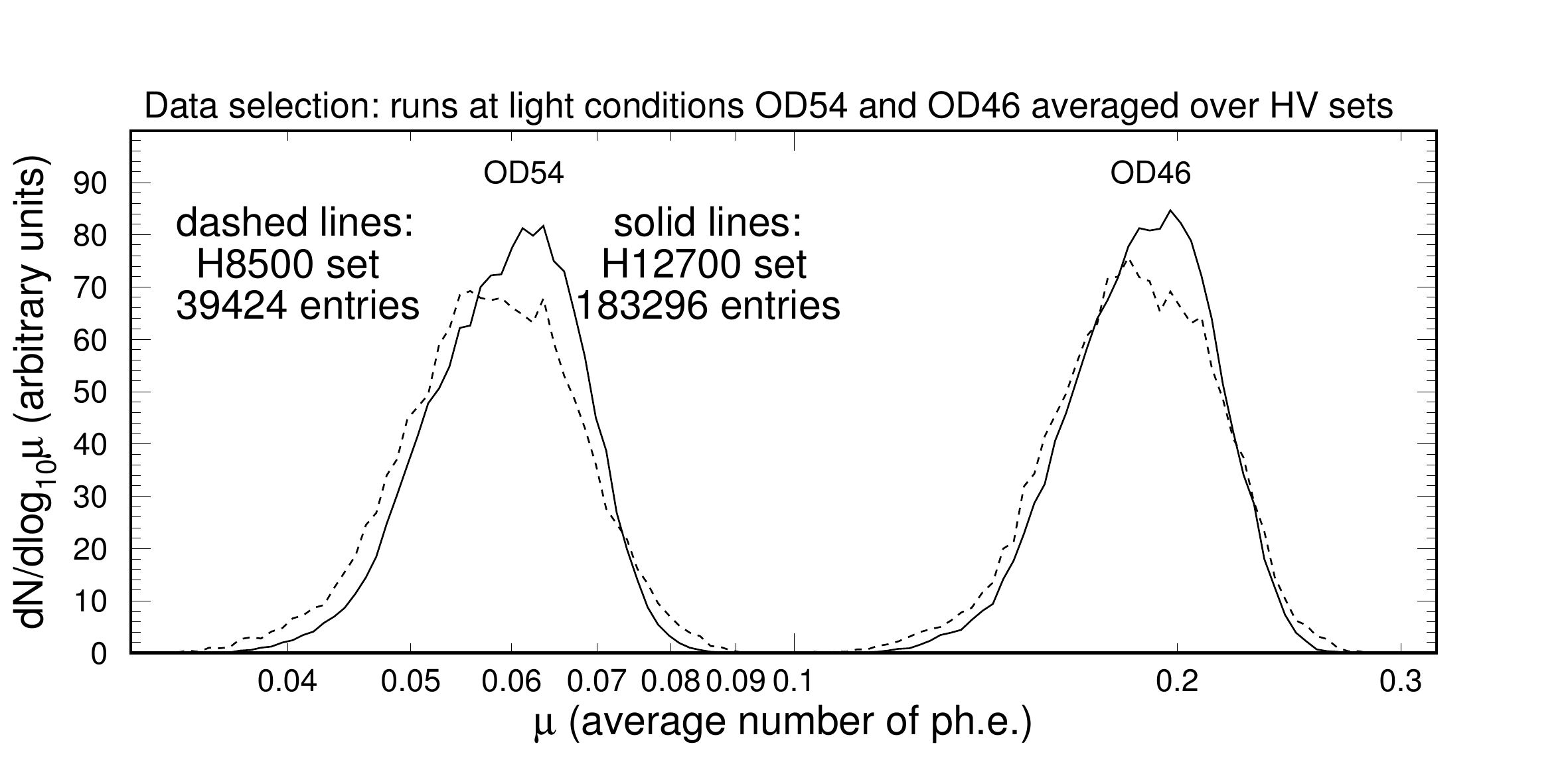}}\\
  \subfloat[Distributions of $\nu$ value for the two MAPMT data
    sets]{
    \includegraphics[clip=true,trim=0 10 0
      27,width=.48\textwidth,keepaspectratio]
                    {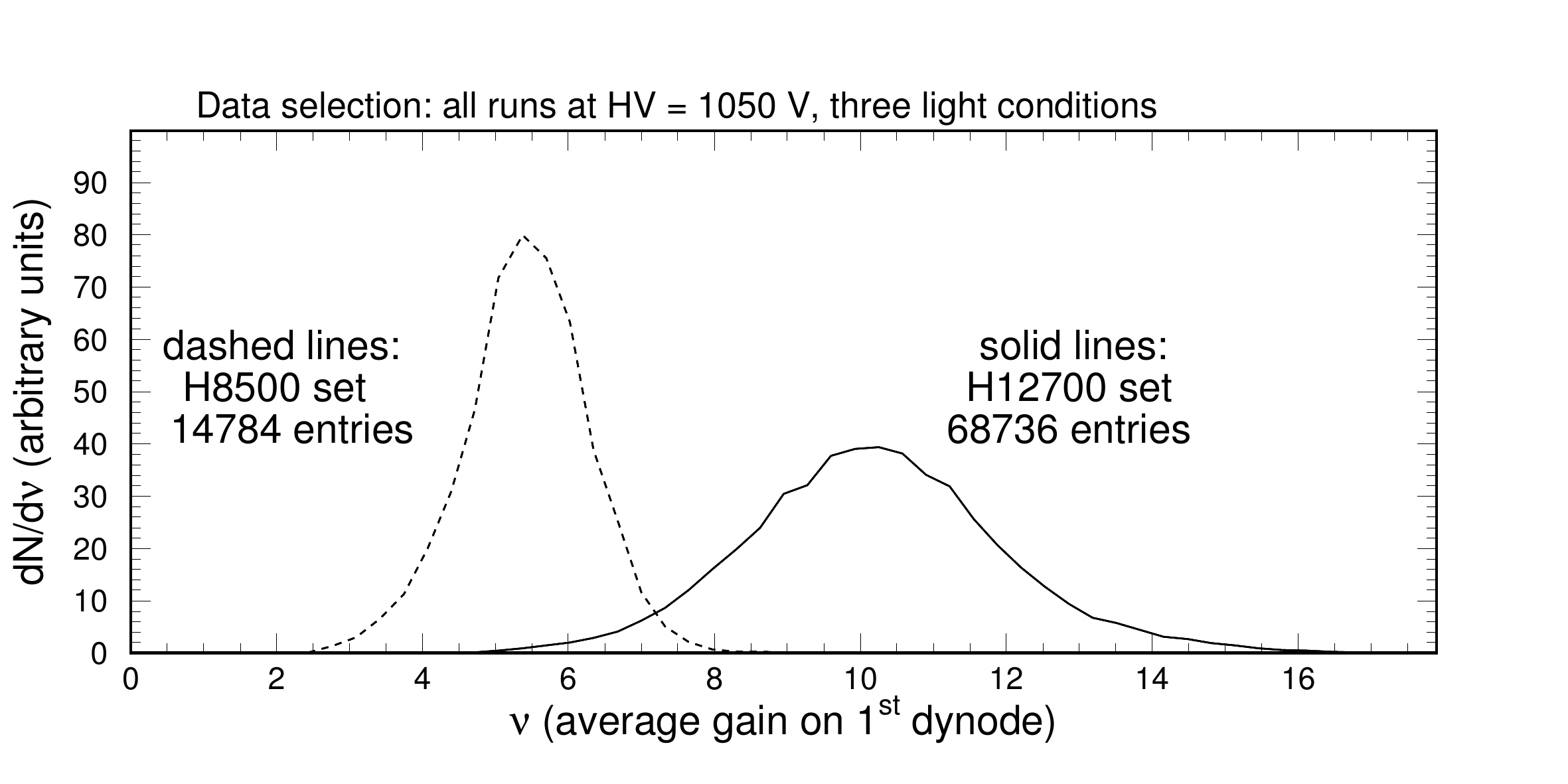}}\hfill
  \subfloat[Light-averaged distributions of $\nu$ value for the
    two MAPMT data sets, in the two HV settings]{
    \includegraphics[clip=true,trim=0 10 0
      27,width=.48\textwidth,keepaspectratio]
                    {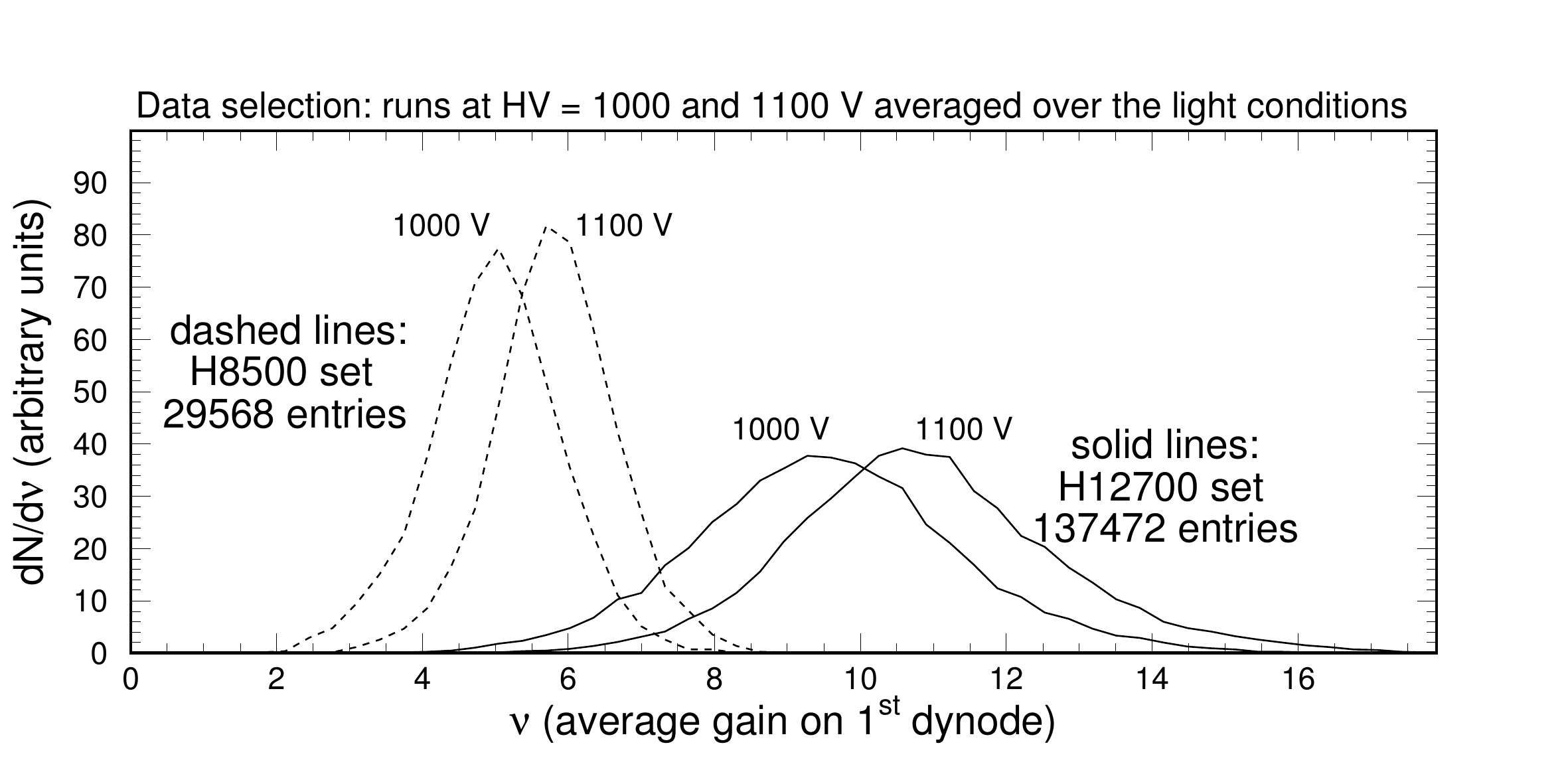}}\\
  \subfloat[Distributions of the light detection efficiency value $\mu
    \varepsilon$ for the two MAPMT data sets at HV = 1050~V]{
    \includegraphics[clip=true,trim=0 10 0
      27,width=.48\textwidth,keepaspectratio]
                    {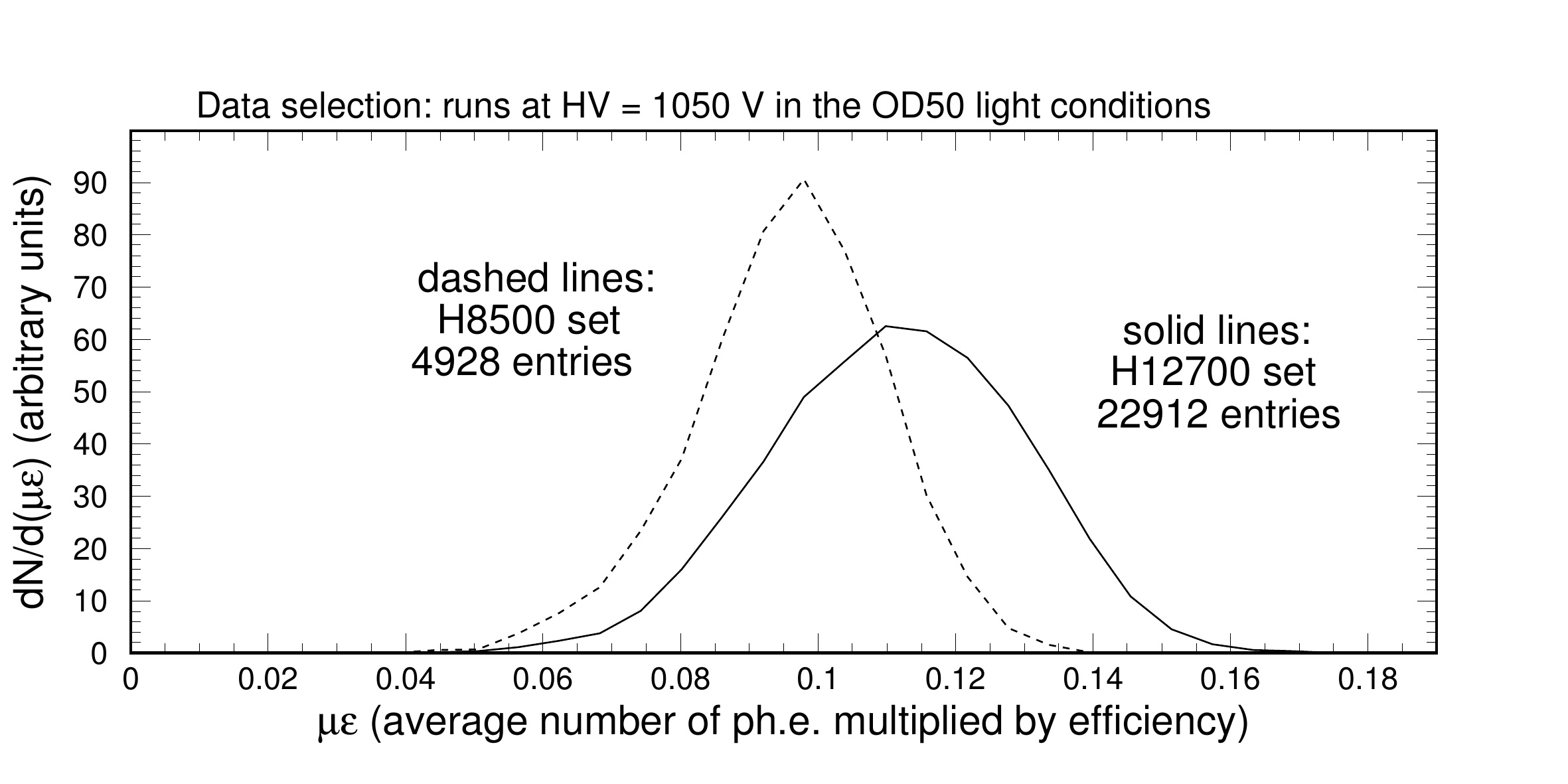}}\hfill
  \subfloat[Same as (g) for the HV settings at 1000~V and
    1100~V]{
    \includegraphics[clip=true,trim=0 10 0
      27,width=.48\textwidth,keepaspectratio] 
                    {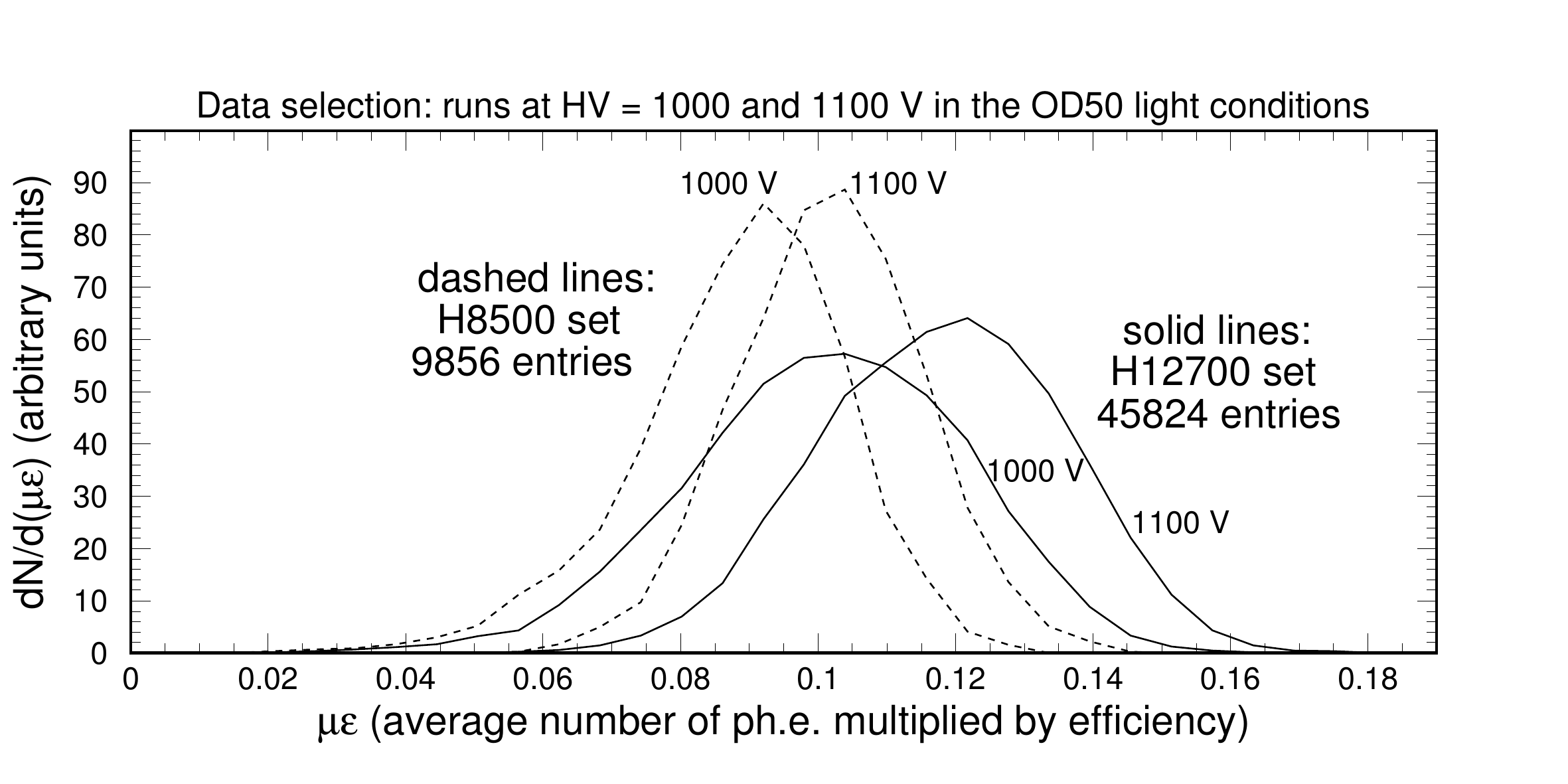}}
\caption{Distributions of $scale$ and $\mu$ parameters, and
  $\nu$ and $\mu \varepsilon$ derived values on the number of model
  parameterizations. Dashed lines show the H8500 data set, and solid lines
  the set of parameters for the H12700 MAPMTs. The distributions are
  normalized to equal areas in each plot. The selections of the
  evaluated parameter sets included in the distributions are indicated
  on top of the panels.}
\label{fig:smne}
\end{figure*}

Fig.~\ref{fig:smne} presents distributions of $scale$ and $\mu$
model parameters, and the derived values of $\nu$ and $\mu
\varepsilon$ for the analyzed data sets. Dashed lines show the H8500,
and solid lines the set of parameters for the H12700 MAPMTs. The
distributions are normalized to equal areas in the plots. The
selections of the parameter sets included in the distributions are
indicated on top of the panels.

Fig.~\ref{fig:smne}a shows the distributions of $scale$ parameter
measured for all anodes of H8500 and H12700 MAPMTs at 1050~V. As it
has already been illustrated in the panels (a) and (b) of
Fig.~\ref{fig:Passport}, the extracted $scale$ parameters do not
depend on the light conditions to a very high degree of accuracy.  The
distributions of the $scale$ parameter in Fig.~\ref{fig:smne}a
accumulated for the different light conditions are practically
identical and are superimposed on top of each other in the plot. While
the spread of the values is quite broad, the H8500 set exhibits
$scale$ values on average about 20\% larger than the H12700 set at the
same HV. Apparently, as compared to the H8500 MAPMT, the lower number
of the amplification stages in the H12700 devices is almost
compensated by the new design features allowing greater amplification
at each stage. This is further illustrated in Fig.~\ref{fig:smne}b
where the $scale$ parameter distributions are shown for HV = 1000~V
and HV = 1100~V, averaged over the light conditions, and plotted using
the logarithmic scale in abscissa to better see the similarities
between the distributions at different applied voltages.

Fig.~\ref{fig:smne}c shows the distributions of $\mu$ parameter
measured for all anodes of H8500 and H12700 MAPMTs at the intermediate
light condition ``OD50'', and all high voltages.  The distributions
indicate on a rather small ($<10$\%) difference in the photoefficiency
and/or photoelectron collection ability between the two types of
MAPMT, showing the slight advantage for the H12700 devices. It may
also be seen in the plot, that the evaluated parameters $\mu$
practically do not depend on HV applied. This observation illustrates
the good level of factorization between the $scale$ and $\mu$
parameters of the model.  The values of these parameters evaluated in
one set of the test conditions are applicable to the tests at
different HV and light. While the stability of the extracted $scale$
parameter is observed to be within the small statistical errors of
under 1\%, the distributions on $\mu$ may indicate on the presence of
a slight (1-2\%) dependence of $\mu$ on the applied HV.  However, this
small effect was difficult to evaluate and analyze in more
detail. Averaging over the sets of tests at different HV allowed us to
further illustrate the differences between the H8500 and H12700 data
in the distributions on $\mu$ measured at different light conditions,
presented in Fig.~\ref{fig:smne}d.

Fig.~\ref{fig:smne}, panels (e) and (f) are similar to panels (a) and
(b) in the same figure, but showing the derived value of the $\nu$
parameter for the same sets of conditions. According to the model, the
set of the SPE parameters of the photon detector do not depend on the
light conditions during the tests. This condition is taken into
account during the ``global fit'' procedure, leading to the $\nu$
independence of the light conditions. Thus, Figs.~\ref{fig:smne}(e,f)
illustrate the difference of the derived $\nu$ values between the H8500
and H12700 MAPMTs, and also its dependence on the HV applied. While
the HV-dependence is relatively week, the difference between the two
types of MAPMTs is quite dramatic, indicating that the average number
of the second-stage electrons knocked out of the first dynode is
almost twice as large in a H12700 MAPMT compared to H8500 in the same
conditions.

Panels (g) and (h) in Fig.~\ref{fig:smne} illustrate the comparison
between the H8500 and H12700 sets of MAPMTs in terms of their ultimate
efficiencies of detecting light. At the same signal thresholds in
channels ADC, H12700 MAPMTs have some advantage in the probability of
detecting light, in spite of generally smaller SPE signals (the
$scale$ parameter). The advantage is due to a somewhat larger photon
conversion efficiency (the $\mu$ parameter), and better shapes of the
SPE spectra with much larger $\nu$ value for H12700 devices,
corresponding also to a better collection efficiency for them.

\begin{figure}[hbt] 
 \centering 
  \includegraphics[clip=true,trim=0 10 0
  27,width=.50\textwidth,keepaspectratio] {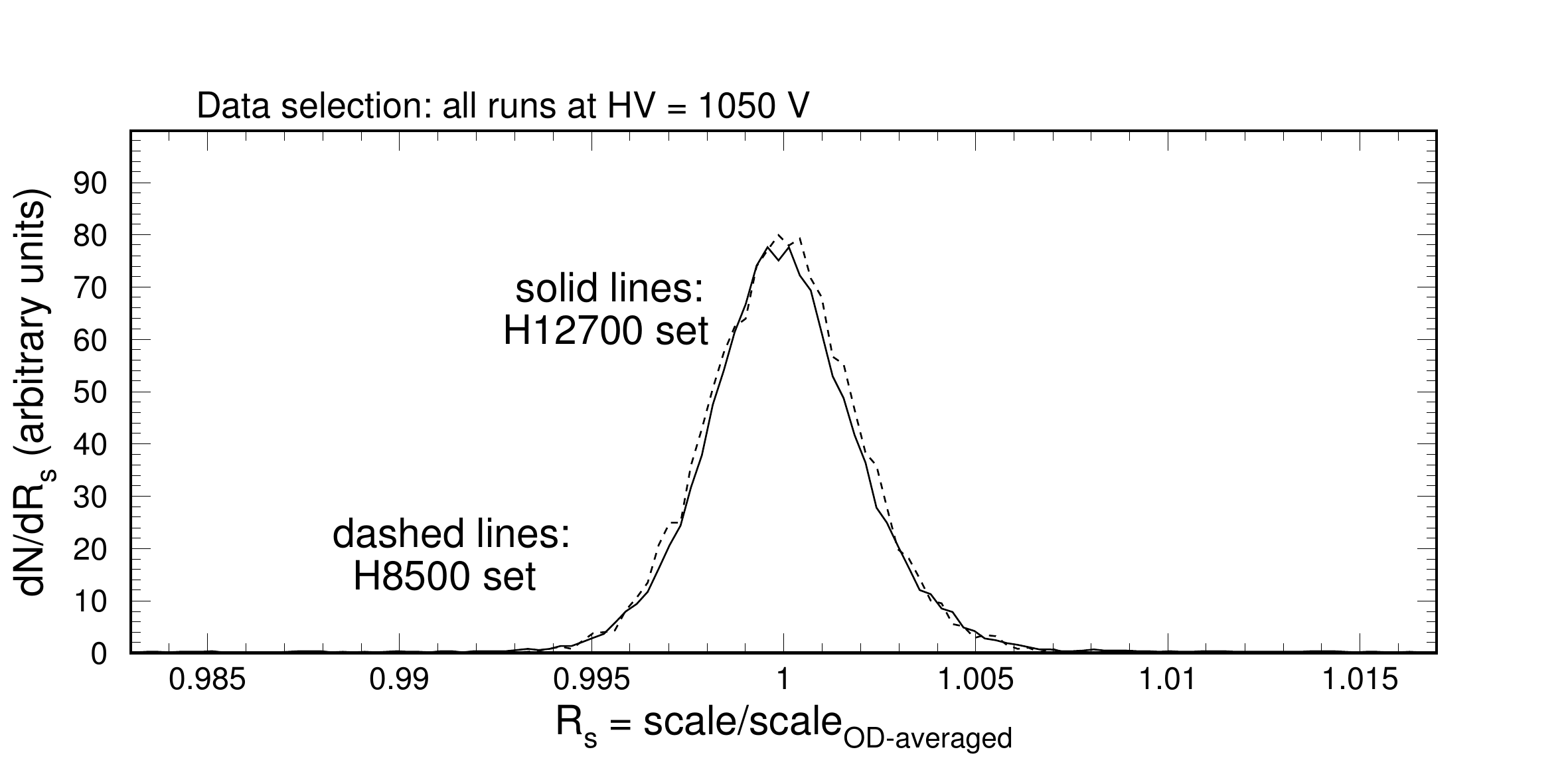}
\caption{Distributions of the stability evaluator for the $scale$
  parameter, corresponding to the relative statistical error in the
  extracted value of the $scale$ for the two MAPMT data sets. See text
  for details.}
\label{fig:Rs}
\end{figure}

\begin{figure}[hbt] 
 \centering 
  \includegraphics[clip=true,trim=0 10 0
  27,width=.50\textwidth,keepaspectratio] {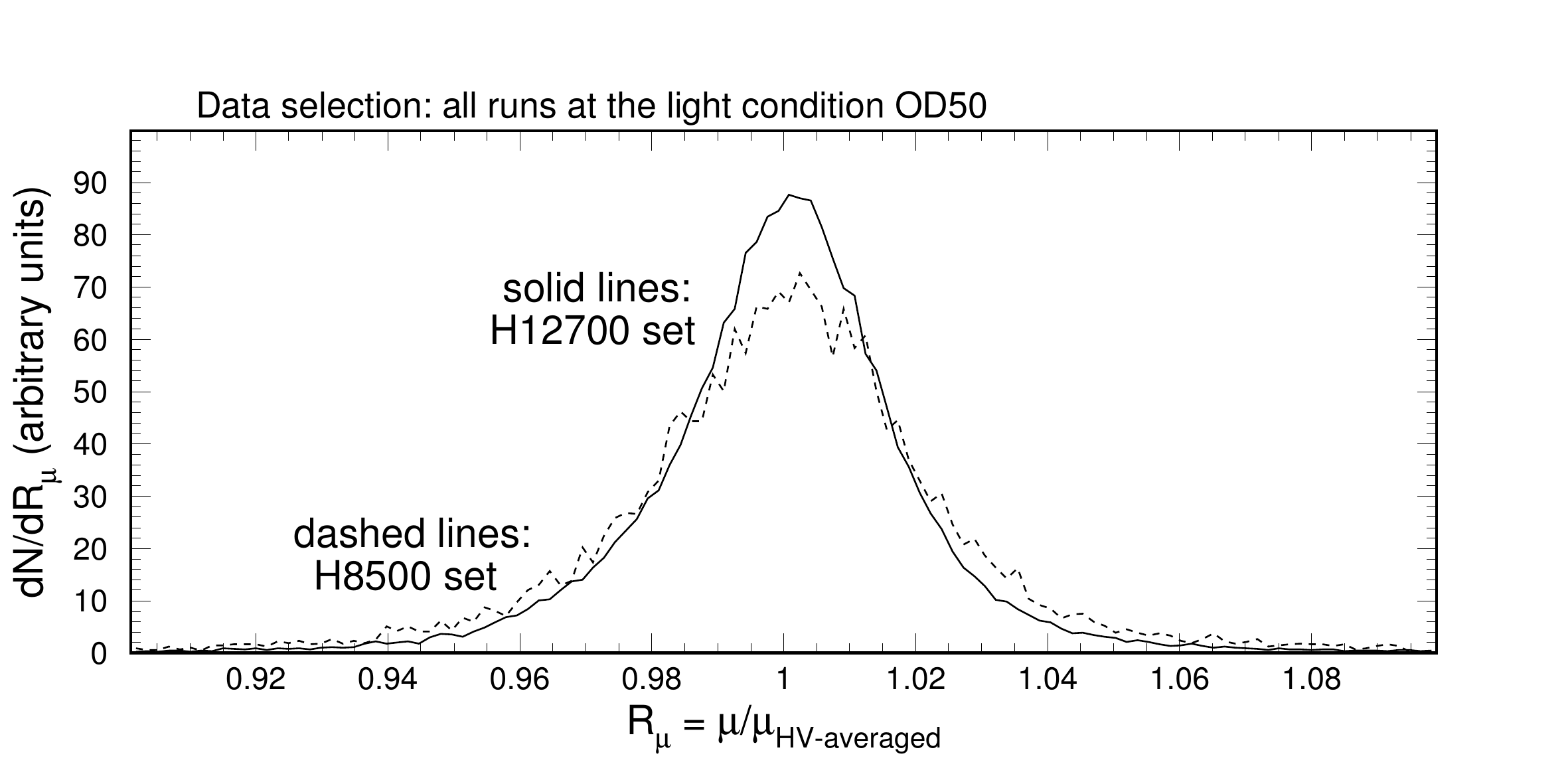}
\caption{Distributions of the stability evaluator for the $\mu$
  parameter, corresponding to the combination of the relative
  statistical error in the extracted value of the $\mu$, and the
  observed weak dependence of $\mu$ on the applied HV, for the two
  MAPMT data sets. See text for details.}
\label{fig:Rm}
\end{figure}

Fig.~\ref{fig:Rs} and Fig.~\ref{fig:Rm} illustrate the levels of
relative stability achieved in the evaluation of the major SPE
parameters $scale$ and $\mu$, by plotting the ratios of individually
evaluated parameters to the values of the same parameters averaged
over the measurements in different conditions, in which the model
ideally should give the same values ($scale$ measured in the three
light conditions in the case of Fig.~\ref{fig:Rs}, and $\mu$ measured
at four values of applied HV in the case of Fig.~\ref{fig:Rm}). While
the distribution of the stability evaluator for the $scale$ parameter
$R_s = scale /{\langle scale \rangle}_{\rm{OD-averaged}}$ is indeed
ultimately good (the spread is about 0.5\% FWHM), the corresponding
spread in the distribution of $R_{\mu} = \mu /{\langle \mu
  \rangle}_{\rm{HV-averaged}}$ is about 4\% FWHM. The latter
observation may indicate, apart from the statistical differences
between the parameters, to an additional weak dependence of the
average number of photoelectrons $\mu$ on the applied high voltage.

The extracted SPE characteristics for each anode in the whole
studied set of multianode photomultipliers were stored in a general
MAPMT parameter database. The accumulated data will facilitate and
improve the detector selection process, and will help to model the
detector response and efficiency. The SPE spectral functions extracted
in such analysis may serve as objective internal characteristics of
each photon detector (each anode of a MAPMT in this case) at an
abstract level, independent of the test conditions. For an extended
experimental setup, the set of such functions describing each detector
may be used to evaluate overall detector performance in current
working conditions that could be different from the test environment.

%
%
%
%

\section{Conclusion}

The new computational model for description of the photomultiplier
response functions has been developed, implemented, and tested in real
applications.  Important features of the model include the ability to
approximate the true single-photoelectron spectra from different
photomultiplier tubes with a variety of parameterized spectral shapes,
reflecting the variability in the design and in the individual
parameters of the detectors.  The new techniques were developed in the
process of building the model, such as the method of decomposition of
the SPE spectra into a series of elementary Poisson probability
density functions, and the use of convolution algebra to build the
multi-photoelectron amplitude distributions describing measured
spectra.

The ``predictive power'' of the model has been tested by demonstrating
that the SPE spectral parameters, obtained in the real measurements,
may describe well the amplitude distributions measured at different
levels of irradiation of the same photodetector. Thus, the model
allowed us to extract the characteristic parameters of the devices
independently of the test measurement conditions. In that way the set
of parameters obtained in one or several test runs at certain running
conditions could serve to obtain predicted detector response and
efficiency for a wider set of running conditions, for a varying level
of light during the real runs, and/or for a different amplitude
resolution of the measurement system.

The SPE spectral parameterization information in experimental physics
or industrial photon detector setups may be utilized to make an
educated selection of the devices that would work best for a
particular purpose, make choices for the characteristics of the
readout electronics necessary for a particular setup, and create new
software tools simulating expected behavior of the photon detectors in
real installations for use in the data analysis procedures.

%
%
%
%

\section*{Acknowledgements}

This material is based upon work supported by the U.S. Department of
Energy, Office of Science, Office of Nuclear Physics under contract
DE-AC05-06OR23177.  The author thanks Alex Vlassov, Valery Kubarovsky,
and Bogdan Wojtsekhowski for their stimulating interest in the work
and useful suggestions.  The work wouldn't be possible without the
data sets accumulated by different experimental groups and kindly
presented to me for analysis by their Authors: Valery Kubarovsky,
Andrey Kim, Simona Malace, Brad Sawatzky, Hakob Voskanyan, Youri
Sharabian, Will Phelps and Rachel Montgomery.








%
%
%
%


\end{document}